\documentclass[12pt]{amsart}
\usepackage{ifthen,amsmath,amssymb,latexsym,epsf,amsthm,amscd,times,diagrams,graphicx,color}
\usepackage[nobysame]{amsrefs}
\newboolean{draft}
\setboolean{draft}{false}
\ifthenelse{\boolean{draft}}{\usepackage{showkeys}}{}

\theoremstyle{SmallCaps}
\newtheorem{dfn}{Definition}[section]
\newtheorem{cor}{Corollary}[section]
\newtheorem{lem}{Lemma}[section]
\newtheorem{pro}{Proposition}[section]

\newtheorem{thm}{Theorem}[section]
\newcommand{\dref}[1]{Definition~\ref{#1}}
\newcommand{\cref}[1]{Corollary~\ref{#1}}
\newcommand{\lref}[1]{Lemma~\ref{#1}}
\newcommand{\pref}[1]{Proposition~\ref{#1}}

\newcommand{\tref}[1]{Theorem~\ref{#1}}
\def\please{\renewcommand\labelstyle{\scriptstyle}}
\def\R{\mathbb{R}}
\def\C{\mathbb{C}}
\def\Z{\mathbb{Z}}
\def\N{\mathbb{N}}
\def\NN{\mathcal{N}}

\def\P{\mathbb{P}}

\newcommand{\onatop}[2]{\genfrac{}{}{0pt}{}{#1}{#2}}

\def\configspace{\;\raisebox{-2.0cm}{\epsfysize=8.0cm\epsfbox{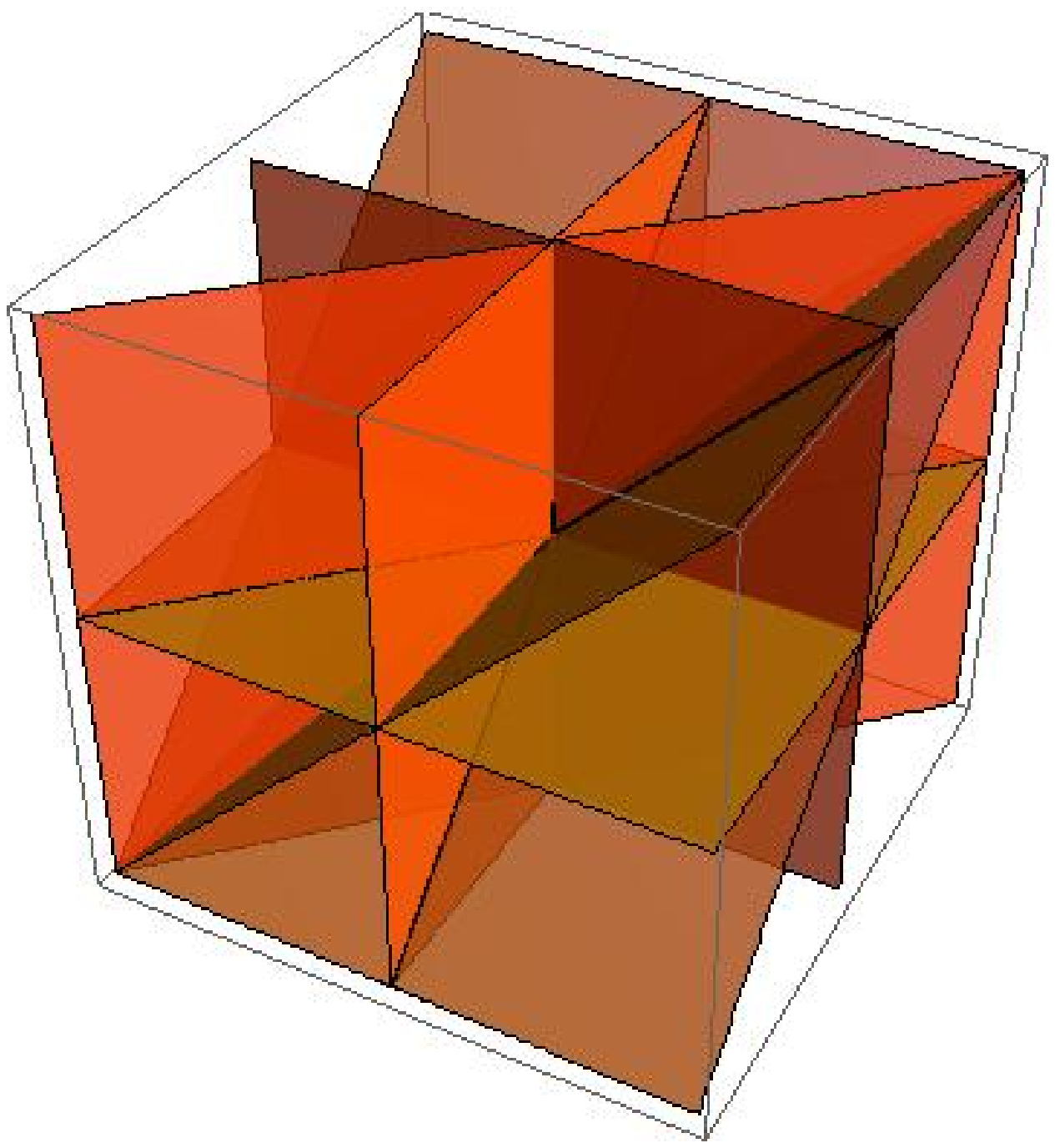}}\;}
\def\fmone{\;\raisebox{-2.0cm}{\epsfysize=8.0cm\epsfbox{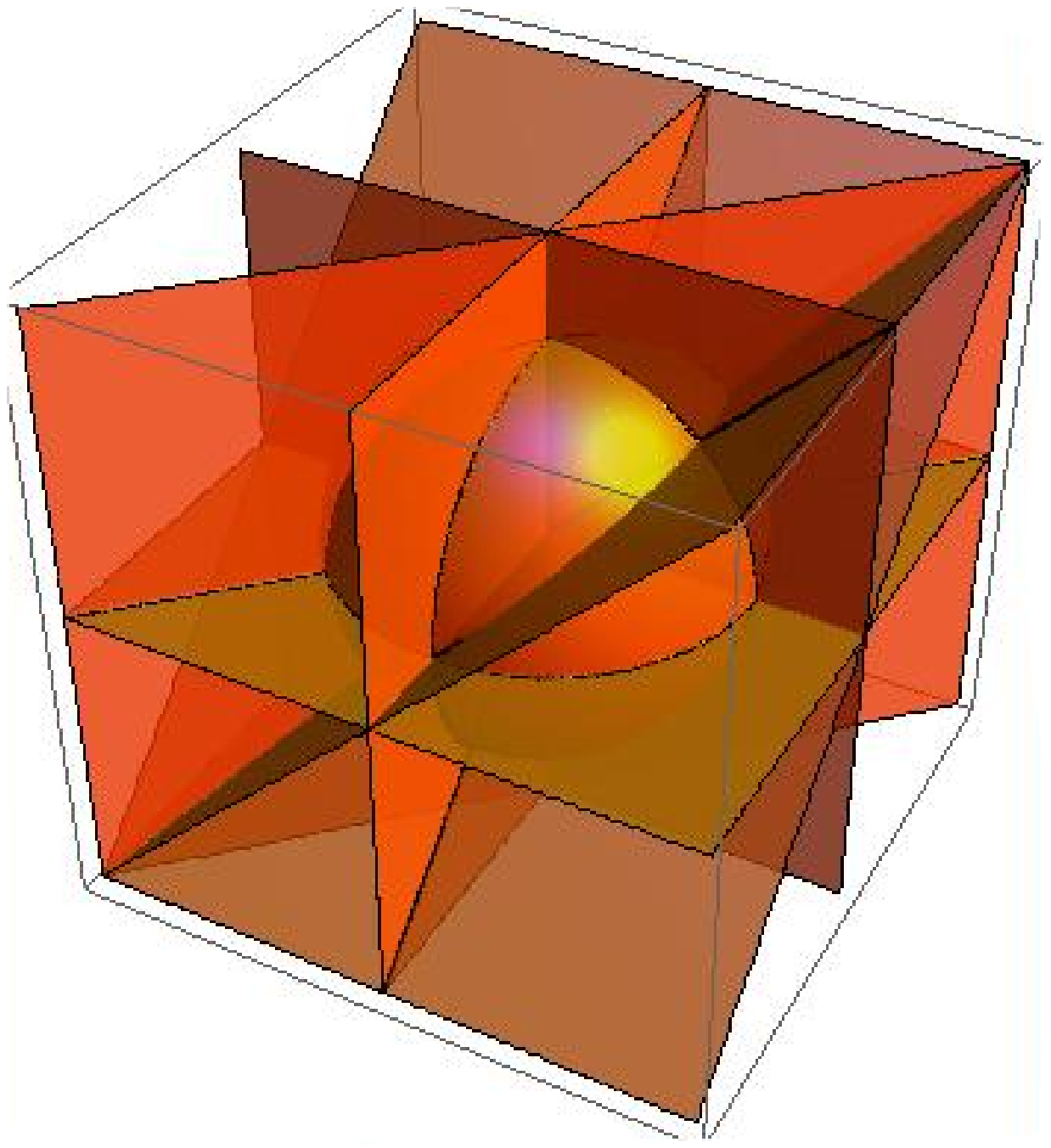}}\;}
\def\fmtwo{\;\raisebox{-2.0cm}{\epsfysize=8.0cm\epsfbox{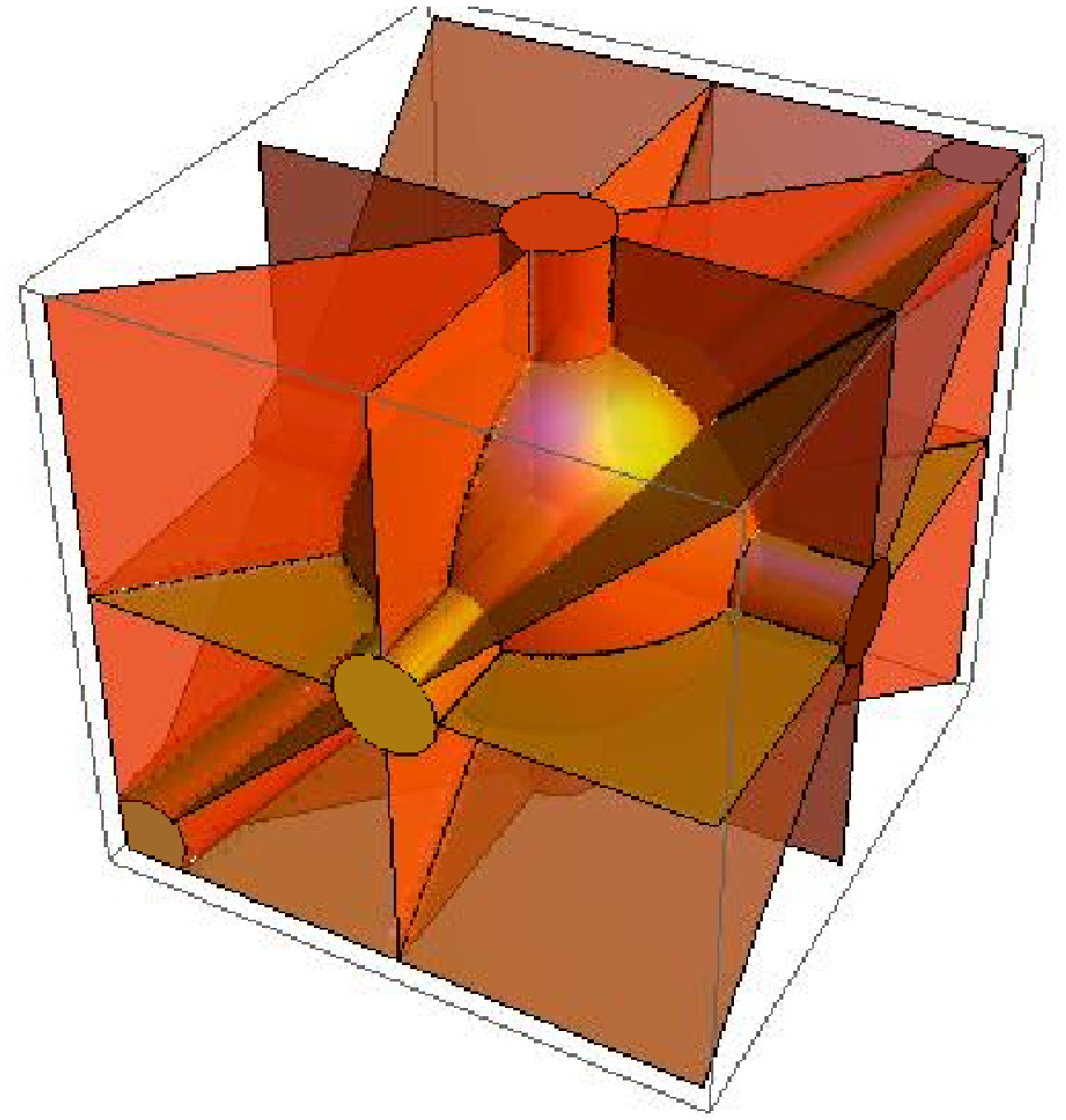}}\;}
\def\gammaone{\;\raisebox{-0.7cm}{\epsfysize=1.8cm\epsfbox{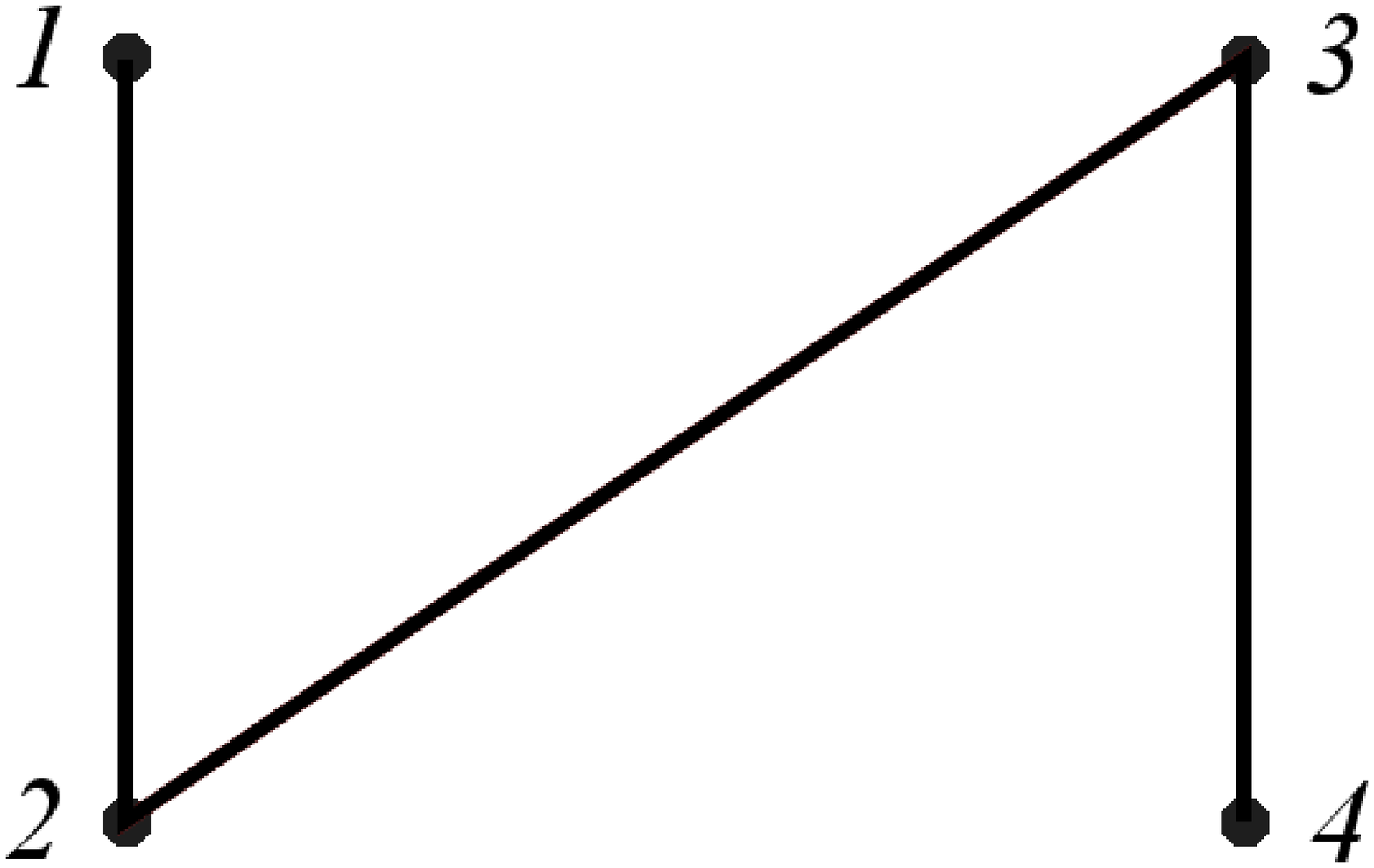}}\;}
\def\gammatwo{\;\raisebox{-0.8cm}{\epsfysize=2cm\epsfbox{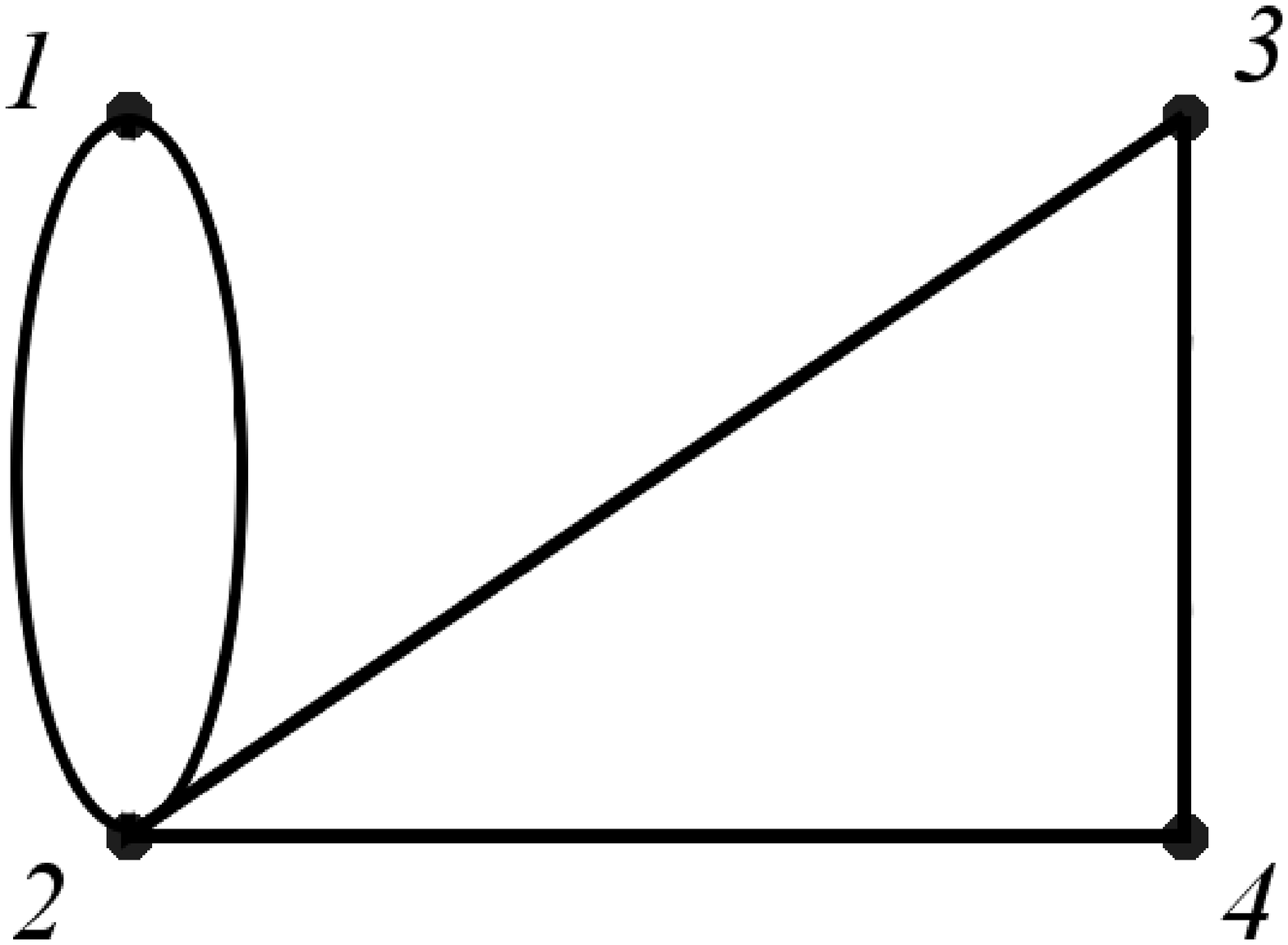}}\;}
\def\gammathree{\;\raisebox{-0.8cm}{\epsfysize=2cm\epsfbox{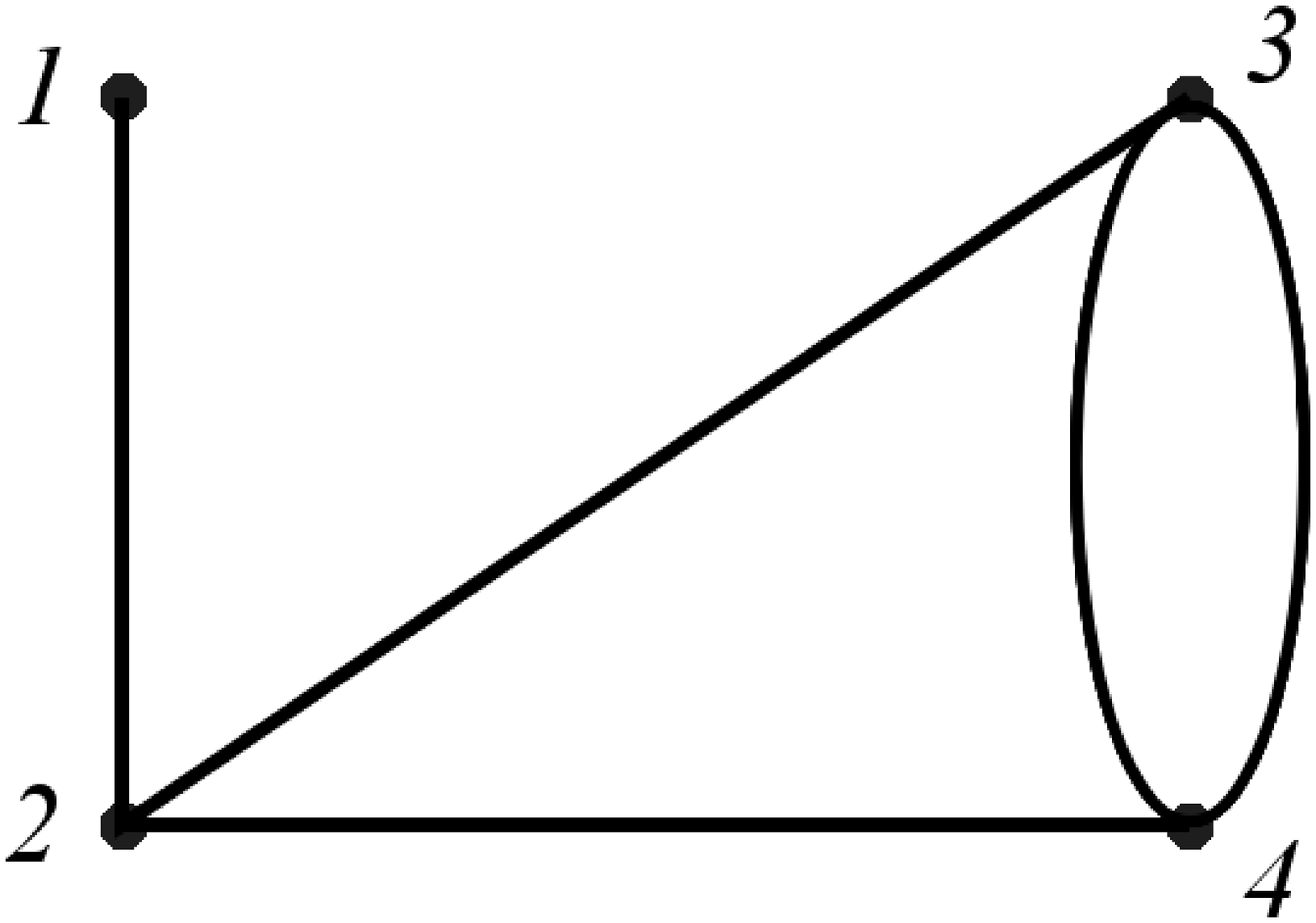}}\;}
\def\gammafour{\;\raisebox{-0.8cm}{\epsfysize=2cm\epsfbox{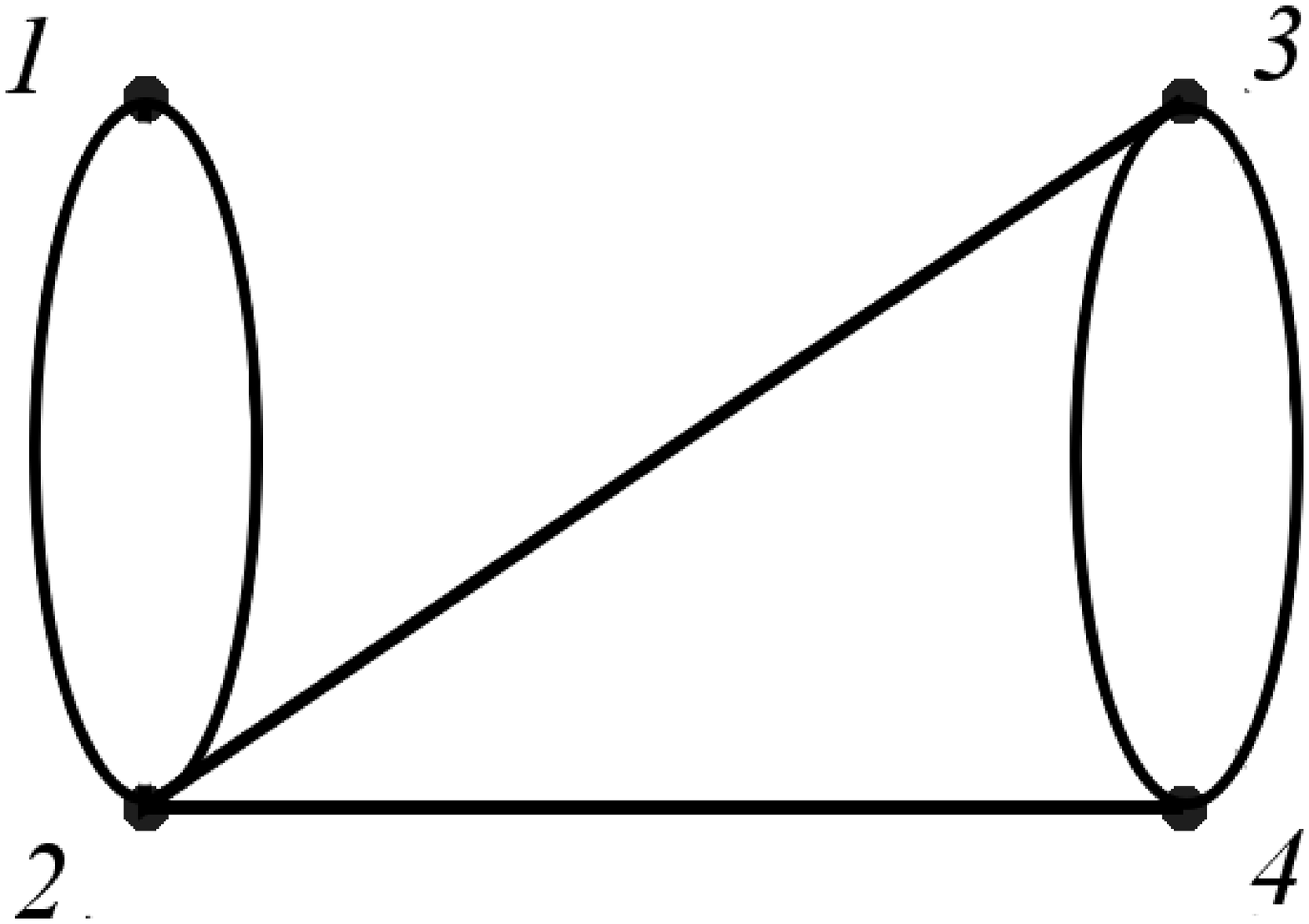}}\;}
\def\gammafive{\;\raisebox{-0.75cm}{\epsfysize=1.9cm\epsfbox{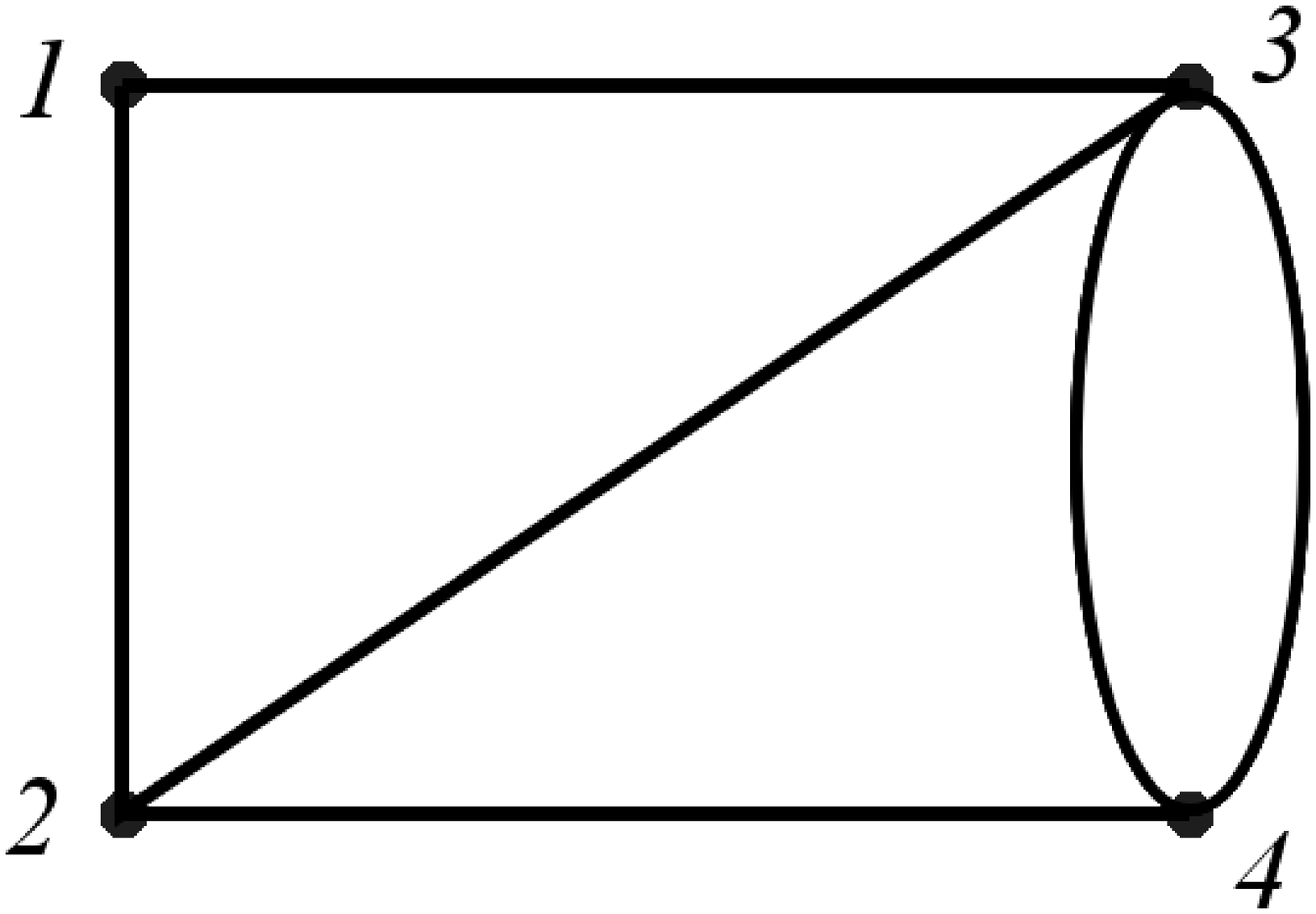}}\;}
\def\gammasix{\;\raisebox{-0.8cm}{\epsfysize=1.8cm\epsfbox{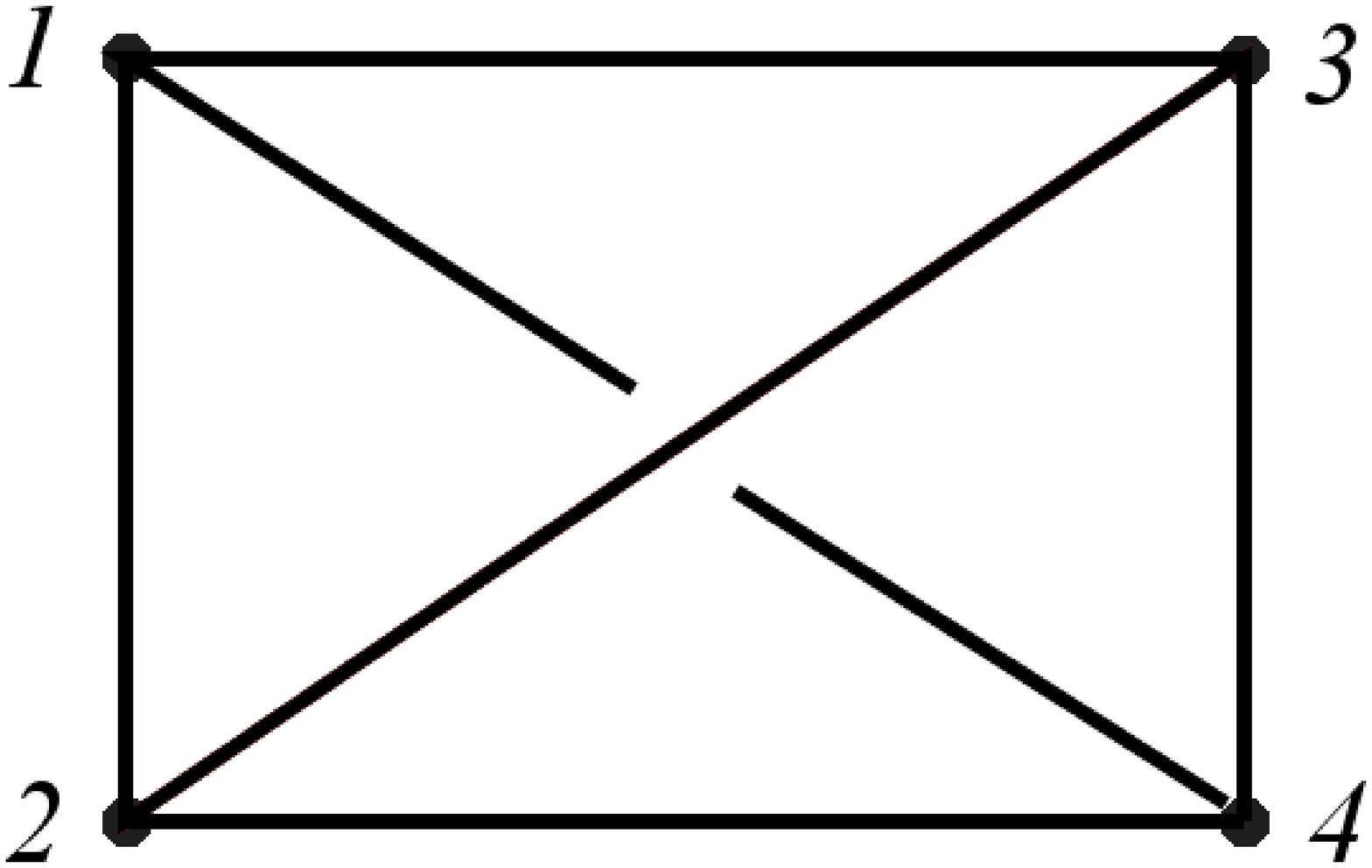}}\;}
\def\gammadisj{\;\raisebox{-0.8cm}{\epsfysize=2cm\epsfbox{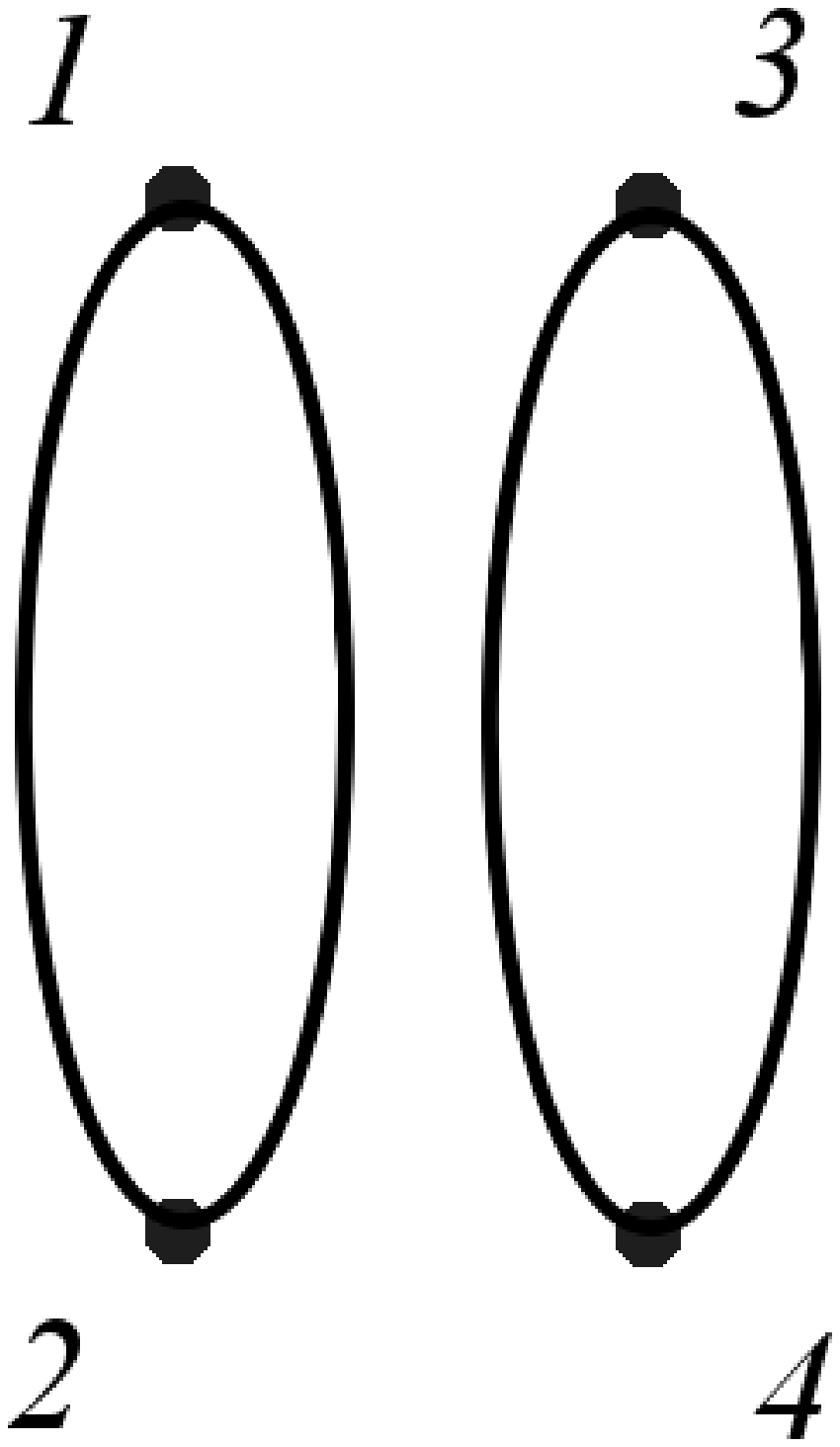}}\;}

\def\gammareduzwei{\;\raisebox{-0.8cm}{\epsfysize=2cm\epsfbox{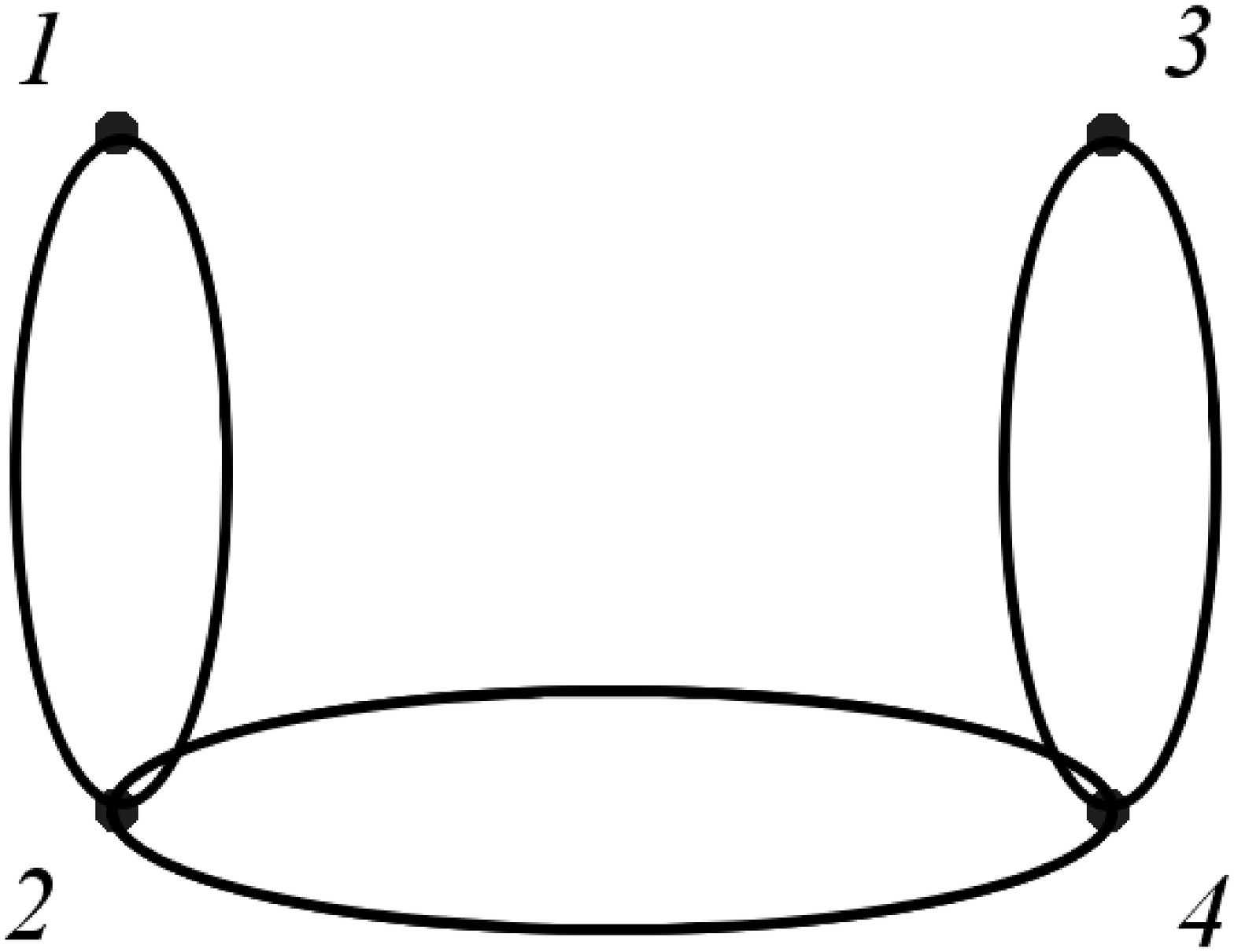}}\;}
\def\gammaeiner{\;\raisebox{-0.8cm}{\epsfysize=2cm\epsfbox{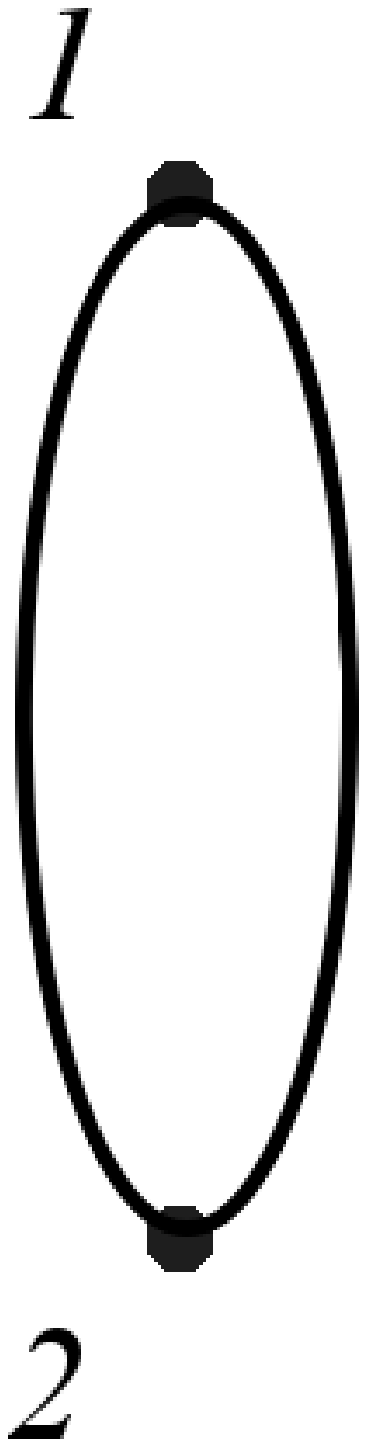}}\;}
\def\pfade{\;\raisebox{-1.0cm}{\epsfysize=3.5cm\epsfbox{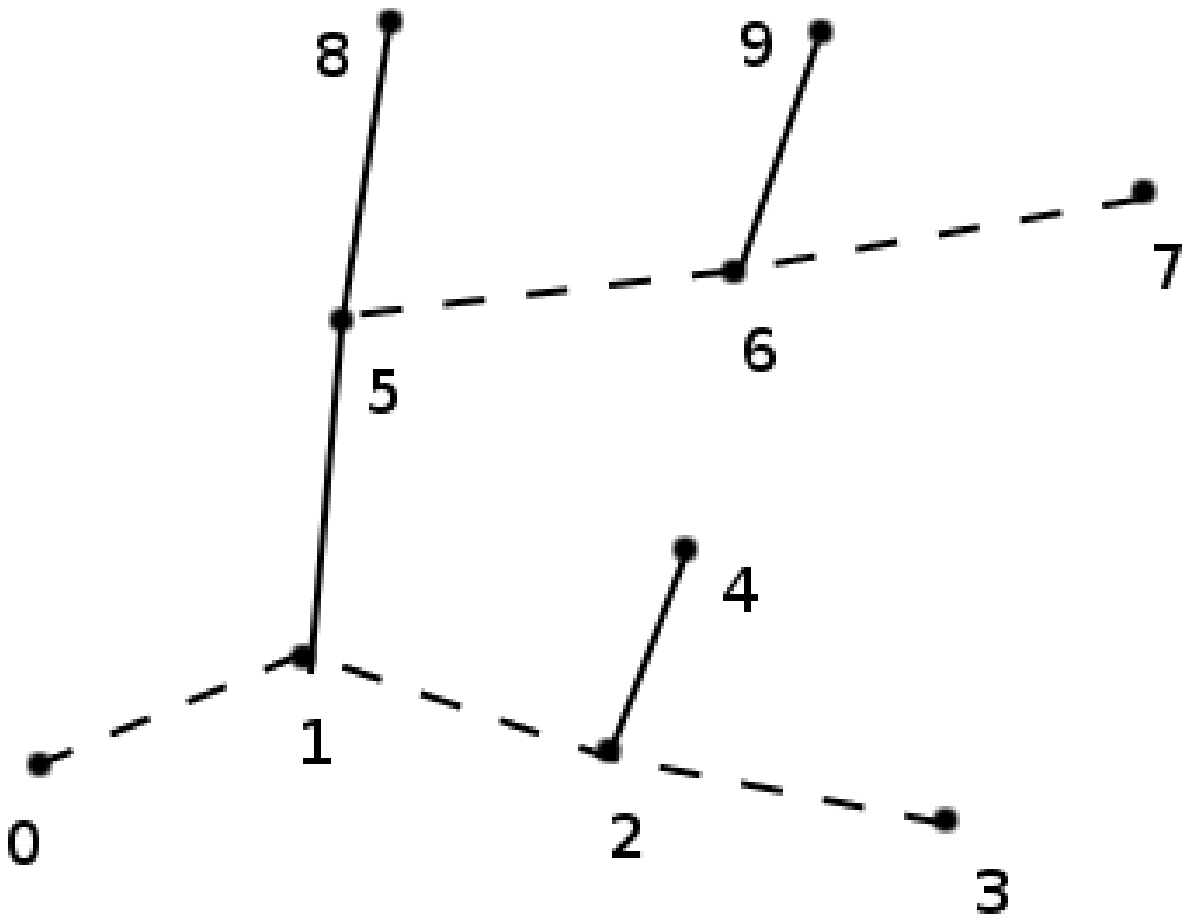}}\;}

\begin{document}
\title{Renormalization and resolution of singularities}
\author{C.~Bergbauer, R.~Brunetti and D.~Kreimer}


\address{Freie Universit\"at Berlin\\Institut f\"ur Mathematik\newline New address: SFB 45\\Universit\"at Mainz\\Institut f\"ur Mathematik}
\email{bergbau@math.fu-berlin.de}

\address{Dipartimento di Matematica\\ Universit\`a di Trento}
\email{brunetti@science.unitn.it}

\address{IHES (CNRS) and Boston University Center for Mathematical Physics}
\email{kreimer@ihes.fr}

\ifthenelse{\boolean{draft}}{\date{\today\, (\bf{DRAFT VERSION: PLEASE DO NOT CIRCULATE})}}{\date{}}
\begin{abstract} Since the seminal work of Epstein and Glaser it is well established
that perturbative renormalization of ultraviolet divergences in position space amounts
to extension of distributions onto diagonals. For a general Feynman graph the relevant diagonals form a nontrivial arrangement of linear subspaces. One may therefore ask if renormalization
becomes simpler if one resolves this arrangement to a normal crossing divisor. In this
paper we study the extension problem of distributions onto the wonderful models of de Concini and
Procesi, which generalize the Fulton-MacPherson compactification of configuration spaces. We show
that a canonical extension onto the smooth model coincides with the usual Epstein-Glaser
renormalization. To this end we use an analytic regularization for position
space. The 't Hooft identities relating the pole coefficients may be recovered from the stratification, and
Zimmermann's forest formula is encoded in the geometry of the compactification. Consequently one subtraction
along each irreducible component of the divisor suffices to get a finite result using local
counterterms. As a corollary, we identify the Hopf algebra of at most logarithmic Feynman graphs in
position space, and discuss the case of higher degree of divergence.
\end{abstract}
\maketitle
\setcounter{tocdepth}{1} \tableofcontents
\section{Introduction}\label{s:intro}
The subject of perturbative renormalization in four-dimensional interacting quantum field
theories looks back to a successful history. Thanks to the achievements of Bogoliubov,
Hepp, Zimmermann, Epstein, Glaser, 't Hooft, Veltman, Polchinski, Wilson -- to mention
just some of the most prominent contributors --, the concept seems in principle well-understood;
and the predictions made using the renormalized perturbative expansion match
the physics observed in the accelerators with tremendous accuracy. However, several decades
later, our understanding of realistic interacting
quantum field theories is still everything but satisfying. Not only is it extremely
difficult to perform computations beyond the very lowest orders, but also the transition
to a non-perturbative framework and the incorporation of gravity pose enormous conceptual
challenges.\\\\
Over the past fifteen years, progress has been made, among others, in the following
three directions. In the algebraic approach to quantum field theory, perturbation theory
was generalized to generic (curved) space-times by one of the authors and Fredenhagen
\cite{BF}, see also \cite{HollandsWald}. On the other hand, Connes and one of the authors introduced
infinite-dimensional Hopf- and Lie algebras \cite{Kreimer,CK2} providing a deeper conceptual understanding of the
combinatorial and algebraic aspects of renormalization, also beyond perturbation theory.
More recently, a conjecture concerning the appearance of a very special class of periods
\cite{Broadhurst-Kreimer-96-2,BroadKreimer3,BB} in all Feynman integrals computed so far, has initiated a
new area of research \cite{BEK,BlochJapan,BlKr} which studies the perturbative expansion from a
motivic point of view. The main purpose of this paper is to contribute to the three approaches mentioned,
by giving a description of perturbative renormalization of short-distance divergences
using a resolution of singularities. For future applications to curved spacetimes it is
most appropriate to do this in the position space framework of Epstein and Glaser \cite{EG,BF}. However
the combinatorial features of the resolution allow for a convenient transition to
the momentum space picture of the Connes-Kreimer Hopf algebras, and to the residues of \cite{BEK,BlKr}
in the parametric representation. Both notions are not immediately obvious in the original Epstein-Glaser
literature. \\\\
Let us present some of the basic ideas in a nutshell. Consider,
in euclidean space-time $M=\R^4,$ the following Feynman graph
\begin{equation*}
\Gamma = \gammaeiner.
\end{equation*}
The Feynman rules, in position space, associate to $\Gamma$ a
distribution
\begin{equation*}
u_\Gamma(x_1,x_2)= u_0^2(x_1-x_2).
\end{equation*}
where $u_0(x)$ is the Feynman propagator, in the massless case
$u_0(x)=1/x^2,$ the $x$ are 4-vectors with coordinates
$x^0,\ldots,x^3,$ and $x^2$ the euclidean square
$x^2=(x^0)^2+\ldots+(x^3)^2.$ Note that since $u_\Gamma$
depends only on the difference vector $x_1-x_2,$ we may equally
well consider $\underline{u}_\Gamma(x)=u_\Gamma(x,0).$ Because
of the singular nature of $u_0$ at $x=0,$ the distribution
$u_\Gamma$ is only well-defined outside of the diagonal
$D_{12}=\{x_1=x_2\}\subset M^2.$ In order to extend $u_\Gamma$
from being a distribution on $M^2-D_{12}$ onto all of $M^2,$ one can
introduce an analytic regularization, say
\begin{equation*}
\underline{u}_\Gamma^s(x) = u_0^{2s}(x).
\end{equation*}
Viewing this as a Laurent series in $s,$ we find, in this
simple case,
\begin{eqnarray*}
\underline{u}_\Gamma^s(x) &=& \frac{1}{x^{4s}}= \frac{c\delta_0(x)}{s-1}+ R_s(x)
\end{eqnarray*}
with $c\in\R,$ $\delta_0$ the Dirac measure at $0,$ and
$s\mapsto R_s$ a distribution-valued function holomorphic in a
complex neighborhood of $s=1,$ the important point being that
the distribution $R_s$ is defined \emph{everywhere} on $M^2.$
The usual way of renormalizing $\underline{u}_\Gamma$ is to
subtract from it a distribution which is equally singular at
$x=0$ and cancels the pole, for example
\begin{equation*}
\underline{u}_{\Gamma,R} = \left.\left(\underline{u}^s_\Gamma - \underline{u}^s_\Gamma [w_0] \delta_0 \right)\right|_{s=1}.
\end{equation*}
Here $w_0$ is any test function which satisfies $w_0(0)=1$ for
then $\frac{\delta_0}{s-1}[w_0]=\frac{1}{s-1}.$ Consequently
\begin{equation*}
\underline{u}_{\Gamma,R} = R_{1}-R_{1}[w_0]\delta_0
\end{equation*}
which is well-defined also at $0.$ The distribution
$\underline{u}_{\Gamma,R}$ is considered the solution to
the renormalization problem for $\Gamma,$ and different choices
of $w_0$ give rise to the renormalization group. Once the graph
$\Gamma$ is renormalized, there is a canonical way to
renormalize the graph
\begin{equation*}
\Gamma'=\gammadisj
\end{equation*}
which is simply a disjoint union of two copies of $\Gamma$.
Indeed,
\begin{equation*}
u_{\Gamma'}(x_1,x_2,x_3,x_4)= u_0^2(x_1-x_2) u_0^2(x_3-x_4)= (\underline{u}_\Gamma\otimes \underline{u}_\Gamma)(x_1-x_2,x_3-x_4).
\end{equation*}
In other words, $u_{\Gamma'}$ is a cartesian product, and one
simply renormalizes each factor of it separately:
$(u_{\Gamma',R})(x_1,\ldots,x_4)=
\underline{u}_{\Gamma,R}^{\otimes 2}(x_1-x_2,x_3-x_4).$ This
works not only for disconnected graphs but for instance also
for
\begin{equation*}
\Gamma''=\gammareduzwei
\end{equation*}
which is connected but (one-vertex-) reducible, to be defined
later. Indeed,
\begin{eqnarray*}
u_{\Gamma''}(x_1,x_2,x_3,x_4) &=& u_0^2(x_1-x_2) u_0^2(x_2-x_4)u_0^2(x_3-x_4)\\
&=&\underline{u}_\Gamma^{\otimes 3}(x_1-x_2,x_2-x_4,x_3-x_4)
\end{eqnarray*}
Again, one simply renormalizes every factor of $u_{\Gamma''}$ on
its respective diagonal. This is possible because the diagonals
$D_{12},$ $D_{24}$ and $D_{34}$ are pairwise perpendicular in
$M^4.$
Consider now a graph which is not of this kind:
\begin{equation*}
\Gamma'''=\gammafive
\end{equation*}
\begin{equation*}
u_{\Gamma'''}(x_1,\ldots,x_4)= u_0(x_1-x_2)u_0(x_1-x_3)u_0(x_2-x_3)u_0(x_2-x_4)u_0^2(x_3-x_4).
\end{equation*}
By the usual power counting we see that $u_{\Gamma'''}$ has
non-integrable singularities at $D_{34}=\{x_3=x_4\},$ at
$D_{234}=\{x_2=x_3=x_4\}$ and at
$D_{1234}=\{x_1=x_2=x_3=x_4\}.$ These three linear subspaces of
$M^4$ are nested $(D_{1234}\subset D_{234} \subset D_{34})$
instead of pairwise perpendicular. In the geometry of $M^4$ it
does not seem possible to perform the three necessary
subtractions separately and independently one of another. For
if a test function has support on some of say $D_{1234},$ its
support intersects also $D_{234}\setminus D_{1234}$ and
$D_{34}\setminus D_{234}.$ This is one of the reasons why much
literature on renormalization is based on recursive or
step-by-step methods.
If one instead transforms $M^4$ to another smooth manifold
$\beta: Y\rightarrow M^4$ such that the preimages under $\beta$
of the three linear spaces $D_{34},D_{234},D_{1234}$ look
locally like an intersection of three cartesian coordinate hyperplanes $y_1y_2y_3=0,$
one can again perform the three renormalizations separately,
and project the result back down to $M^4.$ For this procedure
there is no recursive recipe needed -- the geometry of $Y$
encodes all the combinatorial information. The result is the
same as from the Epstein-Glaser, BPHZ or Hopf algebra methods,
and much of our approach just a careful geometric rediscovery of existing ideas.\\\\
In section \ref{s:arrangements} the two subspace arrangements associated
to a Feynman graph are defined, describing the locus of
singularities, and the locus of non-integrable singularities,
respectively. In section \ref{s:propagators} an analytic
regularization for the propagator is introduced. Some necessary
technical prerequisites for dealing with distributions and
birational transformations are made, and the important notion
of residue density for a primitive graph is defined. The rest
of the paper is devoted to a more systematic development.
Section \ref{s:models} describes the De Concini-Procesi
''wonderful'' models for the subspace arrangements and provides
an explicit atlas and stratification for them in terms of
nested sets. Different models are obtained by varying the
so-called building set, and we are especially interested in the
minimal and maximal building set/model in this class. Section
\ref{s:poles} examines the pullback of the regularized Feynman
distribution onto the smooth model and studies relations
between its Laurent coefficients with respect to the regulator. In
section \ref{s:renormalization} it is shown that the proposed
renormalization on the smooth model satisfies the physical
constraint of locality: the subtractions made can be packaged
as local counterterms into the Lagrangian. For the model
constructed from the minimal building set, this is satisfied by
construction. From the geometric features
of the smooth models one arrives quickly at an analogy with the
Hopf algebras of Feynman graphs, and a section relating the
two approaches concludes the exposition.
As a technical simplication in the main part of the paper only
massless scalar euclidean theories are considered, and only
Feynman graphs with at most logarithmic singularities. The
general case is briefly discussed in section \ref{ss:quadratic}.
Questions of renormalization conditions, and the renormalization
group, are left for future research.\\\\
This research is motivated by a careful analysis of Atiyah's
paper \cite{Atiyah} -- see also \cite{BG}; and \cite{BB2} for a
first application to Feynman integrals in the parametric
representation -- the similarity of the Fulton-MacPherson
stratification with the Hopf algebras of perturbative
renormalization observed in \cite{Kreimer3,Diplom}, and recent
results on residues of primitive graphs and periods of mixed
Hodge structures \cite{BlochJapan,BEK}. Kontsevich has
pointed out the relevance of the Fulton-MacPherson compactification
for renormalization long ago \cite{Kontsevichtalk}, and a real
(spherical) version had been independently developed by him
(and again independently by Axelrod and Singer \cite{AS}) in
the context of Chern-Simons theory, see for example
\cite{KontsevichEMS}. In the parametric representation, many
related results have been obtained independently in the recent
paper \cite{BlKr}, which provides also a description of
renormalization in terms of limiting mixed Hodge structures.
That is beyond our scope. \\\\
An earlier version of this paper has been presented in \cite{Bergbauerthesis}.
\subsubsection*{Acknowledgements} All three authors thank the Erwin Schr\"odinger Institut
for hospitality during a stay in March and April 2007, where this paper has been initially
conceived. C.~B.~ is grateful to K.~ Fredenhagen, H.~
Epstein, J.~ Gracey, H.~ Hauser, F.~ Vignes-Tourneret, S.~ Hollands,
F.~ Brown, E.~ Vogt, C.~ Lange, S.~ Rosenberg, S.~ M\"uller-Stach, S.~ Bloch and H.~ Gangl
for general discussion.  C.~B.~ is funded by the Deutsche Forschungsgemeinschaft, first under
grant VO 1272/1-1 and presently in the SFB TR 45. He was supported by a Junior Fellowship while at ESI. 
He thanks the IHES, the Max-Planck-Institut
Bonn and Boston University for hospitality and support during several stays in 2007 and 2008.
D.~K.~ is supported by CNRS and in parts by NSF grant DMS-0603781. C.~B.~ and D.~K.~
thank the ESI for hospitality during another stay in March 2009.
\section{Subspace arrangements associated to Feynman graphs}\label{s:arrangements}
Let $U\subseteq \R^k$ be an open set. By $\mathcal{D}(U)=C^\infty_0(U)$ we denote the
space of test functions with compact support in $U,$ with the usual topology.
$\mathcal{D}'(U)$ is the space of distributions in $U.$ See \cite{Hoermander} for a
general reference on distributions. We work in Euclidean
spacetime $M=\R^d$ where $d\in 2+2\N=\{4,6,8,\ldots\}$ and
use the (massless) propagator distribution
\begin{equation}\label{eq:dfnu0}
u_0(x)=\frac{1}{x^{d-2}} = \frac{1}{((x^0)^2+\ldots+(x^{d-1})^2)^\frac{d-2}{2}}
\end{equation}
which has the properties
\begin{equation}\label{eq:homogenu0}
u_0(\lambda x ) = \lambda^{2-d} u_0(x), \quad\lambda\in \R\setminus \{0\}
\end{equation}
and
\begin{equation}\label{eq:singsuppu0}
\operatorname{sing}\operatorname{supp} u_0 = \{0\}.
\end{equation}
The singular support of a distribution $u$ is the set of points
having no open
neighborhood where $u$ is given by a $C^\infty$ function. \\\\
Let now $\Gamma$ be a Feynman graph, that is a finite graph,
with set of vertices $V(\Gamma)$ and set of edges $E(\Gamma).$
We assume that $\Gamma$ has no loops (a loop is an edge that
connects to one and the same vertex at both ends). The Feynman
distribution is given by the distribution
\begin{equation}\label{eq:FGD}
u_\Gamma (x_1,\ldots,x_n) = \prod_{i<j} u_0(x_i-x_j)^{n_{ij}}
\end{equation}
on $M^n\setminus \cup_{i<j}D_{ij}$ where $D_{ij}$ is the diagonal
defined by $x_i=x_j$ and $n_{ij}$ is the number of edges
between the vertices $i$ and $j$ (For this equation we assume
that the vertices are numbered $V(\Gamma)=\{1,\ldots,n\}).$
A basic observation is that $u_\Gamma$ may be rewritten as the
restriction of the distribution $u_0^{\otimes |E(\Gamma)|}\in
\mathcal{D'}(M^{|E(\Gamma)|})$ to the complement of a subspace
arrangement, contained in a vector subspace of $M^{|E(\Gamma)|},$ as follows.
\subsection{Configurations and subspace arrangements of singularities}\label{ss:singarrangements}
It is convenient to adopt a more abstract point of view as in
\cite{BEK}. Let $E$ be a finite set and
$\R^E$ the real vector space spanned by $E.$ If $V$ is a vector space we write
$V^\vee$ for its dual space. Similarly, if $x\in V$ we write $x^\vee$ for the dual linear form.
An inclusion of a
linear subspace $i_W: W\hookrightarrow \R^E$ is called a
\emph{configuration}. Since $\R^E$ comes with a canonical basis,
a configuration defines an arrangement of up to $|E|$ linear
hyperplanes in $W:$ namely for each $e\in E$ the subspace
annihilated by the linear form $e^\vee i_W,$ unless this linear
form equals
zero. Note that different basis vectors $e\in E$ may give one and the same hyperplane.\\\\
Given a connected graph $\Gamma,$ temporarily impose an
orientation of the edges (all results will be
independent of this orientation). This defines for a vertex
$v\in V(\Gamma)$ and an edge $e\in E(\Gamma)$ the integer
$(v:e)=\pm 1$ if $v$ is the final/initial vertex of $e,$ and
$(v:e)=0$ otherwise. The (simplicial) cohomology of $\Gamma$ is
encoded in the sequence
\begin{equation}\label{eq:exactseq}
0 \longrightarrow \R \stackrel{c}{\longrightarrow} \R^{V(\Gamma)} \stackrel{\delta}{\longrightarrow} \R^{E(\Gamma)} \longrightarrow H^1(\Gamma,\R) \longrightarrow 0
\end{equation}
with $c(1)=\sum_{v\in V(\Gamma)}v,$ $\delta(v) = \sum_{e\in
E(\Gamma)} (v:e)e.$ This sequence defines two configurations:
the inclusion of $\operatorname{coker}c$ into $\R^{E(\Gamma)},$
and dually the inclusion of $H_1(\Gamma,\R)$ into
$\R^{E(\Gamma)\vee}.$ We are presently interested in
the first one, which corresponds to the position space picture.\\\\
It will be convenient to fix a basis $V_0$ of
$\operatorname{coker}c.$ For example, the choice of a vertex
$v_0\in V(\Gamma)$ (write $V_0=V(\Gamma)\setminus \{v_0\})$ provides an
isomorphism $\phi: \R^{V_0}\rightarrow \operatorname{coker}c$
sending the basis element $v\in V_0$ to $v+\operatorname{im}c.$
We then have a configuration
\begin{equation}\label{eq:config}
i_\Gamma=\delta\phi: \R^{V_0} \hookrightarrow \R^{E(\Gamma)}.
\end{equation}
Each $e\in E(\Gamma)$ defines a linear form $e^\vee i_\Gamma\in
(\R^{V_0})^\vee.$ It is non-zero since $\Gamma$ has no loops.
Consider instead of $(\R^{V_0})^\vee$ the vector space
$(M^{V_0})^\vee$ where $M=\R^d.$ For each $e\in E(\Gamma)$ there
is a $d$-dimensional subspace
\begin{equation}\label{eq:ae}
A_e = (\operatorname{span} e^\vee i_\Gamma)^{\oplus d}
\end{equation}
of $(M^{V_0})^\vee.$ We denote this collection of
$d$-dimensional subspaces of $(M^{V_0})^\vee$ by
\begin{equation}\label{eq:cgamma}
\mathcal{C}(\Gamma)=\{A_e: e\in E(\Gamma)\}.
\end{equation}
Note that the $A_e$ need not be pairwise distinct nor linearly
independent. By duality $\mathcal{C}(\Gamma)$ defines an
arrangement of codimension $d$ subspaces in $M^{V_0}$
\begin{equation}\label{eq:singarr}
M^{V_0}_{sing}(\Gamma)=\bigcup_{e\in E(\Gamma)} A_e^\perp
\end{equation}
where $A_e^\perp$ is the linear subspace annihilated by $A_e.$
The image of $c^{\oplus d}$ in $M^{V(\Gamma)}$ is the thin
diagonal $\Delta.$ It is in the kernel of all the $e^\vee
i_\Gamma,$ and therefore it suffices for us to work in the
quotient space $\operatorname{coker}c.$ By construction
$A_e^\perp=D_{jl}+\Delta$ where $j$ and $l$ are the boundary vertices
of $e.$ In particular, if $\Gamma=K_n$ is the complete graph on
$n$ vertices, then it is clear that $M^{V_0}_{sing}(K_n)$ is
the large diagonal $\bigcup_{j<l}D_{jl}+\Delta.$
The composition $\Phi:M^{V(\Gamma)}\rightarrow M^{V(\Gamma)}/\Delta \rightarrow M^{V_0}$ is given by $\Phi(x_1,\ldots,x_n)=(x_1-x_n,\ldots,x_{n-1}-x_n),$ $x_i\in M,$ where again a numbering $V(\Gamma)=\{1,\ldots,n\},$ $v_0=n,$ of the vertices is assumed. \\\\
For a distribution $u$ on $M^{V}$
constant along $\Delta$ we write $\underline{u}=\Phi_\ast u$
for the pushforward onto $M^{V_0}.$ We usually write
$(x_1,\ldots,x_n)$ for a point in $M^{\{1,\ldots,n\}},$ where
$x_i$ is a $d$-tuple of coordinates $x_i^0,\ldots,x_i^{d-1}$
for $M.$ Similarly, if $f\in (\R^{V_0})^\vee$ then
$f^0,\ldots,f^{d-1}$ are the obvious functionals on $M^{V_0}$
such that $f^{\oplus d}=(f^0,\ldots,f^{d-1}).$
\subsection{Subspace arrangements of divergences}\label{ss:divarrangements}
Now we seek a refinement of the collection
$\mathcal{C}(\Gamma)$ in order to sort out singularities where
$u_\Gamma$ is locally integrable and does not require an
extension. In a first step we stabilize the collection
$\mathcal{C}(\Gamma)$ with respect to sums. Write
\begin{equation}\label{eq:csing}
\mathcal{C}_{sing}(\Gamma) = \left\{\sum_{e\in E'} A_e; \,\emptyset\subsetneq E'\subseteq E(\Gamma)\right\}.
\end{equation}
This is again a collection of non-zero subspaces of
$(M^{V_0})^\vee.$ A subset $E'$ of $E(\Gamma)$ defines
a unique subgraph $\gamma$ of $\Gamma$ (not necessarily
connected) with set of edges $E(\gamma)=E'$ and set of vertices $V(\gamma)=V(\Gamma).$ Each subgraph $\gamma$ of
$\Gamma$ determines an element
\begin{equation}\label{eq:agamma}
A_\gamma = \sum_{e\in E(\gamma)} A_e
\end{equation}
of $\mathcal{C}_{sing}(\Gamma).$ The map $\gamma\mapsto
A_\gamma$ is in general not one-to-one.
\begin{dfn}\label{dfn:saturated} A subgraph $\gamma\subseteq \Gamma$ is called \emph{saturated} if
$A_{\gamma}\subsetneq A_{\gamma'}$ for all subgraphs
$\gamma'\subseteq \Gamma$ such that $E(\gamma)\subsetneq
E(\gamma').$
\end{dfn}
It is obvious that for any given $\gamma$ there is always a
saturated subgraph, denoted $\gamma_s,$ with
$A_\gamma=A_{\gamma_s}.$ Also, $A_e\cap A_{\gamma_s}=\{0\}$ for
all $e\in E(\Gamma)\setminus E(\gamma_s).$
\begin{dfn}\label{dfn:amlog}
A graph $\Gamma$ is called \emph{at most logarithmic} if all
subgraphs $\gamma\subseteq\Gamma$ satisfy the condition
$d \dim H_1(\gamma)-2|E(\gamma)|\le 0.$
\end{dfn}
\begin{dfn}\label{dfn:divergent} A subgraph $\gamma\subseteq \Gamma$ is called \emph{divergent} if
$d \dim H_1(\gamma)=2|E(\gamma)|.$
\end{dfn}
\begin{pro}\label{pro:saturated}
Let $\Gamma$ be at most logarithmic. If $\gamma\subseteq\Gamma$
is divergent then it is saturated.
\end{pro}
\emph{Proof.} Assume that $\gamma$ satisfies the equality and
is not saturated. Then there is an $e\in E(\gamma_s)\setminus
E(\gamma).$ Since $\gamma$ and $\gamma\cup \{e\}$ have the same
number of components but $\gamma\cup\{e\}$ one more edge, it
follows from (\ref{eq:exactseq}) that $\dim
H_1(\gamma\cup\{e\})=\dim H_1(\gamma)+1.$ Consequently,
 $d \dim H_1(\gamma\cup\{e\})=2|E(\gamma\cup \{e\})|+2$ in contradiction to $\Gamma$ being
at most logarithmic. \hfill $\Box$ \\\\
Let $\Gamma$ be at most logarithmic. We define
\begin{equation}\label{eq:cdiv}
\mathcal{C}_{div}(\Gamma) = \left\{A_\gamma;\,\emptyset\subsetneq \gamma \subseteq \Gamma,\, \gamma\mbox{ divergent}\right\}
\end{equation}
as a subcollection of $\mathcal{C}_{sing}(\Gamma).$ It is
closed under sum (because $\dim H_1(\gamma_1\cup\gamma_2)\ge
\dim H_1(\gamma_1)+\dim H_1(\gamma_2)).$ It does not contain the
space $\{0\}.$ In the dual, the arrangement
\begin{equation}\label{eq:divarr}
M^{V_0}_{div}(\Gamma)=\bigcup_{\begin{array}{cc}\scriptstyle \emptyset\subsetneq \gamma\subseteq \Gamma\\
\scriptstyle d \dim H_1(\gamma)=2|E(\gamma)|\end{array}}A_\gamma^\perp
\end{equation}
in $M^{V_0}$ describes the locus where extension is necessary:
\begin{pro}\label{pro:singdiv} Let $\Gamma$ be at most logarithmic. Then the largest open subset of $M^{V_0} \subset M^{E(\Gamma)}$ to which $u_0^{\otimes |E(\Gamma)|}$ can be restricted
is the complement of $M^{V_0}_{div}(\Gamma).$ The restriction equals $\underline{u}_\Gamma$ there, and the singular support of $\underline{u}_\Gamma$ is the complement of $M^{V_0}_{div}(\Gamma)$ in $M^{V_0}_{sing}(\Gamma).$
\end{pro}
\emph{Proof.} Recall the map $i_\Gamma$ defining the
configuration (\ref{eq:config}). It provides an inclusion
$i_\Gamma^{\oplus d}: M^{V_0}\hookrightarrow M^{E(\Gamma)}.$
Wherever defined, $\underline{u}_\Gamma$ may be written
$\underline{u}_\Gamma(x_1,\ldots,x_{n-1}) = \prod_{e\in
E(\Gamma)} u_0\left(\sum_{v} (v:e)x_v\right)$ with
$V_0=\{1,\ldots,n-1\}.$ Since $i_\Gamma(v)=\sum_e(v:e)e,$ in
coordinates $i_\Gamma(\xi_1,\ldots,\xi_{n-1})$ $=\left(\sum_v
(v:e) \xi_v\right)_{e\in E(\Gamma)},$ it is clear that
$\underline{u}_\Gamma=(i_\Gamma^{\oplus d})^\ast u_0^{\otimes
|E(\Gamma)|}$ wherever it is defined. As by
(\ref{eq:singsuppu0}), $\operatorname{sing}\,\operatorname{supp} u_0=\{0\},$
the singular support of $u_0^{\otimes |E(\Gamma)|}$ is the
locus where at least one $d$-tuple of coordinates vanishes:
$x_e^0=\ldots=x_e^{d-1}=0$ for some $e\in E(\Gamma).$ Its
preimage under $i_\Gamma^{\oplus d}$ is the locus annihilated
by one of the $A_e,$ whence the last statement. For the first
statement, we have to show that for a compact subset $K\subset
M^{V_0}$ the integral
$\underline{u}_\Gamma|_K[\underline{1}]=\int_K
\underline{u}_\Gamma(x)dx$ converges if and only if $K$ is
disjoint from all the $A_\gamma^\perp,$ for
$\gamma\subseteq\Gamma$ such that $d\dim H_1(\gamma)=
2|E(\gamma)|.$ Assume that $K\cap
\left(A_\gamma^\perp\setminus \bigcup_{\gamma_s\subsetneq \gamma'}
A_{\gamma'}^\perp\right)\neq\emptyset$ for some $\gamma.$ Write
$\underline{u}_\Gamma = \prod_{e\in E(\gamma_s)} u_0(\sum_v
(v:e)x_v) f$ where $f=\prod_{e\in E(\Gamma)\setminus E(\gamma_s)}
u_0(\sum_v (v:e)x_v).$ The distribution $f$ is $C^\infty$ on
$A_{\gamma_s}^\perp\setminus \bigcup_{\gamma_s\subsetneq\gamma'}
A_{\gamma'}^\perp$ since $A_e\cap A_{\gamma_s}=\{0\}$ for all
$e\in E(\Gamma)\setminus E(\gamma_s).$ The integral $\int_{K}
\underline{u}_\Gamma(x)dx$ is over a $d(n-1)$-dimensional
space. The subspace $A_{\gamma_s}^\perp$ is given by $\dim
A_{\gamma_s}$ equations. Each single $u_0(x)$ is of order
$o(x^{2-d})$ as $x\rightarrow 0,$ and there are $|E(\gamma_s)|$
of them in the first factor of $u_\Gamma.$ Hence the integral
is convergent only if $\dim A_{\gamma_s}>(d-2)|E(\gamma_s)|,$
which is the same as $2|E(\gamma_s)|>d\dim H_1(\gamma_s).$
Conversely if this is the case for all $\gamma'_s\subseteq
\gamma_s$ then the integral is convergent. Our restriction to
saturated subgraphs $\gamma_s$ is justified by
\pref{pro:saturated}. \hfill $\Box$  \\\\
From now on through the end of section \ref{s:poles}, $\Gamma$ is a fixed, connected, at most
logarithmic graph. The general case where linear, quadratic, etc.~
divergences occur is discussed in section \ref{ss:quadratic}.
\subsection{Subspaces and polydiagonals}\label{ss:polydiagonals}
Let again $\gamma\subseteq\Gamma,$ that is $E(\gamma)\subseteq
E(\Gamma)$ and $V(\gamma)=V(\Gamma).$ Recall from the end of
section \ref{ss:singarrangements} that
\begin{equation}\label{eq:agammaperp}
\Phi^{-1}(A_\gamma^\perp) = \bigcap_{e\in E(\gamma)} D_{e}
\end{equation}
with the diagonals $D_e = D_{jl}$ for $j$ and $l$ boundary vertices of $e.$ An intersection $\bigcap_{e\in E(\gamma)} D_e$ of diagonals is called a \emph{polydiagonal}. \\\\
Just as in  (\ref{eq:exactseq}) we have an exact sequence
\begin{equation}\label{eq:newexactseq}
0 \longrightarrow H^0(\gamma,\R) \stackrel{c_\gamma}{\longrightarrow} \R^{V(\Gamma)} \stackrel{\delta_\gamma}{\longrightarrow} \R^{E(\gamma)} \longrightarrow H^1(\gamma,\R) \longrightarrow 0
\end{equation}
with $c_\gamma$ sending each generator of $H^0(\gamma,\R)$
(i.~e.~, a connected component $C$ of $\gamma$) to the sum of
vertices in this component, $1_C\mapsto \sum_{v\in C}v$ and
$\delta_\gamma (v) = \sum_{e\in E(\gamma)} (v:e)e.$ It is then
a matter of notation to verify
\begin{pro}\label{pro:agammaimdeg}
\begin{equation}\label{eq:agammaimdeg}
\Phi^{-1}(A_\gamma^\perp) = \ker\delta_\gamma^{\oplus d}.
\end{equation} \hfill $\Box$
\end{pro}
A polydiagonal $\Phi^{-1}(A_\gamma^\perp)$ corresponds
therefore to a partition $cc(\gamma)$ on the vertex set
$V(\Gamma)$ as follows: $cc(\gamma)=\{Q_1,\ldots,Q_k\}$ with
pairwise disjoint \emph{cells} $Q_1,\ldots,Q_k\subseteq
V(\Gamma)$ such that the vectors
\begin{equation}\label{eq:cc}
\sum_{v\in Q_i} v,\quad i=1,\ldots,k,
\end{equation}
generate $\ker \delta_\gamma.$ \\\\
In other words, $cc(\gamma)$ is the equivalence
relation/partition ''connected by $\gamma$'' on the set
$V(\Gamma).$  If $\Gamma=K_n$ is the complete graph on $n$
vertices, this correspondence is clearly a bijection
\begin{equation}\label{eq:bijection}
\{A_\gamma^\perp: \gamma\subseteq K_n\} \stackrel{\cong}{\rightarrow} \{\mbox{ Partitions of }V(K_n)\}.
\end{equation}
The next proposition refines this statement. Recall our index
notation from the end of section \ref{ss:singarrangements}.
\begin{pro} \label{pro:spanningbasis} Let $\gamma,t\subseteq \Gamma.$ Then the set
\begin{equation}\label{eq:spanningbasis}
\mathcal{B}=\left\{(e^\vee i_\Gamma)^j: e\in E(t), j=0,\ldots,d-1\right\}
\end{equation}
is a basis of $A_\gamma$ if and only if $t$ is a spanning
forest for $cc(\gamma),$
\end{pro}
where a spanning forest is defined as follows.
\begin{dfn}\label{dfn:spanningforestcc} Let $\gamma,t\subseteq \Gamma.$ Then $t$ is a
\emph{spanning forest for} $cc(\gamma)$ if the map $\delta_t:
\R^{V(\Gamma)}\rightarrow \R^{E(t)}$ as in (\ref{eq:newexactseq})
is surjective and $\ker\delta_t=\ker\delta_\gamma.$
\end{dfn}
\begin{dfn}\label{dfn:spanningforest} Let $\gamma,t\subseteq \Gamma$ and $t$ be a spanning forest for $cc(\gamma).$ If  $t\subseteq \gamma$ then $t$ is a \emph{spanning forest of} $\gamma.$ If $\gamma$ is connected (consequently so is $t$) then $t$ is called a \emph{spanning tree} of $\gamma.$ \end{dfn}
In other words, a spanning forest of $\gamma$ is a subgraph of $\gamma$ without cycles that has the
same connected components. A spanning forest for $cc(\gamma)$ has the same property but needs not be a subgraph of $\gamma.$ \\\\
\emph{Proof of \pref{pro:spanningbasis}.} By
\pref{pro:agammaimdeg}, $A_\gamma=A_t$ if and only if $\ker
\delta_\gamma=\ker\delta_t.$ It remains to show that the set
(\ref{eq:spanningbasis}) is linearly independent if and only if
$\delta_t$ is surjective. Since $\ker\delta_\Gamma\subseteq
\ker\delta_t$ the map $\delta_t$
is surjective if and only if $i_t=\delta_t\phi: \R^{V_0}\rightarrow \R^{E(t)}$ (see (\ref{eq:config})) is surjective, which in turn is equivalent to (\ref{eq:spanningbasis}) having full rank, as $e^\vee i_\Gamma=e^\vee i_t$ for $e\in E(t).$ \hfill $\Box$ \\\\
We also note two simple consequences for future use. Recall
our definition of a subgraph $\gamma$ of $\Gamma:$ If $\Gamma$
is a graph with set of vertices $V(\Gamma)$ and set of edges
$E(\Gamma),$ a subgraph $\gamma$ is given by a subset
$E(\gamma)\subseteq E(\Gamma)$ of edges. By definition
$V(\gamma)=V(\Gamma).$ However, we define
$V_{\operatorname{eff}}(\gamma)$ to be the subset of vertices
in $V(\gamma)$ which are not isolated -- a vertex $v$ is not
isolated if it is connected to another vertex through $\gamma.$
By abuse of language we say a proper subgraph $\gamma$ of $\Gamma$ is \emph{connected} if it is connected as a graph with
vertex set
$V_{\operatorname{eff}}(\gamma)$ and edge set $E(\gamma),$ in other words, not taking the
isolated vertices into account.
\begin{pro} \label{pro:intersection} Let $\gamma_1,\gamma_2\subseteq \Gamma,$
and assume $\gamma_1$ connected. Then
\begin{equation}\label{eq:intersectiona}
A_{\gamma_1}\cap A_{\gamma_2} = A_\gamma
\end{equation}
for any subgraph $\gamma$ of $\Gamma$ satisfying
\begin{equation}\label{eq:intersectioncc}
\operatorname{cc}(\gamma_1)\cap \operatorname{cc}(\gamma_2)=\operatorname{cc}(\gamma).
\end{equation}
\end{pro}
The intersection $P_1\cap P_2$ of partitions $P_1,P_2$ on the
same set $V(\Gamma)$ is defined by $P_1\cap P_2 = \{Q_1\cap
Q_2: Q_1\in P_1, Q_2\in P_2\}.$ It is easily seen
that this is a partition on $V(\Gamma)$ again. We write $0$ for the full partition $\{\{v\}: v\in V(\Gamma)\}.$ \\\\
\emph{Proof.} It is clear from \pref{pro:agammaimdeg} that
\begin{equation*}
\Phi^{-1}((A_{\gamma_1}\cap A_{\gamma_2})^\perp)=
\ker\delta_{\gamma_1}^{\oplus d}+\ker\delta_{\gamma_2}^{\oplus d},
\end{equation*}
and one needs a partition $cc(\gamma)$ whose cells provide a
basis as in (\ref{eq:cc}) but now for the space
$\ker\delta_{\gamma_1}+\ker\delta_{\gamma_2}.$ Let
$cc(\gamma_i)=\{Q_1^i,\ldots,Q_{l_i}^i\}.$ Since
\begin{equation*}
\sum_{v\in Q_k^1} v \in \operatorname{span}(\sum_{v\in Q_k^1\cap Q_1^2} v, \ldots,\sum_{v\in Q_k^1\cap Q_{l_2}^2}v
),
\end{equation*}
and similarly for $1$ and $2$ interchanged, the vectors $\sum_{v\in Q_k^1\cap Q_m^2}v$ generate $\ker\delta_{\gamma_1}+\ker\delta_{\gamma_2}.$
Consequently, $\ker\delta_{\gamma_1}+\ker\delta_{\gamma_2}\subseteq \ker\delta_{\gamma}.$ In order to have equality, it suffices
to show that the dimensions of both sides match. Since $\gamma_1$ is connected, we can assume $Q_1^1 = \{1,\ldots,i\},$
$Q_2^1=\{i+1\},\ldots,Q^1_{n-i+1}=\{n\}.$ In that case clearly $\dim \ker \delta_\gamma=\dim H_0(\gamma)=|cc(\gamma_1)\cap cc(\gamma_2)|=|cc(\gamma_1)|+\left|cc(\gamma_2)|_{\{1,\ldots,i\}}\right|-1$
where $P|_I$ denotes the partition $\{Q\cap I,Q\in P\}.$ On the other hand one verifies that $\dim(\ker \delta_{\gamma_1}+\ker\delta_{\gamma_2})$ is the same.\hfill $\Box$\\\\
Apart from the intersection of partitions as defined above, it
is useful to have the notion of a union of partitions. Let
$cc(\gamma_1),cc(\gamma_2)$ be partitions on $V(\Gamma).$ One
defines most conveniently
\begin{equation}\label{eq:defunion}
cc(\gamma_1)\cup cc(\gamma_2) = cc(\gamma_1\cup\gamma_2).
\end{equation}
From the description before (\ref{eq:bijection}) it is clear
that the right hand side in this definition depends only on $cc(\gamma_1)$ and
$cc(\gamma_2)$ but not on $\gamma_1$ and $\gamma_2$ themselves.
We immediately have
\begin{pro} \label{pro:union}
Let $\gamma_1,\gamma_2,\gamma\subseteq\Gamma.$ Then
\begin{equation}\label{eq:uniona}
A_{\gamma_1}+A_{\gamma_2}=A_\gamma
\end{equation}
if and only if
\begin{equation}\label{eq:unioncc}
\operatorname{cc}(\gamma_1)\cup \operatorname{cc}(\gamma_2)=\operatorname{cc}(\gamma).
\end{equation} \hfill $\Box$
\end{pro}
It will be convenient later to have an explicit description of
the dual basis $\mathcal{B}^\vee,$ for $\mathcal{B}$ as in
\pref{pro:spanningbasis}, that is the corresponding basis of
$M^{V_0}.$ Recall our choice (above equation (\ref{eq:config}))
of a vertex $v_0$ in order to work modulo the thin diagonal.
Recall also that the edges are oriented. Given a spanning tree
$t$ of $\Gamma,$ we say $e\in E(t)$ \emph{points to} $v_0$ if
the final vertex of $e$ is closer to $v_0$ in $t$ than the
initial vertex of $e.$ Otherwise we say that $e$ \emph{points
away from} $v_0.$ Furthermore, erasing the edge $e$ from $t$
separates $t$ into two connected components. The one \emph{not}
containing $v_0$ is denoted $t_1,$ and we write
$V_{\hat e,0}=V_{\operatorname{eff}}(t_1)$ for the set of its non-isolated vertices.
\begin{pro}\label{pro:dualadapted}
Let $\mathcal{B}^\vee=\{b^j_e: e\in E(t),j=0,\ldots,d-1\}$ be
the basis of $M^{V_0}$ dual to a basis $\mathcal{B}$ of
$(M^{V_0})^\vee$ as in \pref{pro:spanningbasis} , that is $(e^\vee
i_\Gamma)^j(b^k_{e'})=\delta_{e,e'}\delta_{j,k}.$  Then
\begin{equation*}
b_e = (-1)^{Q_e} \sum_{v\in V_{\hat e,0}} v.
\end{equation*}
($V_{\hat e,0},$ being a subset of the basis $V_0$ of $\R^{V_0},$ is also
contained in $\R^{V_0}).$ We define $Q_e=\pm 1$ if $e$ points
to/away from $v_0.$
\end{pro}
\emph{Proof.} Write $b_{e'}= \sum_{v\in V_0} \beta^{e'}_v v.$
We require
\begin{equation*}
\delta_{e,e'}=(e^\vee i_\Gamma)(b_{e'})=
(e^\vee \delta \phi)(b_{e'})= \sum_{v\in V_0} \beta^{e'}_v (v:e).
\end{equation*}
Now fix an $e.$ Write $v_{in}(e),v_{out}(e)$ for the initial
and final vertex of $e,$ respectively. We have
$\beta^e_{v_{in}(e)}-\beta^e_{v_{out}(e)}=1$ and
$\beta^e_{v_{in}(e')}=\beta^e_{v_{out}(e')}$ for the other
edges $e'$ except the one $e'_0$ leading to $v_0,$ for which
$\beta^e_{v_{in}(e'_0)}=0$ or $\beta^e_{v_{out}(e'_0)}=0,$
depending on the direction of $e'_0.$ Thus starting from $v_0$
and working one's way along the tree $t$ in order to determine
the $\beta^e_v,$ all the $\beta^e_v=0$ until one reaches the
edge $e,$ where $\beta^e_{v}$ jumps up or down to $1$ or $-1,$
depending on the orientation of $e,$ and stays constant then
all beyond $e.$ \hfill $\Box$ \\\\
Let us now describe the map $i_\Gamma^{\oplus d}:
M^{V_0}\rightarrow M^{E(\Gamma)}$ in such a dual basis
$\mathcal{B}^\vee.$ Let $x\in \R^{V_0},$ write $x=\sum_{e\in
E(t)} x_e b_e$ with $b_e = (-1)^{Q_e}\sum_{v\in V_{\hat e,0}} v$ as in
\pref{pro:dualadapted}. Write $[v_i,v_j]\subseteq E(t)$ for the
unique path in $t$ connecting the vertices $v_i$ and $v_j.$ It
follows that
\begin{equation*}
i_\Gamma(x)= \sum_{e\in E(\Gamma)} \sum_{v\in V_0} \sum_{e'\in [v_0,v]} (-1)^{Q_{e'}}x_{e'} (v:e)e.
\end{equation*}
For a given $e,$ only two vertices $v$ contribute to the sum,
namely the boundaries $v_{in}(e)$ and $v_{out}(e)$ of $e.$ All
the terms $(-1)^{Q_{e'}}x_{e'}$ for $e'$ on the path from $v_0$
to $v_{in}(e)$ cancel since they appear twice, once with a
negative sign $(v_{in}(e):e),$ once with a positive sign
$(v_{out}(e):e).$ What remains are the terms on the path in $t$
from $v_{in}(e)$ to $v_{out}(e)$. We write $e'\leadsto e$ if
$e'\in [v_{in}(e),v_{out}(e)]\subset E(t).$ Then
\begin{equation}\label{eq:beta}
i_\Gamma(x)= \sum_{e\in E(\Gamma)} \sum_{e'\leadsto e} x_{e'} e = \sum_{e\in E(t)} x_e e  + \sum_{e\in E(\Gamma)\setminus E(t)}\sum_{e'\leadsto e} x_{e'} e.
\end{equation}
Note that in the second sum there may be terms with only one
$x_{e'}$ contributing, namely when $A_e=A_{e'}.$

\section{Regularization, blowing up, and residues of primitive graphs} \label{s:propagators}
The purpose of this section is first to review a few standard
facts about distributions and simple birational
transformations. See \cite{Hoermander} for a general reference
on distributions. In the second part, the important notion of
residue of a primitive Feynman graph is introduced by raising
$u_\Gamma$ to a complex power $s$ in the neighborhood of $s=1$
and considering the residue at $s=1$ as a distribution
supported on the exceptional divisor of a blowup.
\subsection{Distributions and densities on manifolds}\label{ss:densities}
We recall basic notions that can be looked up, for example, in
\cite[Section 6.3]{Hoermander}. When one wants to define the
notion of distributions on a manifold one has two choices: The
first is to model a distribution locally according to the idea
that distributions are supposed to generalize $C^\infty$ functions,
so they should transform like $u_i=(\psi_j\psi_i^{-1})^\ast
u_j$ where $\psi_i,\psi_j$ are two charts. On the other hand,
distributions are supposed to be measures, that is one wants
them to transform like $\tilde u_i=|\det \operatorname{Jac}
\psi_j\psi_i^{-1}|(\psi_j\psi_i^{-1})^\ast \tilde
u_j.$ The latter concept is called a distribution density. \\\\
By a manifold we mean a paracompact connected $C^\infty$ manifold
throughout the paper. Let $\mathcal{M}$ be a manifold of
dimension $m$ with an atlas $(\psi_i,U_i)$ of local charts
$\psi_i: M_i\rightarrow U_i\subset \R^m.$
\begin{dfn} \label{dfn:distribution} A \emph{distribution} $u$ on $\mathcal{M}$ is a collection
$u=\{u_i\}$ of distributions $u_i\in\mathcal{D}'(U_i)$
satisfying
\begin{equation*}
u_i = (\psi_j\psi_i^{-1})^\ast u_j
\end{equation*}
in $\psi_j(M_i\cap M_j).$ The space of distributions on
$\mathcal{M}$ is denoted $\mathcal{D}'(\mathcal{M}).$
\end{dfn}
\begin{dfn} \label{dfn:density} A \emph{distribution density} $\tilde u$ on $\mathcal{M}$ is a
collection $\tilde u=\{\tilde u_i\}$ of distributions $\tilde
u_i\in\mathcal{D}'(U_i)$ satisfying
\begin{equation*}
\tilde u_i = |\det \operatorname{Jac} \psi_j\psi_i^{-1}|
(\psi_j\psi_i^{-1})^\ast \tilde u_j
\end{equation*}
in $\psi_j(M_i\cap M_j).$ The space of distribution densities on
$\mathcal{M}$ is denoted $\mathcal{\tilde D}'(\mathcal{M}).$ A density is
called $C^\infty$ if all $\tilde u_i$ are $C^\infty.$ The space of
$C^\infty$ densities on $\mathcal{M}$ with compact support is denoted $\tilde
C^\infty_0(\mathcal{M}).$
\end{dfn}
\begin{pro} $\left.\right.$ \label{pro:distanddens}
\begin{enumerate}
\item[(i)] ${C_0^\infty}'(\mathcal{M})=\mathcal{\tilde
    D}'(\mathcal{M}).$
\item[(ii)] $\tilde {C_0^\infty} ' (\mathcal{M}) =
    \mathcal{D}'(\mathcal{M}).$
\item[(iii)] A nowhere vanishing
    $C^\infty$ density $\alpha$
    provides isomorphisms $u\mapsto u\alpha$ between
    $\mathcal{D}'(\mathcal{M})$ and $\mathcal{\tilde
    D}'(\mathcal{M}),$ and $C^\infty_0(\mathcal{M})$ and
    $\tilde C^\infty_0(\mathcal{M}),$ respectively.
\end{enumerate} \hfill $\Box$
\end{pro}
$C^\infty$ densities are also called \emph{pseudo $m$-forms}. If
the manifold is oriented, every pseudo $m$-form is also a
regular $m$-form, and conversely an $m$-form $\omega$
gives rise to two pseudo $m$-forms: $\omega$ and $-\omega.$ In
a nonorientable situation we want to work with distribution
densities and write them like pseudo forms $u(x)|dx|.$
\subsection{Distributions and birational transformations}\label{ss:birational}
Let $\mathcal{M}$ be a manifold of dimension $m$ and
$z\in \mathcal{M}$ a point in it. We work in local coordinates
and may assume $\mathcal{M}=\R^m$ and $z=0.$ Blowing up $0$
means replacing $0$ by a real projective space
$\mathcal{E}=\P^{m-1}(\R)$ of codimension 1.
The result is again a manifold as follows. \\\\
Let $Y=(\mathcal{M}\setminus \{0\})\sqcup \mathcal{E}$ as a set. Tangent
directions at $0$ shall be identified with elements of
$\mathcal{E}.$ Let therefore $Y'$ be the subset of
$\mathcal{M}\times \mathcal{E}$ defined by $z_iu_j=z_ju_i,$
$1\le i,j\le m$ where $z_1,\ldots ,z_m$ are the affine
coordinates of $\R^m$ and $u_1,\ldots,u_m$ are homogeneous
coordinates of $\mathbb{P}^{m-1}(\R).$ The set $Y'$ is a
submanifold of $\mathcal{M}\times \mathcal{E}.$ On the other
hand, there is an obvious bijection $\lambda: Y\rightarrow Y'$
whose restriction onto $\mathcal{M}\setminus \{0\}\subset Y$ is a
diffeomorphism onto its image. Pulling back along $\lambda$ the
differentiable structure induced on $Y'$ defines a
differentiable structure on all of $Y.$ The latter is called the
\emph{blowup of} $\mathcal{M}$ at $\{0\}.$ The submanifold
$\mathcal{E}$ of $Y$ is called the \emph{exceptional divisor.}
There is a proper $C^\infty$ map $\beta: Y\rightarrow \mathcal{M}$
which is the identity on $\mathcal{M}\setminus \{0\}$ and sends $\mathcal{E}$ to $0.$ Viewed as a map from $Y'\subset \mathcal{M}\times\mathcal{E},$ $\beta$ is simply the projection onto the first factor. \\\\
Note that if $m$ is even (which is the case throughout the
paper) then $Y$ is not orientable but $\mathcal{E}$ is. If $m$
is odd then $Y$ is orientable but $\mathcal{E}$ is not. Indeed
$Y$ is a fiber bundle $\tau:Y\rightarrow \mathcal{E}$
over $\mathcal{E}$ with fiber $\R$ -- the tautological line bundle.
For example, for $m=2,$ $Y$ is the open M\"obius strip.
\\\\
%
%
Let $m$ be even from now on. For $U_i=\R^m,$ $i=1,\ldots,m,$
one defines maps $\rho_i: U_i\rightarrow \mathcal{M}\times \mathcal{E},$
\begin{eqnarray}\label{eq:simplechart}
(y_1,\ldots,y_m) &\mapsto & ((z_1,\ldots,z_m),[z_1,\ldots,z_m]) \nonumber\\
                 && z_i = (-1)^i y_i, \\
                 && z_k = y_iy_k,\,k\neq i \nonumber
\end{eqnarray}
where $z_i$ are coordinates on $\mathcal{M}$ and at the same time
homogeneous coordinates for $\mathcal{E}.$ Clearly $\rho_i$
maps into $Y$ and onto the affine chart of $\mathcal{E}$ where
$z_i\neq 0.$ Let $\psi_i = \rho_i^{-1}$ on $\rho_i(U_i).$ Then
$(\psi_i,U_i)$ furnish an atlas for $Y.$ We note for future
reference the transition maps
\begin{eqnarray} \label{eq:chartchange}
\psi_j\psi_i^{-1}: &&U_i\setminus\{y_j=0\}\rightarrow U_j\setminus \{y'_i=0\} \nonumber\\
 (y_1,\ldots,y_m) &\mapsto& (y'_1,\ldots,y'_m) \nonumber\\
              && y'_i = (-1)^{i+j}/y_j, \\
              && y'_j = (-1)^j y_iy_j,\nonumber \\
              && y'_k= (-1)^j y_k/y_j, \,k\neq i,j \nonumber
\end{eqnarray}
and the determinants of their derivatives
\begin{equation}\label{eq:jacmain}
\det\operatorname{Jac}\psi_j\psi_i^{-1}= (-1)^{j+1}y_j^{1-m}.
\end{equation}
Note that the atlas $(\psi_i,U_i)$ is therefore not even
oriented on the open set $Y\setminus \mathcal{E}$ diffeomorphic to
$\mathcal{M}\setminus \{0\}.$
For the exceptional divisor $\mathcal{E}=\P^{m-1}(\R),$ which is
given in $U_i$ by the equation $y_i=0,$ we use induced charts
$(\phi_i,V_i)$ with coordinates
$y_1,\ldots,\widehat{y_i},\ldots,y_m$ (in this very order)
where $\widehat{y_i}$ means omission. The transition map
\begin{eqnarray} \label{eq:chartchange2}
\phi_j\phi_i^{-1}: &&V_i\setminus \{y_j=0\}\rightarrow V_j\setminus \{y'_i=0\} \nonumber\\
 (y_1,\ldots,\widehat{y_i},\ldots,y_m) &\mapsto& (y'_1,\ldots,\widehat{y'_j},\ldots,y'_m) \nonumber\\
              && y'_i = (-1)^{i+j}/y_j, \\
              && y'_k= (-1)^j y_k/y_j, \,k\neq i,j \nonumber
\end{eqnarray}
has Jacobian determinant
\begin{equation}\label{eq:jacinduced}
\det\operatorname{Jac}\phi_j\phi_i^{-1}= y_j^{-m}>0.
\end{equation}
The induced atlas $(V_i,\phi_i)$ is therefore an oriented one.
The tautological bundle $\tau$ is given in local coordinates by
$\tau: (y_1,\ldots,y_m)\mapsto
(y_1,\ldots,\widehat{y_i},\ldots,y_m).$ \\\\
Similarly one defines blowing up along a submanifold:
The submanifold is replaced by its projectivized normal bundle.
Assume the submanifold is given in local coordinates by
$z_1=\ldots=z_k=0.$ Then a natural choice of coordinates for
the blowup is given again by (\ref{eq:simplechart}), applied
only to the subset of
coordinates $z_1,\ldots,z_k.$ See for instance \cite[Section 3]{Mather} for details.\\\\
The \emph{blowdown map} $\beta: Y\rightarrow \mathcal{M}$ is surjective, proper
and $C^\infty$ everywhere but open (i.~e.~ has surjective
differential) only away from the exceptional divisor. It is
called the blowdown map. It will be useful to be able to push
distributions forward along $\beta$ and to pull distributions back along $\beta|_{Y\setminus\mathcal{E}}.$ \\\\
In general, let $f: U\rightarrow V$ be a surjective proper
$C^\infty$ map between open sets $U$ of $\R^{m_1}$ and $V$ of $\R^{m_2}.$
Let $u$ be a distribution on $U.$ The pushforward of $u$ by
$f,$ denoted $f_\ast u,$ is the distribution on $V$ defined by
$(f_\ast u)[\phi]=u[f^\ast \phi]$ where $\phi$ is a test
function on $V$ and $f^\ast\phi$ is its pullback along $f:$
$f^\ast\phi=\phi\circ f.$ If $u$ has compact support the
requirement that $f$ be proper can be dropped. Similarly, for
$f:\mathcal{M}\rightarrow \mathcal{N}$ a surjective proper
$C^\infty$ map between manifolds $\mathcal{M}$ and $\mathcal{N}$
with atlases $(\psi_i,U_i)$ and $(\theta_i,V_i),$ let $u$ be a
distribution density on $\mathcal{M}.$ Then $f_\ast u$ defined
by
\begin{equation*}
(f_\ast u)_i = (\theta_i f\psi^{-1}_k)_\ast u_k
\end{equation*}
in $V_i\cap (\theta_if\psi_k^{-1})(U_k),$ is a distribution
density on $\mathcal{N}.$ Let now $f:\mathcal{M}\rightarrow
\mathcal{N}$ a surjective $C^\infty$ map between manifolds
$\mathcal{M}$ and $\mathcal{N}.$ It need not be proper. Let
$u\in\mathcal{\tilde D}'(\mathcal{M})$ and $\phi\in
C^\infty_0(\mathcal{M}).$ The density $u[\phi]_f
\in\mathcal{\tilde D}'(\mathcal{N})$ is defined by
\begin{equation}\label{eq:partialint}
u[\phi]_f = f_\ast (\phi u).
\end{equation}
Note that $\phi u$ has compact support so the pushforward is
well-defined although $f$ is not necessarily proper. If $u$ is
given by a locally integrable function $u(z)$ on
$\mathcal{M}=\R^{m}$ and $f$ is the projection onto
$\mathcal{N}=\{z_1,\ldots,z_i=0\}\subseteq \R^{m},$ $i<m,$
this notion corresponds to integrating out the orthogonal
complement $\{z_{i+1},\ldots,z_m=0\}$ of $\mathcal{N}$ in $\R^m:$
\begin{equation*}
u[\phi]_f(z_{i+1},\ldots,z_m) = \int (u\phi)(z_1,\ldots,z_m) dz_1\ldots dz_i.
\end{equation*}
The reverse operation of pulling back distributions along
$C^\infty$ maps is only possible under certain conditions, see
\cite[Sections 6.1, 8.2, etc.]{Hoermander} for a general
exposition. Here we only need the following simple case: Let
$U_1,U_2\subseteq \R^n$ open and $f:U_1\rightarrow U_2$ a
$C^\infty$ and everywhere open map. Then there is a unique
continuous linear map $f^\ast: \mathcal{D'}(U_2)\rightarrow
\mathcal{D'}(U_1)$ such that $f^\ast u=u\circ f$ if $u\in
C^0(U_2).$
See \cite[Theorem 6.1.2]{Hoermander} for a proof of this
statement. It can obviously be generalized to the case of a
submersion $f:\mathcal{M}\rightarrow \mathcal{N}$
of manifolds.\\\\
$\mathcal{M}=\R^m$ and its open subsets being orientable, distributions
can be identified with distribution densities there, see
\pref{pro:distanddens} (iii).
If $\beta$ is the blowdown map, by the pullback $\beta^\ast
\tilde u$ of a distribution density $\tilde u\in\mathcal{\tilde
D}'(\mathcal{M}\setminus \{0\})$ obviously the pullback along the
diffeomorphism $\beta|_{Y\setminus \mathcal{E}}$ is understood. The
result is a distribution density on $Y\setminus \mathcal{E}.$
\subsection{Analytic regularization}\label{ss:regularization}
As a first step toward understanding $u_\Gamma^s$ as a
distribution-valued meromorphic function of $s$ in a
neighborhood of $s=1,$ we study distributions $u$ on $\R\setminus \{0\}$ of
the form $u(z)=|z|^{-a}$ where $a\in\Z.$ Clearly if $a<1$ then
$u\in L^1_{loc}(\R).$ The case $a\ge 1$ can be handled using analytic continuation with respect to the
exponent: Let $a\in \mathbb{N}$ be fixed. We
extend $u^s=|z|^{-as}$ meromorphically to the area $\Re s>1$ as
follows. Let $n=\lfloor a/2\rfloor.$
\begin{eqnarray}\label{eq:usphi}
u^s[\phi] &= &\int_{0}^1
z^{-as}(\phi(z)+\phi(-z))dz +\int_{\R\setminus [-1,1]} |z|^{-as} \phi(z)dz \nonumber\\
&=& \int_{0}^1 z^{-as} \left(\phi(z)+\phi(-z) -2\left(\phi(0)+\ldots+\frac{z^{2n}\phi^{(2n)}(0)}{(2n)!}\right)\right) dz \\
&&+ \int_{\R\setminus [-1,1]} |z|^{-as} \phi(z)dz +2 \sum_{k=0}^{n} \frac{\phi^{(2k)}(0)}{(2k)!((2k+1)-as)}.\nonumber
\end{eqnarray}
This holds for $\Re s<1+\frac{1}{a}.$ See \cite[Section
I.�3]{GS} for the complete argument. There will be more poles
beyond the half-plane $\Re s <1+\frac{1}{a}$ but
they are not relevant for our purposes.
\begin{dfn}\label{dfn:canext} The \emph{canonical regularization} of $|z|^{-a}$ is the
distribution-valued meromorphic function in $s\in
(-\infty,1+\frac{1}{a})+i\R$ given by
\begin{equation}\label{eq:xsa}
|z|_{ext}^{-as}=2 \sum_{k=0}^{n} \frac{\delta_0^{(2k)}}{(2k)!((2k+1)-as)} +
|z|_{fin}^{-as}
\end{equation}
where $n=\lfloor a/2\rfloor$ and
\begin{eqnarray}\label{eq:xfin}
|z|_{fin}^{-as}[\phi] &=& \int_{0}^1 z^{-as} \left(\phi(z)+\phi(-z) -2\left(\phi(0)+\ldots+\frac{z^{2n}\phi^{(2n)}(0)}{(2n)!}\right)\right) dz \nonumber\\
&&+ \int_{\R\setminus [-1,1]} |z|^{-as} \phi(z)dz.
\end{eqnarray}
\end{dfn}
The function $s\mapsto |z|_{fin}^{-as}$ is holomorphic in
$(-\infty,1+\frac{1}{a})+i\R.$ When the context allows, we
simply write $|z|^{-as}$ for $|z|^{-as}_{ext}$ again. Let $f\in
C^\infty(\R).$ Since $s\mapsto f^s[\phi]$ is holomorphic, it
makes sense to define the canonical regularization for
$|z|^{-a}f$ also:
\begin{equation}\label{eq:mult}
(|z|^{-a}f)^s_{ext} = |z|^{-as}_{ext} \cdot f^s.
\end{equation}
This does not work for $f\in L^1_{loc}(\R).$ For example, $|z|^{-(a+b)s}_{ext}\neq |z|^{-as}_{ext}|z|^{-bs}_{ext}.$ \\\\
Unfortunately, the term ''regularization'' is used for two
different notions in the mathematics and physics literature,
respectively. They must be carefully distinguished. While in the
mathematics literature, the ''regularized'' distribution is
usually understood to be $|z|_{fin}^{-a},$ a physicist calls
this the ''renormalized'' distribution, and
refers to the mapping $s\mapsto |z|^{-as}$ as a regularization (in fact, one out of many possible regularizations). The latter is also our convention. \\\\
We finally note the special case $a=1,$
\begin{equation}\label{eq:xsa1}
|z|_{ext}^{-s} = -\frac{2\delta_0}{s-1}+|z|_{fin}^{-s},
\end{equation}
\begin{equation}\label{eq:xsa2}
|z|_{fin}^{-s}[\phi] = \int_{-1}^{1} |z|^{-s} (\phi (z)-\phi(0))dz+\int_{\R\setminus [-1,1]} |z|^{-s} \phi(z)dz.
\end{equation}
And, for future reference, in the area $\Re
s<\frac{2+(D-1)}{D},$
\begin{equation}\label{eq:xsa3}
|z|_{ext}^{D-Ds-1} = - \frac{2}{D}\frac{\delta_0}{s-1}+|z|_{fin}^{D-Ds-1}
\end{equation}
where $D\in 2\N.$
\subsection{Primitive graphs, their residues and renormalization} \label{ss:residues}
We consider the blowup $\beta:Y\rightarrow \mathcal{M}$ of $\mathcal{M}$ at 0 as in
section \ref{ss:birational} where now $\mathcal{M}=M^{V_0}$ for
our Feynman graph $\Gamma$ (see section \ref{s:arrangements} for
notation). Let $d_\Gamma=d|V_0|=\dim \mathcal{M}.$
In this section we continue to use the coordinates
$z_1,\ldots,z_{d_\Gamma}$ on $M^{V_0}$ and
$y_1,\ldots,y_{d_\Gamma}$ on the charts $U_i$ for $Y.$ They
are related to the coordinates $x_i^j$ of section 2 by $x_i^j=z_{d(i-1)+j}.$
Recall that
since $Y$ is not orientable (and the induced atlas on
$Y\setminus \mathcal{E}$ is not oriented), top degree $(L^1_{loc})$ forms and
distribution densities can not be identified. We only use forms on the oriented submanifold
$\mathcal{E},$ where the two notions coincide. We write $|dz|$
for the Lebesgue measure on $\mathcal{M}.$
\begin{dfn}\label{dfn:primitive}
A connected Feynman graph $\Gamma$ is called \emph{primitive}
if $\mathcal{C}_{div}(\Gamma)=\{A_\Gamma\}.$
\end{dfn}
Recall the notion of saturated subgraph from \dref{dfn:saturated} and \pref{pro:saturated}.
\begin{lem} \label{lem:infraprep}
Let $\Gamma$ be primitive. Let $t$ be a spanning tree for
$\Gamma$ and $t'$ a subgraph of $t.$ Then
\begin{equation*}
d|E(t')|\le (d-2)(|E(\Gamma)|-|E((t\setminus t')_s)|)
\end{equation*}
and equality holds if and only if $t'=t.$
\end{lem}
\emph{Proof.} Clearly $\dim A_t=\dim A_{t'}+\dim A_{t\setminus t'}$ and
$\dim A_{t'}=d|E(t')|.$ Since $\Gamma$ is divergent,
$(d-2)|E(\Gamma)|=\dim A_t.$
Since $\Gamma$ has no proper divergent subgraphs,
$(d-2)|E((t\setminus t')_s)|< \dim A_{(t\setminus t')_s}=\dim A_{t\setminus t'}$ for all proper subgraphs $t'$ of $t.$ \hfill $\Box$ \\\\
\begin{lem} \label{lem:deltae}
Let $\delta_\mathcal{E}$ (resp.~ $\frac{1}{|y_\mathcal{E}|})$
be collections of distributions\footnote{We do not
claim that they are distributions or densities on $Y$ themselves
as they do not transform correctly.} in the $U_i$ given by
$(\delta_\mathcal{E})_i=\delta_0(y_i)$ and
$(1/|y_\mathcal{E}|)_i=\frac{1}{|y_i|}$ in $U_i.$ Let $\omega$
be a locally integrable volume form $\omega$ on $\mathcal{E}.$
Then $\omega \delta_\mathcal{E}$ and $\omega/|y_\mathcal{E}|,$
locally
\begin{eqnarray*}
(\omega\delta_\mathcal{E})_i&=&\omega_i(\delta_\mathcal{E})_i=\omega_i(y_1,\ldots,\widehat{ y_i},\ldots,y_{d_\Gamma})\delta_0(y_i),\\
(\omega/|y_\mathcal{E}|)_i&=&\omega_i/|y_\mathcal{E}|_i=\omega_i(y_1,\ldots,\widehat{y_i},\ldots,y_{d_\Gamma})/|y_i|
\end{eqnarray*}
define densities on $Y.$
\end{lem}
\emph{Proof.} By (\ref{eq:jacmain}) and (\ref{eq:jacinduced})
$|\operatorname{det}\operatorname{Jac} \psi_j\psi_i^{-1}|=
\operatorname{det}\operatorname{Jac}\phi_j\phi_i^{-1}\cdot
|1/y_j|$ and both $\delta_0$ and $1/|y_i|$ transform with the
factor $|1/y_j|$ under transition $U_i\rightarrow U_j.$ \hfill
$\Box$
\begin{thm} \label{thm:primitive}
Let $\Gamma$ be primitive.
\begin{enumerate}
\item[(i)] By pullback along the diffeomorphism
    $\beta|_{Y\setminus \mathcal{E}},$ the distribution density
    $\tilde u_\Gamma=\underline{u}_\Gamma|dz|$ furnishes a
    strictly positive density $\tilde w_\Gamma$ on
    $Y\setminus \mathcal{E},$ given in local coordinates of $U_i$ by
\begin{equation} \label{eq:pulldens}
(\tilde w_\Gamma)_i|dy| = \frac{1}{|y_i|} (f_\Gamma)_i(y_1,\ldots,\widehat{ y_i},\ldots,y_{d_\Gamma}) |dy|
\end{equation}
where $(f_\Gamma)_i \in L^1(V_i).$ The
$(f_\Gamma)_i\,dy_1\wedge\ldots\wedge
\widehat{dy_i}\wedge\ldots\wedge dy_{d_\Gamma}$ in each $V_i$
determine an integrable volume form $f_\Gamma$ on
$\mathcal{E}.$ We may therefore write $\tilde
w_\Gamma=f_\Gamma/|y_\mathcal{E}|.$
\item[(ii)] The meromorphic density-valued function
    $s\mapsto \tilde w_\Gamma^s=\beta^\ast \tilde
    u_\Gamma^s,$
\begin{equation*}
(\tilde w_\Gamma^s)_i |dy|= \frac{(f_\Gamma)_i^s|dy|}{|y_i|^{d_\Gamma s-(d_\Gamma-1)}}
\end{equation*}
has a simple pole at $s=1.$ Its residue is the density
\begin{equation}\label{eq:residuedens}
\operatorname{res}_{s=1} \tilde w_\Gamma^s=-\frac{2}{d_\Gamma}\delta_\mathcal{E}f_\Gamma,
\end{equation}
supported on the exceptional divisor. Pushing forward along
$\beta$ amounts to integrating a projective integral over
the exceptional divisor:
\begin{equation}\label{eq:residueint}
\beta_\ast (\operatorname{res}_{s=1} \tilde w_\Gamma^s) = -\frac{2}{d_\Gamma}\delta_0|dz|\int_\mathcal{E} f_\Gamma =-\frac{2}{d_\Gamma}\delta_0 \int_{V_i} (f_\Gamma)_i dy_1\ldots \widehat{dy_i}\ldots dy_{d_\Gamma}
\end{equation}
for any $i.$
\item[(iii)] Let $\mu\in \mathcal{D}(\R^{d_\Gamma})$ with $\mu(0)=1,$
and $\nu = \beta^\ast \mu.$ Let
    $\tau:Y\rightarrow\mathcal{E}$ be the tautological
    bundle. 
Then
\begin{equation}\label{eq:simpleren}
\tilde w_{\Gamma,R}^s=\tilde w_\Gamma^s-\tilde w_\Gamma^s [\nu]_\tau \delta_\mathcal{E}
\end{equation}
defines a density-valued function on $Y$ holomorphic in a neighborhood of $s=1.$ Also
$\beta_\ast \tilde w^s_{\Gamma,R}= (\underline{u}_\Gamma^s-\underline{u}_\Gamma^s[\mu]\delta_0)|dz|=\underline{u}^s_{\Gamma,R}|dz|.$
\end{enumerate}
\end{thm}
The density (\ref{eq:residuedens}) is called \emph{residue
density}, the volume form $f_\Gamma$ \emph{residue form}, and
the complex number
\begin{equation}\label{eq:defresidue}
\operatorname{res}\Gamma = -\frac{2}{d_\Gamma}\int_\mathcal{E}f_\Gamma
\end{equation}
\emph{residue of} $\Gamma.$ The distribution
$\underline{u}_{\Gamma,R}=\underline{u}_{\Gamma,R}^s|_{s=1}$ is defined on all of
$M^{V_0}$ and said to be the \emph{renormalized
distribution.} \\\\
\emph{Proof of \tref{thm:primitive}.} (i) For
(\ref{eq:pulldens}) observe that in local coordinates of $U_i$ the map $\beta$ is
given by $\rho_i,$ see (\ref{eq:simplechart}). The Lebesgue
measure $|dz|$ on $M^{V_0}$ pulls back to
$|y_i|^{d_\Gamma-1}|dy|$ in $U_i.$ By (\ref{eq:homogenu0}), $\underline{u}_\Gamma$
scales like $\lambda^{(2-d)|E(\Gamma)|}$ as $z_i\rightarrow \lambda z_i$ for all $i.$
Since $\Gamma$ is divergent, $d_\Gamma=(2-d)|E(\Gamma)|,$ which
explains the factor $1/|y_i|$ in (\ref{eq:pulldens}) and that
$(f_\Gamma)_i$ does not depend on $y_i.$ That
$(f_\Gamma)_i\in L^1_{loc}(V_i)$ follows from \pref{pro:singdiv},
where $M^{V_0}_{div}=A_\Gamma^\perp=\{0\},$ and
$\beta|_{Y\setminus \mathcal{E}}$ being a diffeomorphism. In order to
show that $(f_\Gamma)_i\in L^1(V_i)$ one uses \lref{lem:infraprep}
as follows: Choose a spanning tree $t$ for $\Gamma$ such that
the coordinate $z_i$ equals $(e_0^\vee i_\Gamma)^{j_0}$ for some
$e_0\in E(t)$ and $j_0\in\{0,\ldots,d-1\}$ (see \pref{pro:spanningbasis}). Write
$x_e^j=(e^\vee i_\Gamma)^j$ for $e\in E(t), j=0,\ldots,d-1.$ In
this basis $\underline{u}_\Gamma$ is given by
$\underline{u}_\Gamma(\{x_e^j\})=\prod_{e\in E(\Gamma)}
u_0(\sum_{e'\leadsto e} x_{e'}^j)$ (see (\ref{eq:beta})).
Therefore, if the coordinates $y_e^j,\,e\in E(t')\,j=0,\ldots,d-1$ defined by
$t'$ a proper subforest of $t$ not containing $e_0$ go to $\infty,$ then there are
exactly $E(t_s)\setminus E((t\setminus t')_s)$ factors of $u_0$ the argument of
which goes to $\infty.$ \lref{lem:infraprep} shows that the
integration over that subspace converges. One verifies that all
subspaces susceptible to infrared divergences are of this form.
Therefore $(f_\Gamma)_i\in L^1(V_i).$ Finally, the $(f_\Gamma)_i$
produce a factor $y_i^{-d_\Gamma}$ under transition between charts. By
(\ref{eq:jacinduced}) this makes $f_\Gamma$ a density on
$\mathcal{E}.$ Since $\mathcal{E}$ is
oriented, a strictly positive density is also a strictly positive ($L^1_{loc})$ volume form. \\
(ii) The simple pole and (\ref{eq:residuedens}) follow from
(\ref{eq:pulldens}) by (\ref{eq:xsa3}), the local expressions
matched together using \lref{lem:deltae}. For
(\ref{eq:residueint}) let $\phi\in\mathcal{D}(M^{V_0}).$ Then
$\beta_\ast (\operatorname{res}_{s=1}\tilde w^s_\Gamma)[\phi]=
\operatorname{res}_{s=1}\tilde w^s_\Gamma[\beta^\ast \phi].$ The
distribution density $\operatorname{res}_{s=1} \tilde w^s_\Gamma,$ being
supported on $\mathcal{E},$ depends only on
$\beta^\ast\phi|_{\mathcal{E}}=\phi(0).$ By the results of (i),
$\int_\mathcal{E} f_\Gamma$ is a projective integral and it
suffices to integrate inside one chart, say $U_i.$ There
$\operatorname{res}_{s=1}\tilde w^s_\Gamma[\beta^\ast
\phi]=-\frac{2}{d_\Gamma}\int_{U_i}\delta_0(y_i)f_\Gamma(y)\phi(\rho_i(y))dy$
$=-\frac{2}{d_\Gamma}\phi(0)\int_{V_i}f_\Gamma(y)dy $ $=
-\frac{2}{d_\Gamma}\phi(0)\int_\mathcal{E}f_\Gamma.$  \\
(iii) There is no pole at $s=1$ since $\nu|_\mathcal{E}=1.$
The $(\tilde w^s_{\Gamma,R})_i$ furnish a density by \lref{lem:deltae}:
The Jacobian of $\delta_\mathcal{E}$ cancels the one of $[\ldots]_\tau.$
For the last statement, let again $(\psi_i,U_i)_{i=1,\ldots,d_\Gamma}$ be
the chosen atlas for $Y$ and $(\phi_i,V_i)_{i=1,\ldots,d_\Gamma}$ the
induced atlas for $\mathcal{E}.$ Since $\mathcal{E}$ is
compact, there exists a partition of unity
$(\xi_i\phi_i)_{i=1,\ldots,d_\Gamma}$ on $\mathcal{E}$
subordinate to the $V_i$ such that $\xi_i\in \mathcal{D}(V_i),$
$\xi_i\ge 0$ and $\sum_i (\xi_i\phi_i)(x)=1$ for all $x\in
\mathcal{E}.$ Let $\tau: Y\rightarrow \mathcal{E}.$ Then
$(\xi_i\phi_i\tau)_{i=1,\ldots,d_\Gamma}$ is a partition of unity
on $Y$ subordinate to $(\psi_i,U_i)_{i=1,\ldots,d_\Gamma}$
(however not compactly supported). We fix such a partition of
unity $(\xi_i).$ In $U_i$ we write $y$ for $(y_1,\ldots,y_{d_\Gamma})$
and $\widehat{y_i}$ for
$(y_1,\ldots,\widehat{y_i},\ldots,y_{d_\Gamma}),$ for example
$\xi_i(y)=\xi_i(\widehat{y_i})$ since it is constant along
$y_i.$ We also write
$u(y_i,y_i\widehat{y_i})=u(y_iy_1,\ldots,y_i,\ldots,y_iy_{d_\Gamma})$
for convenience. Let $f\in\mathcal{D}(M^{V_0}).$
\begin{eqnarray*}
\beta_\ast(\tilde w^s_{\Gamma,R})[f] &=& \beta_\ast(\tilde w^s_{\Gamma}-\tilde w^s_{\Gamma}[\nu]_\tau\delta_\mathcal{E})[f]\\
&=&\sum_{i}(\tilde w^s_{\Gamma}-\tilde w_\Gamma^s[\nu]_\tau
\delta_\mathcal{E})_i[\xi_i\beta^\ast f] \\
&=& \sum_i \int_{U_i} (\tilde w_\Gamma^s(y)-\int_{\R} \tilde w_\Gamma^s(z_i,\widehat{y_i})\mu(z_i,z_i\widehat{y_i})dz_i
\delta_0(y_i))\\
&&\times \xi_i(y)
f(y_i,y_i\widehat{y_i})dy\\
&=& \sum_i \int_{U_i} \tilde w_\Gamma^s(y)\xi_i(y) f(y_i,y_i\widehat{y_i}) \\
&&- \tilde w_\Gamma^s(y)\mu(y_i,y_i\widehat{y_i}) \xi_i(0,\widehat{y_i}) f(0)dy  \\
&=& \sum_i (\beta_\ast \tilde w_\Gamma^s-\beta_\ast \tilde w_\Gamma^s[\xi_i\nu]\delta_0)[f].
\end{eqnarray*}\hfill $\Box$\\\\
The following corollary concerns infrared divergences of a
graph $\Gamma.$ Those are divergences which do not occur at the
$A_\gamma^\perp$ but as the coordinates $z_i$ of $M^{V_0}$
approach $\infty,$ in other words, if one attempts to integrate
$\underline{u}_\Gamma$ against a function which is not compactly supported.
\begin{cor}\label{cor:infraredsafe}
Let $\Gamma$ be at most logarithmic and primitive. Then
$\underline{u}_\Gamma$ is not (globally) integrable on
$M^{V_0}\setminus M^{V_0}_{div}(\Gamma).$ However $(\chi
\underline{u}_\Gamma)[\underline{1}_L\otimes \mu]$ is well-defined, if
$\mu$ is a test function on a non-zero subspace of $M^{V_0},$
$\underline{1}_L$ the constant function on the orthogonal
complement $L,$ and $\chi$ the characteristic function of the
complement of an open neighborhood of $M^{V_0}_{div}(\Gamma)$
in $M^{V_0}.$
\end{cor}
\emph{Proof.} This follows from part (i) of \tref{thm:primitive}. \hfill $\Box$ \\\\
The renormalized distribution
$\underline{u}_{\Gamma,R}=\underline{u}_{\Gamma,R}^s|_{s=1}$ obtained from the theorem
depends of course on $\mu.$ Write $\underline{u}_{\Gamma,R}$ for one using
$\mu$ and $\underline{u}'_{\Gamma,R}$ for another one using $\mu',$ then
the difference $\underline{u}_{\Gamma,R}-\underline{u}'_{\Gamma,R}$ is supported on $0$
and of the form $c\delta_0$ with $c\in\R.$ This one-dimensional
space of possible extensions represents the renormalization ambiguity. \\\\
Here is an example. Let $M=\R^4.$ For
\begin{equation*}
\Gamma = \gammaeiner
\end{equation*}
we have
\begin{equation*}
\underline{u}_\Gamma(x)=u_0^2(x)=1/x^4,
\end{equation*}
the latter a distribution on $M^{V_0}\setminus \{0\}=M\setminus \{0\}.$ Pulling
back along $\beta,$
\begin{equation*}
(\beta^\ast \tilde u_\Gamma)_i |dy| = (\psi_i^{-1})^\ast \beta^\ast \tilde u_\Gamma |dy| = \frac{|dy|}{|y_i|(1+\sum_{j\neq i} y_j^2)^2}
\end{equation*}
in $U_i\setminus\{y_i=0\},$ $i=0,\ldots,3.$ As $\tilde u_\Gamma$ was
not defined at $0,$ $(\beta^\ast\tilde u_\Gamma)_i$ is not
defined at $\mathcal{E},$ given locally by $\{y_i=0\}.$ Raising
to the power $s$ gives
\begin{eqnarray*}
(\beta^\ast \tilde u^s_\Gamma)_i |dy| &=& \frac{|dy|}{|y_i|^{4s-3}(1+\sum_{j\neq i} y_j^2)^{2s}}\\
&= &\left(\frac{-\delta_0(y_i)}{2(s-1)}+o(s-1)^0\right) \frac{|dy|}{(1+\sum_{j\neq i} y_j^2)^{2s}}
\end{eqnarray*}
Therefore the residue density at $s=1$ is given, in this chart,
by
\begin{equation*}
\operatorname{res}_{s=1} (\beta^\ast \tilde u_\Gamma)^s_i |dy| = -\frac{1}{2}\delta_0(y_i) \frac{1}{(1+\sum_{j\neq i} y_j^2)^{2}}|dy|.
\end{equation*}
The residue is given as a projective integral by
\begin{equation*}
\operatorname{res} \Gamma =-\frac{1}{2}\int_\mathcal{E}\frac{\sum_i (-1)^i Y_i dY_1\wedge\ldots\wedge \widehat{dY_i}\wedge\ldots\wedge dY_4}{Y^4}
\end{equation*}
where $Y_1,\ldots,Y_4$ are homogeneous coordinates. In any of
the charts $V_i,$ and for the integration one chart suffices,
\begin{equation*}
\operatorname{res}\Gamma = -\frac{1}{2}\int_{V_i}\frac{dy_1\wedge\ldots\wedge \widehat{dy_i}\wedge\ldots\wedge dy_4}{(1+\sum_{j\neq i} y_j^2)^{2}}.
\end{equation*}
As mentioned before, there is a 1-dimensional space of possible
extensions $u_{\Gamma,R}$ due to the choice of $\mu$ that needs
to be made. There is no canonical $\mu.$ However from practice
in momentum space the following choice is useful. In momentum
space, the ill-defined Fourier transform of $u_0^2$ is
\begin{equation*}
(\mathcal{F} u_0)^{\ast 2}:\quad p\mapsto \int \frac{d^4k}{k^2(k-p)^2}.
\end{equation*}
A regularization or cutoff is now being understood in the
integral. It can be renormalized, for example, by subtracting
the value at $p^2=m^2$ where $m>0$ has the meaning of an energy
scale.
\begin{equation*}
(\mathcal{F} u_0)_R^{\ast 2}:\quad p\mapsto \int \frac{d^4k}{k^2(k-p)^2}-\left.\int \frac{d^4k}{k^2(k-p)^2}\right|_{p^2=m^2}
\end{equation*}
This prescription has the advantage that it is useful for
calculations beyond perturbation theory. The Fourier transform
of the distribution $\delta(p^2-m^2)$ is a Bessel function
$\mu(x)$ (with noncompact support), which can be approximated
by a sequence $\mu_n\rightarrow\mu$ of test functions $\mu_n$
with compact support. Since $m>0,$ $\mu\neq \underline{1},$ and
infrared divergences do not occur (as long as the position space test
function has compact support, i.~e.~ one does not evaluate the Fourier
transform at $p^2=0$). \\\\
In the case of primitive graphs, the renormalization operation
described above can be performed, and the residue be defined,
while on $M^{V_0},$ without blowing up. For general graphs
however blowing up provides an advantage, as will be shown in
section \ref{s:renormalization}: All divergences can be removed
at the same time while observing the physical principle of
locality. This concludes our discussion of primitive
divergences, and we start with the general theory for arbitrary
graphs.
\section{Models for the complements of subspace arrangements}\label{s:models}
In section \ref{s:arrangements} a description of the singular
support of $\underline{u}_\Gamma$ and of the locus where $\underline{u}_\Gamma$ fails
to be locally integrable was given as subspace arrangements in
a vector space. In general both $M^{V_0}_{sing}(\Gamma)$ and
$M^{V_0}_{div}(\Gamma)$ will not be cartesian products of
simpler arrangements. In this section we describe birational
models for $M^{V_0}$ where the two subspace arrangements are
transformed into normal crossing divisors. For this purpose it
is convenient to use results of De Concini and Procesi
\cite{deCPro} on more general subspace arrangements. See also
the recent book \cite{deCProBook} for a general introduction to
the subject. Although for the results of the present paper only
the smooth models for the divergent arrangements
$M^{V_0}_{div}(\Gamma)$ are needed, it is very instructive,
free of cost, and useful for future application to primitive
graphs, to develop the smooth models for the singular
arrangements $M^{V_0}_{sing}(\Gamma)$ at the same time.
\subsection{Smooth models and normal crossing divisors}\label{ss:modelsgeneral}
Consider for a finite dimensional real vector space $V$ a
collection $\mathcal{C}=\{A_1,\ldots,A_m\}$ of subspaces $A_i$
of $V^\vee$ and the corresponding arrangement
$V_\mathcal{C}=\bigcup_{A\in \mathcal{C}} A^\perp$ in $V.$ In order to explain our language, let us temporarily also consider
the corresponding arrangement in $V(\C)=V\otimes\C,$ denoted $V_\mathcal{C}(\C).$  The
problem is to find a smooth complex variety $Y_\mathcal{C}(\C)$ and a
proper surjective morphism $\beta:
Y_\mathcal{C}(\C)\rightarrow V(\C)$ such that
\begin{enumerate}
\item[(1)] $\beta$ is an isomorphism outside of
    $\beta^{-1}(V_\mathcal{C})(\C).$
\item[(2)] The preimage $\mathcal{E}(\C)$ of $V_\mathcal{C}(\C)$ is
    a divisor with normal crossings, i.~e.~ there are local
    coordinates $z_1,\ldots,z_n$ for $Y_\mathcal{C}(\C)$ such
    that $\beta^{-1}(V_\mathcal{C})(\C)$ is given in the chart
    by the equation $z_1\cdot\ldots \cdot z_k=0.$
\item[(3)] $\beta$ is a composition of blowups along smooth
    centers.
\end{enumerate}
Such a map $\beta:Y_\mathcal{C}(\C)\rightarrow V(\C)$ is called a
\emph{smooth model for} $V_\mathcal{C}(\C).$ Since $\beta$ is a
composition of blowups, it is a birational equivalence.
By the
classical result of Hironaka it is clear that for much more
general algebraic sets such a model always
exists in characteristic 0. For the special case of subspace
arrangements $V_\mathcal{C}$ a comprehensive and very useful
treatment is given in \cite{deCPro}. It will be instructive to
not only consider one smooth model, but a family of smooth
models $Y_\mathcal{P}$ constructed below along the lines of \cite{deCPro}. \\\\
The arrangement $V_\mathcal{C}$ is defined over $\R$ (in the case of the graph arrangements even over $\Z$) and therefore the real locus $Y_\mathcal{P}(\R)$ a
real $C^\infty$ manifold. We will only be working with the real loci in this paper
and simply write $Y_\mathcal{P}$ for $Y_\mathcal{P}(\R),$ $\mathcal{E}$ for $\mathcal{E}(\R)$ and so on. Also in the real context
we simply call $Y_\mathcal{P}$ the smooth model, $\mathcal{E}$ the exceptional divisor, and speak of birational maps, isomorphisms etc.~
without further justification.\\\\
By
abuse of language, a smooth model may be seen as a
''compactification'' of the complement of the arrangement, for
if $K\subset V$ is compact, then $\beta|_{\beta^{-1}(K)}$ is a
compactification of
$(V\setminus V_\mathcal{C})\cap K$ since $\beta$ is proper. \\\\
In the following we construct the smooth models of De Concini
and Procesi for the special case of $V=M^{V_0}$ and
$\mathcal{C}=\mathcal{C}_{sing}(\Gamma)$ or
$\mathcal{C}=\mathcal{C}_{div}(\Gamma).$
\subsection{The Wonderful Models}\label{ss:wonderful}
For a real vector space $V$ write $\mathbb{P}(V)$ for the projective
space of lines in $V.$ For any subspace $U$ of $V$ there is an
obvious map $V\setminus U\rightarrow V/U\rightarrow \mathbb{P}(V/U).$
The smooth models of De Concini and Procesi, called ''wonderful
models'', are defined as the closure $Y_\mathcal{P}$ of the
graph of the map
\begin{equation}\label{eq:closure}
V\setminus V_\mathcal{C} \rightarrow \prod_{A\in \mathcal{P}} \mathbb{P}(V/A^\perp)
\end{equation}
(the closure taken in $V\times
\prod_{A\in\mathcal{P}}\mathbb{P}(V/A^\perp))$ where
$\mathcal{P}$ is a subset of $\mathcal{C},$ subject to certain
conditions, to be defined below. The set $\mathcal{P}$ controls
what the irreducible components of the divisor $\mathcal{E}$
are, and how they intersect. In other words, one gets different
smooth models as one varies the subset $\mathcal{P}.$ We assume
that the collection $\mathcal{C}$ is closed under sum. The
following definition describes the most basic combinatorial
idea for the wonderful models.
\begin{dfn}\label{dfn:building}
A subset $\mathcal{P}$ of $\mathcal{C}$ is a \emph{building
set} if every $A\in \mathcal{C}$ is the direct sum
$A=\bigoplus_i B_i$ of the maximal elements $B_i$ of
$\mathcal{P}$ that are contained in $A,$ such that, in
addition, for every $C\in\mathcal{C}$ with $C\subseteq A$ also
$C=\bigoplus_i (C\cap B_i).$ Elements of a building set are
called \emph{building blocks}.
\end{dfn}
Our definition is a slight specialization of the one in
\cite[Theorem (2) in �2.3]{deCPro}. In their notation, our
building sets $\mathcal{P}$ are those for which
$\mathcal{C}=\mathcal{C}_{\mathcal{P}}$ (see
\cite[�2.3]{deCPro}). Note that a building set is not in
general closed under sum again. \dref{dfn:building} singles out
subsets $\mathcal{P}$ of $\mathcal{C}$ for which taking the
closure of (\ref{eq:closure}) makes sense. Indeed one has
\begin{thm}[De Concini, Procesi]\label{thm:decPro} If $\mathcal{P}$ is a building set, then the closure $Y_\mathcal{P}$
of the graph of (\ref{eq:closure}) provides a smooth model for
the arrangement $V_\mathcal{C}.$ Its divisor $\mathcal{E}$ is
the union of smooth irreducible components $\mathcal{E}_A,$ one
for each $A\in\mathcal{P}.$ \hfill $\Box$
\end{thm}
\subsection{Irreducibility and building sets} \label{ss:building}
Let us now turn toward the building sets and the wonderful
models for $V=M^{V_0}$ and
$\mathcal{C}=\mathcal{C}_{sing}(\Gamma)$ or
$\mathcal{C}_{div}(\Gamma).$ We review some basic notions from
\cite{deCPro} and apply them to the special case of graph
arrangements.
\begin{dfn} \label{dfn:irreducible}
For an $A\in \mathcal{C}$ a \emph{decomposition} of $A$ is a
family of non-zero $A_1,\ldots,A_k\in \mathcal{C}$ such that
$A=A_1\oplus\ldots\oplus A_k$ and, for every $B\subset A,
B\in \mathcal{C},$ also $B\cap A_1,\ldots,B\cap A_k\in
\mathcal{C}$ and $B=(B\cap A_1)\oplus\ldots\oplus (B\cap A_k).$
If $A$ admits only the trivial decomposition it is called
\emph{irreducible}. The set of irreducible elements is denoted
$\mathcal{F(C)}.$
\end{dfn}
By induction on the dimension each $A\in\mathcal{C}$ has
a decomposition into irreducible subspaces (This decomposition can be seen to be unique \cite[Prop.~2.1]{deCPro}).\\\\
It is easily seen that $A$ is irreducible if and only if there
are no $A_1,A_2\in \mathcal{C}$ such that $A=A_1\oplus A_2$ and
$B=(B\cap A_1)+(B\cap A_2)$ for all $B\subset A, B\in
\mathcal{C}.$ For if $A=A_1\oplus A_2\oplus A_2'$ is a
decomposition of $A,$ then $A=A_1\oplus (A_2\oplus A_2')$ is a
decomposition of $A$ into two terms
since $(B\cap A_2)\oplus(B\cap A_2')\subseteq B\cap(A_2\oplus A_2').$ This observation can be improved as follows.
\begin{lem}\label{lem:irredmin}
For $A\in \mathcal{C}$ to be irreducible it is
\begin{enumerate}
\item[(i)] \emph{sufficient} that for all $A_1,A_2\in \mathcal{C}$ one of which is \emph{irreducible}, satisfying $A=A_1\oplus A_2$ there is a $B\in\mathcal{C},$ $B\subset A,$
such that $B\neq (B\cap A_1)+(B\cap A_2),$ and
\item[(i)] \emph{necessary} that for all $A_1,A_2\in \mathcal{C}$ with $A=A_1\oplus A_2$ there is an \emph{irreducible} $B\in\mathcal{C},$ $B\subset A,$ such that
$B\neq (B\cap A_1)+(B\cap A_2).$
\end{enumerate}
\end{lem}
\emph{Proof.} (i) This follows from the existence of a decomposition into irreducible elements (remark after the definition). \\
(ii) Let $A=A_1\oplus A_2$ and $B\subset A, B\in\mathcal{C}.$ Let us say $B$ \emph{disturbs} if $B\neq (B\cap A_1)+(B\cap A_2).$
Assume $B$ disturbs.
Let $B=B'\oplus B_r$ be a decomposition with $B'$ irreducible. If neither $B'$ nor $B_r$ disturbed, then neither would $B,$ for $B=B'+B_r=(B'\cap A_1)$ $+(B'\cap A_2)$ $+(B_r\cap A_1)$ $+(B_r\cap A_2)$
$\subseteq (B'+
B_r)\cap A_1+(B'+B_r)\cap A_2$ $=B\cap A_1+B\cap A_2.$ Consequently $B'$ or (using induction $B\rightarrow B_r$) an irreducible component of $B_r$ is an irreducible disturbing element. \hfill$\Box$ \\\\
We now describe the irreducible elements of
$\mathcal{C}_{sing}(\Gamma),\mathcal{C}_{div}(\Gamma).$
Recall from section \ref{ss:polydiagonals} that a subgraph $\gamma$ is
called connected if it is connected with respect to the set of non-isolated vertices
$V_{\operatorname{eff}}(\gamma).$
For two partitions $P_1,P_2$ on a given set
write $P_1\le P_2$ if $Q\in P_1$ implies
$Q\subseteq Q'$ for some $Q'\in P_2.$ Write $P_1< P_2$ if
$P_1\le P_2$ and $P_1\neq P_2.$
\begin{dfn}\label{dfn:irredgraph}
Let $\mathcal{G}$ be a collection of subgraphs of $\Gamma.$ A
subgraph $\gamma$ of $\Gamma$ is called \emph{irreducible wrt.~
}$\mathcal{G}$ if for all subgraphs
$\gamma_1,\gamma_2\in\mathcal{G},$ one of them assumed connected, -- defining partitions
$P_1=cc(\gamma_1),P_2=cc(\gamma_2)$ on $V(\gamma)$ -- such that
$P_1\cup P_2=\operatorname{cc}(\gamma)$ and $P_1\cap P_2=0$
there exists a connected subgraph $g\in\mathcal{G}$ with
$\operatorname{cc}(g)\le\operatorname{cc}(\gamma)$ which is not
the union of a subgraph in $P_1$ with a subgraph in $P_2.$ (A
subgraph in $P_i$ is a subgraph $g_i$ of $\Gamma$ such that
$cc(g_i)\cap P_i=cc(g_i).)$ Otherwise $\gamma$ is called \emph{reducible}.
\end{dfn}
It follows from the definition that all subgraphs with only two connected
vertices $(|V_{\operatorname{eff}}(\gamma)|=2)$ are irreducible
(because there are no such $P_1$ and $P_2$ at all). Also, every
irreducible graph is connected. Indeed, let $\gamma$ be
irreducible wrt.~ $\mathcal{G}$ and $\gamma$ have for example two
components $\gamma=\gamma_1\sqcup\gamma_2.$ Taking
$P_1=\operatorname{cc}(\gamma_1)$ and
$P_2=\operatorname{cc}(\gamma_2)$ one arrives at a
contradiction (See also \pref{pro:irreddiv} later for a reason why this
argument works for $\mathcal{G}$ the set of divergent graphs).
Note that the notion of irreducibility of $\gamma$ wrt.~ $\mathcal{G}$ depends only on $\operatorname{cc}(\gamma)$ and $\mathcal{G}.$ \\\\
It turns out that the irreducible graphs are exactly those
which provide irreducible subspaces:
\begin{pro} $\left.\right.$ \label{pro:irrgraphsandspaces}
\begin{equation}\label{eq:irredsing}
\mathcal{F}(\mathcal{C}_{sing}(\Gamma))=\{A_\gamma\in \mathcal{C}_{sing}(\Gamma): \gamma \mbox{ irred.~ wrt.~ all subgraphs of }\Gamma\},
\end{equation}
\begin{eqnarray}\label{eq:irreddiv}
&\mathcal{F}(\mathcal{C}_{div}(\Gamma))=\{A_\gamma\in \mathcal{C}_{div}(\Gamma):& \gamma \mbox{ divergent and irreducible wrt.~}\\
&& \mbox{all divergent subgraphs of }\Gamma\},\nonumber
\end{eqnarray}
\begin{equation}\label{eq:irredkn}
\mathcal{F}(\mathcal{C}_{sing}(K_n))=\{A_\gamma\in \mathcal{C}_{sing}(K_n): \gamma \mbox{ connected }\}.
\end{equation}
\end{pro}
\emph{Proof.} (\ref{eq:irredsing})-(\ref{eq:irreddiv}):
Using the fact that irreducible graphs are connected and \lref{lem:irredmin},
one can apply \pref{pro:intersection} and
\pref{pro:union} to transform the statements $A_\gamma=A_{\gamma_1}\oplus
A_{\gamma_2}$ and $A_g = A_g\cap A_{\gamma_1} + A_g\cap
A_{\gamma_2}$ into
$\operatorname{cc}(\gamma)=\operatorname{cc}(\gamma_1)\cup\operatorname{cc}(\gamma_2),$
$\operatorname{cc}(\gamma_1)\cap \operatorname{cc}(\gamma_2)=0$
and
$\operatorname{cc}(g)=(\operatorname{cc}(g)\cap\operatorname{cc}(\gamma_1))\cup(\operatorname{cc}(g)\cap\operatorname{cc}(\gamma_2)).$ \\
(\ref{eq:irredkn}): Since the connectedness of $\gamma$ is
necessary for $A_\gamma$ to be irreducible, we only need to show sufficiency.
Let therefore $\gamma,\gamma_1,\gamma_2$ be connected subgraphs
of $K_n$ such that
$\operatorname{cc}(\gamma)=\operatorname{cc}(\gamma_1)\cup\operatorname{cc}(\gamma_2)$
and $\operatorname{cc}(\gamma_1)\cap
\operatorname{cc}(\gamma_2)=0.$ Pick an edge $e\in E(K_n)$
which joins a vertex in $V_{\operatorname{eff}}(\gamma_1)$ with
one in $V_{\operatorname{eff}}(\gamma_2).$ This gives an
$A_e\in \mathcal{C}_{sing}(K_n)$ such that $A_e\cap
A_{\gamma_1}= A_e\cap A_{\gamma_2}=\{0\}.$ Consequently
$A_\gamma$ is irreducible.
\hfill $\Box$ \\\\
Recall the definition of a building set, \dref{dfn:building}, which we can now rephrase as follows: All $A\in\mathcal{C}$ have a decomposition (in the sense of \dref{dfn:irreducible}) into the \emph{maximal} building blocks contained in $A.$ \\\\
The irreducible elements $\mathcal{F}(\mathcal{C})$ of a
collection $\mathcal{C}$ are the minimal building set for the
compactification of $V\setminus \bigcup_{A\in \mathcal{C}} A^\perp.$
\begin{pro} \label{pro:allbuildingsets}
The irreducible elements $\mathcal{F}(\mathcal{C}),$ and
$\mathcal{C}$ itself, form building sets in $\mathcal{C},$ and
$\mathcal{F}(\mathcal{C})\subseteq \mathcal{P}\subseteq
\mathcal{C}$ for every building set $\mathcal{P}$ in
$\mathcal{C}.$
\end{pro}
\emph{Proof.} (see also \cite{deCPro}[Proposition 2.1 and
Theorem 2.3 (3)]) Every $A\in\mathcal{C}$
has a decomposition into irreducible elements $B_i$. Assume one
of them is not maximal, say $A=\bigoplus_i B_i$ with
$B_1\subsetneq B\in\mathcal{F(C)}.$ Let $C\in\mathcal{C},$
$C\subset B,$ then $B=\bigoplus_i (B\cap B_i)$ with
$C=\bigoplus_i (C\cap B_i)=\bigoplus_i C\cap (B\cap B_i)$ would
be a nontrivial decomposition of $B.$ Therefore
$\mathcal{F}(\mathcal{C})$ is a building set. Let now
$\mathcal{P}$ be an arbitrary building set, and
$A\in\mathcal{F(C)}.$ There is a decomposition of $A$ into
maximal building blocks, but since $A$ is irreducible the
decomposition is trivial and $A$ is a building block itself.
Consequently $\mathcal{F}(C)\subseteq \mathcal{P}.$ The remaining statements are obvious. \hfill $\Box$ \\\\
We conclude this section with a short remark about reducible
divergent graphs.
\begin{pro}\label{pro:irreddiv}
Let $\gamma\subseteq \Gamma$ be divergent, and let
$A_\gamma=A_{\gamma_1}\oplus\ldots \oplus A_{\gamma_k}$ be a
decomposition in $\mathcal{C}_{div}(\Gamma).$ We may assume
that the $\gamma_i$ are saturated, that is
$\gamma_i=(\gamma_i)_s.$ Then all $\gamma_i$ are divergent
themselves.
\end{pro}
\emph{Proof.} Using (\ref{eq:newexactseq}), we need to conclude
$(d-2)|E(\gamma_i)|=\dim A_{\gamma_i}$ from
$(d-2)|E(\gamma)|=\dim A_{\gamma}.$ Since the $\gamma_i$
decompose $\gamma$ and are saturated, we have a disjoint union
$E(\gamma)=E(\gamma_1)\sqcup\ldots \sqcup E(\gamma_k).$ Also
$\dim A_{\gamma}=\sum_i \dim A_{\gamma_i}.$ Consequently, if we
had an $i$ such that $(d-2)|E(\gamma_i)|\lneq \dim
A_{\gamma_i},$ then there would be a $j$ such that
$(d-2)|E(\gamma_j)|\gneq \dim A_{\gamma_j},$ in contradiction
to $\Gamma$ being at most logarithmic (see \dref{dfn:amlog}).
\hfill$\Box$
\subsection{Nested sets}\label{ss:nested}
Let $\mathcal{P}$ be a building set in $\mathcal{C}.$  We are
now ready to describe the wonderful models $Y_\mathcal{P}.$
Note that $V_\mathcal{C}=V_{\mathcal{F}(\mathcal{C})}$ since
$(A_1\oplus A_2)^\perp=A_1^\perp\cap A_2^\perp.$ Consequently,
using \pref{pro:allbuildingsets},
$V_\mathcal{C}=V_\mathcal{P}.$ The charts for $Y_\mathcal{P}$
are assembled from \emph{nested} sets of subspaces, defined as
follows (see also \cite[Section 2.4]{deCPro})
\begin{dfn}\label{dfn:nested} A subset $\mathcal{N}$ of $\mathcal{P}$ is \emph{nested wrt.~} $\mathcal{P}$ (or $\mathcal{P}$-\emph{nested}) if for any $A_1,\ldots,A_k\in \mathcal{N}$ pairwise non-comparable we have $\sum_{i=1}^k A_i\not\in \mathcal{P}$ (unless $k=1).$
\end{dfn}
Note that in particular the $\mathcal{F}(\mathcal{C})$-nested
sets are sets of \emph{irreducible} subspaces. We now determine
the $\mathcal{P}$-nested sets of
$\mathcal{C}=\mathcal{C}_{sing}(\Gamma),$
$\mathcal{C}_{div}(\Gamma),$ $\mathcal{C}_{sing}(K_n)$ for the
minimal and maximal building sets
$\mathcal{P}=\mathcal{F}(\mathcal{C})$ and
$\mathcal{P}=\mathcal{C},$ respectively. Let $\gamma$ be a
subgraph of $\Gamma.$ Recall from section
\ref{ss:polydiagonals} that $A_\gamma$ depends only on the
partition $cc(\gamma)$ of the vertex set $V(\Gamma).$
\begin{pro} \label{pro:nested} A subset $\mathcal{N}=\{A_{\gamma_1},\ldots,A_{\gamma_k}\}$ is nested in $\mathcal{C}=\mathcal{C}_{sing}(\Gamma)$ (resp.~ $\mathcal{C}_{div}(\Gamma))$
\begin{enumerate}
\item[(i)] wrt.~ $\mathcal{P}=\mathcal{C}$ if and only if
    the set $\{cc(\gamma_1),\ldots,cc(\gamma_k)\}$ is
    linearly ordered by the strict order $<$ of partitions,
\item[(ii)] wrt.~ $\mathcal{P}=\mathcal{F}(\mathcal{C})$ if
    and only if the $\gamma_i$ are irreducible wrt.~ all
    (divergent) subgraphs of $\Gamma,$ and for all
    $I\subseteq \{1,\ldots,k\},$ $|I|\ge 2,$ the graph
    $\bigcup_{i\in I} \gamma_i$ is reducible wrt.~
    (divergent) subgraphs, unless
    $cc(\gamma_i)<cc(\gamma_j)$ for some $i,j\in I.$
\end{enumerate}
\end{pro}
Recall that a union $\bigcup_i \gamma_i$ is reducible for example if the $\gamma_i$ are pairwise disjoint. \\\\
\emph{Proof. } Straightforward from the definitions. \hfill
$\Box$
\begin{pro} \label{pro:nestedkn} A subset $\mathcal{N}=\{A_{\gamma_1},\ldots,A_{\gamma_k}\}$ is nested
in $\mathcal{C}_{sing}(K_n)$ wrt.~ the minimal building set if
and only if the $\gamma_i$ are connected and for $i\neq j$ if
either $V_{\operatorname{eff}}(\gamma_i)\subset
V_{\operatorname{eff}}(\gamma_j),$
$V_{\operatorname{eff}}(\gamma_j)\subset
V_{\operatorname{eff}}(\gamma_i),$ or
$V_{\operatorname{eff}}(\gamma_i)\cap
V_{\operatorname{eff}}(\gamma_j)=\emptyset.$
\end{pro}
\emph{Proof.} Straightforward from (\ref{eq:irredkn}). \hfill $\Box$ \\\\
We recall further notions from \cite[Section 2]{deCPro}. Let
$\mathcal{P}$ be a building set and $\mathcal{N}$ a
$\mathcal{P}$-nested set for $\mathcal{C}.$ For every $x\in
V^\vee\setminus \{0\},$ the set of subspaces in
$\mathcal{N}'=\mathcal{N}\cup \{V^\vee\}$ containing $x$ is
linearly ordered by inclusion and non-empty. Write $p(x)$ for the minimal
element in $\mathcal{N}'.$ This defines a map
$p:V^\vee\setminus \{0\}\rightarrow \mathcal{N}'.$
\begin{dfn} \label{dfn:adapted} A basis $\mathcal{B}$ of $V^\vee$ is \emph{adapted} to $\mathcal{N}$ if, for all $A\in \mathcal{N}$ the set $\mathcal{B}\cap A$ generates $A.$ A \emph{marking} of $\mathcal{B}$ is, for all $A\in \mathcal{N},$ the choice of an element $x_A\in \mathcal{B}$ with $p(x_A)=A.$
\end{dfn}
In the case of arrangements coming from graphs,
$\mathcal{C}=\mathcal{C}_{sing}(\Gamma),\mathcal{C}_{div}(\Gamma),$
particular bases are obtained from spanning forests,
cf.~\pref{pro:spanningbasis}.
\begin{pro} \label{pro:adapted}
Let $t$ be a spanning tree of $\Gamma.$ Then the basis $\mathcal{B}=\{(e^\vee
i_\Gamma)^{j}: \, e\in E(t),\, j=0,\ldots,d-1\}$ of
$(M^{V_0})^\vee$ is adapted to $\mathcal{N}=\{A_{\gamma_1},
\ldots,A_{\gamma_k}\}$ if and only if the graph with edges
$\{e\in E(t): e\le cc(\gamma_i)\}$ is a spanning forest for
$cc(\gamma_i)$ for all $i=1,\ldots,k.$
\end{pro}
\emph{Proof.} Straightforward from \pref{pro:spanningbasis}.\hfill $\Box$\\\\
We call such a spanning forest an \emph{adapted spanning
forest}. Also, a marking of the basis corresponds to a certain
subforest $E(t_M)\subseteq E(t)$ with $k+1$ edges, and a choice
of one out of $d$ upper indices for each edge.
\begin{pro} \label{pro:adaptedexists}
Let $\mathcal{N}$ be a $\mathcal{P}$-nested set for
$\mathcal{C}=\mathcal{C}_{sing}(\Gamma)$ or
$\mathcal{C}_{div}(\Gamma).$ Then there exists an adapted
spanning tree.
\end{pro}
\emph{Proof. } By induction on the dimension: Let
$A_{\gamma_1},\ldots,A_{\gamma_h}$ be the maximal elements in
$\mathcal{N}$ contained in a given $A_\gamma.$ Assume an
adapted spanning forest (see \pref{pro:adapted}) for each of
the $A_{\gamma_i}$ is chosen. The union of these bases is then
a basis $\mathcal{B'}$ for $\bigoplus_i A_{\gamma_i}$ (the sum
is direct because $\mathcal{N}$ is nested and the
$A_{\gamma_i}$ maximal). The set $\{(e^\vee i_\Gamma)^j: e\in
E(\gamma)\}$ is a generating set for $A_\gamma.$ Extending the
basis $\mathcal{B'}$ to a basis for $A_{\gamma}$ using elements of this
generating set provides, by \pref{pro:spanningbasis}, an
adapted spanning forest for $\gamma.$ \hfill $\Box$\\\\
Let us now return to marked bases in general. A marking of an
adapted basis $\mathcal{B}$ provides a partial order on
$\mathcal{B}:$ $y_1\preceq y_2$ if $p(y_1)\subseteq p(y_2)$ and
$y_2$ is marked. This partial order determines a map $\rho:
V\rightarrow V$ as follows. Consider the elements of
$\mathcal{B}=\{y_1,\ldots,y_m\}$ as (nonlinear) coordinates on the
source $V.$ The (linear) coordinates $(x_1,\ldots,x_m)$ of
the image $\rho(y_1,\ldots,y_m)$ are given by
\begin{equation}\label{eq:blowupcoordinates}
x_i = \prod_{y_i\preceq y_j} y_j = \left\{ \begin{array}{ll} y_i\prod_{p(y_i)\subset A} y_A & \mbox{if }y_i\mbox{ is not marked,}\\
\prod_{p(y_i)\subset A} y_A & \mbox{if }y_i\mbox{ is marked.} \end{array}\right.
\end{equation}
The map $\rho,$ and already the partial order $\preceq,$
determine implicitly a sequence of blowups. Indeed
\begin{pro} (see \cite[Lemma 3.1]{deCPro}) \label{pro:blowupcoord}
\begin{enumerate}
\item[(i)] $\rho$ is a birational morphism,
\item[(ii)] $\rho(\{y_A=0\})=A^\perp$ and
\item[(iii)] $\rho$ restricts to an isomorphism
    $V\setminus \bigcup_{A\in\mathcal{N}}\{y_A=0\}\cong
    V\setminus \bigcup_{A\in\mathcal{N}}A^\perp.$
\item[(iv)] Let $x\in V^\vee \setminus  \{0\}$ and
    $p(x)=A\in\mathcal{N}.$ Then $x=x_A P_x(y_i),$ where
    $x_A=\prod_{y_A\preceq y_i} y_i$ and $P_x$ is a
    polynomial depending on the variables $y_i<x_A,$ and
    linear in each variable, that is $\partial^2
    P_x/\partial y_i^2=0.$
\end{enumerate} \hfill $\Box$
\end{pro}
\subsection{Properties of the Wonderful Models} \label{ss:propwonderful}
Recall the definition (\ref{eq:closure}) of the wondeful
models: $Y_\mathcal{P}$ is the closure of $V\setminus V_\mathcal{P}$
in $V \times
\prod_{A\in\mathcal{P}}\mathbb{P}(V/A^\perp).$ The birational
map $\beta:Y_\mathcal{P}\rightarrow V$ is simply the projection
onto the first factor $V.$ Let $\mathcal{N}$ be a
$\mathcal{P}$-nested set in $\mathcal{C},$ and $\mathcal{B}$ an
adapted, marked basis of $V^\vee.$ Both determine a birational
map $\rho: V\rightarrow V$ as defined in
(\ref{eq:blowupcoordinates}). For a given building block $B\in
\mathcal{P}$ set $Z_B =\{P_x=0, x \in B\}\subset V.$ The
composition of $\rho$ with the rational map $V\rightarrow
V/A^\perp\rightarrow \mathbb{P}(V/A^\perp)$ is then defined as
a regular morphism outside of $Z_B.$ Doing this for every
factor in $\prod_{A\in\mathcal{P}}\mathbb{P}(V/A^\perp),$ one
gets an open embedding $j^\mathcal{B}_\mathcal{N}:
U^\mathcal{B}_\mathcal{N}=V\setminus \bigcup_{B\in\mathcal{P}}
Z_B\hookrightarrow Y_\mathcal{P}$ \cite[Theorem 3.1]{deCPro}.
Write
$Y^\mathcal{B}_\mathcal{N}=j^\mathcal{B}_\mathcal{N}(U^\mathcal{B}_\mathcal{N}).$
As $\mathcal{N}$ and the marking of $\mathcal{B}$ vary, one
obtains an atlas
$((j^\mathcal{B}_\mathcal{N})^{-1},U^\mathcal{B}_\mathcal{N})$
for $Y_\mathcal{P}.$ Note that the sign convention of (\ref{eq:simplechart}) in order to make the orientation of
the exceptional divisor explicit is discontinued from here on. It is shown in \cite[Theorem
3.1]{deCPro} that the divisor
$\mathcal{E}=\beta^{-1}(V_\mathcal{P})$ is given locally by
\begin{equation}\label{eq:divisorlocally}
(j^\mathcal{B}_\mathcal{N})^{-1}(\mathcal{E}\cap Y^\mathcal{B}_\mathcal{N})=\left\{\prod_{A\in\mathcal{N}} y_A=0 \right\}.
\end{equation}
\begin{figure}
$$ \configspace $$
\caption{\label{fig:configspace} A picture of $\R^{V_0}_{sing}(K_4).$}
\end{figure}
\begin{figure}
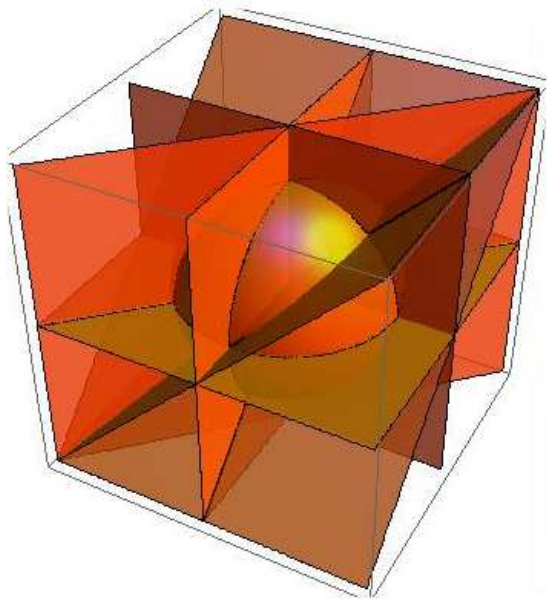

$$ \fmone $$
\caption{\label{fig:fmone} (Spherical) blowup of the origin in $\R^{V_0}_{sing}(K_4),$ where projective spaces are
replaced by spheres. The maximal wonderful
model would proceed by blowing up all strict transforms of lines incident to the exceptional divisor, and finally the strict
transforms of the planes.}
\end{figure}
\begin{figure}
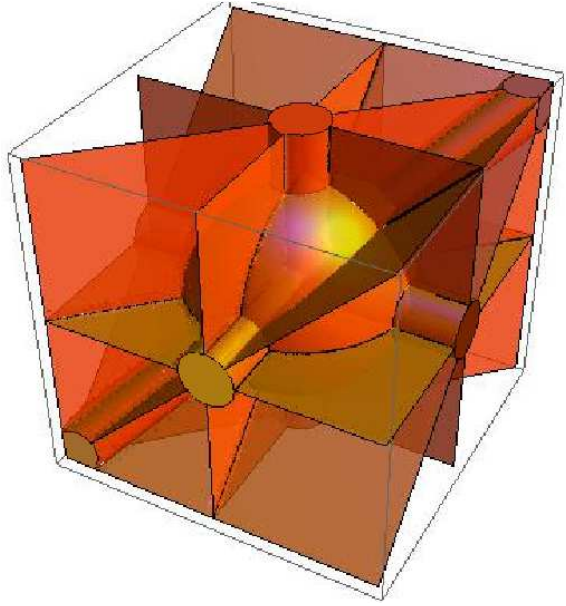

$$ \fmtwo $$
\caption{\label{fig:fmtwo} Minimal (spherical) model of $\R^{V_0}_{sing}(K_4),$ corresponding to the Fulton-MacPherson
compactification of the configuration space of 4 points in $\R.$ After the central blowup, only those
strict transforms of lines are blown up which are not a normal crossing intersection in the first place.}
\end{figure}
\emph{Remarks.} In the case of the complete graph $K_n,$ the
minimal wonderful model
$Y_{\mathcal{F}(\mathcal{C}_{sing}(K_n))}$ is known as the
Fulton-MacPherson compactification \cite{FM}, while the maximal
wonderful model $Y_{\mathcal{C}_{sing}(K_n)}$ has been
described in detail by Ulyanov \cite{Ulyanov}. For any graph,
the benefit of the minimal model is that the divisor is small
in the sense that it has only a minimal number of irreducible
components, whereas the actual construction by a sequence of
blowups is less canonical. On the other hand, for the maximal
model, which has a larger number of irreducible components, one
can proceed in the obvious way blowing up the center and then
strict transforms by increasing dimension. See figures \ref{fig:configspace},
\ref{fig:fmone}, \ref{fig:fmtwo} for an example where $M$ is supposed one-dimensional
in order to be able to draw a picture. Also the
resolution of projective hyperplane arrangements described in \cite{ESV}
and referred to in \cite[Lemma 5.1]{BEK}
proceeds by increasing dimension but corresponds to the minimal wonderful model nonetheless.
This is a special effect due to the fact that the strict transforms of hyperplanes, having codimension 1,
do not need to be blown up. If the subspaces in the arrangement have higher codimension, the blowup sequence
will be different. See \cite{FM,Ulyanov} and \cite[Theorem 3.2]{deCPro} for details.\\\\
\subsection{Examples}\label{ss:modelexamples}
For the fixed vertex set $V=\{1,2,3,4\}$ we consider a series
of graphs on $V$ with increasing complexity. Only some of them
are relevant for renormalization.
\begin{equation*}
\begin{array}{ll}
\Gamma_1 = \gammaone\quad & \Gamma_4 = \gammafour\\
\Gamma_2 = \gammatwo\quad& \Gamma_5 = \gammafive\\
\Gamma_3 = \gammathree\quad& \Gamma_6 = \gammasix
\end{array}
\end{equation*}
For these graphs, we examine the arrangements $M^{V_0}_{sing}$
and $M^{V_0}_{div},$ the irreducible subspaces and nested sets
for the minimal and maximal building set, respectively. Write
$A_{ij}$ for $A_e$ with $e$ an edge connecting the vertices $i$
and $j.$ Note that $A_{12}+A_{23}=A_{13}+A_{23}=A_{12}+A_{13}$ etc.,
and in the examples a choice of basis is made.
\begin{eqnarray*}
\mathcal{C}_{sing}(\Gamma_1) &=& \{A_{12},A_{23},A_{34},\mbox{ and sums thereof}\}\\
\left.\begin{array}{r}
\mathcal{C}_{sing}(\Gamma_2)\\
\mathcal{C}_{sing}(\Gamma_3)\\
\mathcal{C}_{sing}(\Gamma_4)\\
\end{array}\right\} &=& \{A_{12},A_{23},A_{24},A_{34},\mbox{ and sums thereof}\}\\
\mathcal{C}_{sing}(\Gamma_5) &=& \{A_{12},A_{13},A_{23},A_{24},A_{34},\mbox{ and sums thereof}\}\\
\mathcal{C}_{sing}(\Gamma_6) &=& \{A_{12},A_{13},A_{14},A_{23},A_{24},A_{34},\mbox{ and sums thereof}\}
\end{eqnarray*}
The divergent arrangements are determined by the following collections of
dual spaces:
\begin{eqnarray*}
\mathcal{C}_{div}(\Gamma_1) &=& \emptyset \\
\mathcal{C}_{div}(\Gamma_2) &=& \{A_{12}\} \\
\mathcal{C}_{div}(\Gamma_3) &=& \{A_{34},A_{23}+A_{34}\} \\
\mathcal{C}_{div}(\Gamma_4) &=& \{A_{12},A_{34},A_{23}+A_{34},A_{12}+A_{34},A_{12}+A_{23}+A_{34}\}\\
\mathcal{C}_{div}(\Gamma_5) &=& \{A_{34},A_{23}+A_{34},A_{12}+A_{23}+A_{34}\}\\
\mathcal{C}_{div}(\Gamma_6) &=& \{A_{12}+A_{23}+A_{34}\}
\end{eqnarray*}
The irreducible singular subspace collections are
\begin{eqnarray*}
\mathcal{F}(\mathcal{C}_{sing}(\Gamma_1)) &=& \{A_{12},A_{23},A_{34}\}\\
\left.\begin{array}{r}
\mathcal{F}(\mathcal{C}_{sing}(\Gamma_2))\\
\mathcal{F}(\mathcal{C}_{sing}(\Gamma_3))\\
\mathcal{F}(\mathcal{C}_{sing}(\Gamma_4))\\
\end{array}\right\} &=& \{A_{12},A_{23},A_{24},A_{34},A_{23}+A_{34}\}\\
\mathcal{F}(\mathcal{C}_{sing}(\Gamma_5)) &=& \{A_{12},A_{13},A_{23},A_{24},A_{34},\\
&&A_{12}+A_{13},A_{23}+A_{24},A_{12}+A_{23}+A_{34}\}\\
\mathcal{F}(\mathcal{C}_{sing}(\Gamma_6)) &=& \{A_{12},A_{13},A_{14},A_{23},A_{24},A_{34},\\
&&A_{12}+A_{13},A_{12}+A_{14},A_{13}+A_{14},A_{23}+A_{34},\\
&&A_{12}+A_{23}+A_{34}\}
\end{eqnarray*}
\emph{Remark.} Note that these irreducible single subspace collections are in one-to-one correspondence with the terms generated by the core Hopf algebra \cite{BlKr,KreimerSui} if one takes into account the multiplicities generated by a labeling of vertices. A detailed comparison is left to future work.\\\\
The irreducible divergent subspace collections are
\begin{eqnarray*}
\mathcal{F}(\mathcal{C}_{div}(\Gamma_1)) &=& \emptyset\\
\mathcal{F}(\mathcal{C}_{div}(\Gamma_2))&=& \{A_{12}\}\\
\mathcal{F}(\mathcal{C}_{div}(\Gamma_3))&=& \{A_{34},A_{23}+A_{34}\}\\
\mathcal{F}(\mathcal{C}_{div}(\Gamma_4)) &=& \{A_{12},A_{34},A_{23}+A_{34}\}\\
\mathcal{F}(\mathcal{C}_{div}(\Gamma_5)) &=& \{A_{34},A_{23}+A_{34},A_{12}+A_{23}+A_{34}\}\\
\mathcal{F}(\mathcal{C}_{div}(\Gamma_6)) &=& \{A_{12}+A_{23}+A_{34}\}
\end{eqnarray*}
The maximal nested sets of the divergent collection wrt.~ the
minimal building set:
\begin{eqnarray*}
\mbox{ for }\Gamma_1: && \emptyset \\
\mbox{ for }\Gamma_2: && \{A_{12}\}\\
\mbox{ for }\Gamma_3: && \{A_{23}+A_{34},A_{34}\}\\
\mbox{ for }\Gamma_4: && \{A_{12},A_{23}+A_{34},A_{34}\}\\
\mbox{ for }\Gamma_5: && \{A_{12}+A_{23}+A_{34},A_{23}+A_{34},A_{34}\}\\
\mbox{ for }\Gamma_6: && \{A_{12}+A_{23}+A_{34}\}
\end{eqnarray*}
The maximal nested sets of the divergent collection wrt.~ the
maximal building set:
\begin{eqnarray*}
\mbox{ for }\Gamma_1: && \emptyset \\
\mbox{ for }\Gamma_2: && \{A_{12}\}\\
\mbox{ for }\Gamma_3: && \{A_{23}+A_{34},A_{34}\}\\
\mbox{ for }\Gamma_4: && \{A_{12}+A_{23}+A_{34},A_{12}+A_{34},A_{12}\},\\
 && \{A_{12}+A_{23}+A_{34},A_{12}+A_{34},A_{34}\},\\
&& \{A_{12}+A_{23}+A_{34},A_{23}+A_{34},A_{34}\}\\
\mbox{ for }\Gamma_5: && \{A_{12}+A_{23}+A_{34},A_{23}+A_{34},A_{34}\}\\
\mbox{ for }\Gamma_6: && \{A_{12}+A_{23}+A_{34}\}
\end{eqnarray*}

\section{Laurent coefficients of the meromorphic extension}\label{s:poles}
\subsection{The Feynman distribution pulled back onto the wonderful model}\label{ss:pullback}
Recall the definition (\ref{eq:FGD}) of the Feynman
distribution $u_\Gamma = \prod_{i<j} u_0(x_i-x_j)^{n_{ij}}.$ We
write $\underline{u}_\Gamma=\Phi_\ast u_\Gamma$ where $\Phi$ is
the projection along the thin diagonal defined at the end of
section \ref{ss:singarrangements}. It is clear from the
discussion in section \ref{s:arrangements} that
$\underline{u}_\Gamma = (i_\Gamma^{\oplus d})^\ast u_0^{\otimes
|E(\Gamma)|}$ where defined. Let $\beta:Y_\mathcal{P}\rightarrow M^{V_0}$ be
a wonderful model for the arrangement $M^{V_0}_{div}(\Gamma)$
or $M^{V_0}_{sing}(\Gamma).$ The purpose of this section is
to study the regularized pullback $\beta^\ast \underline{\tilde
u}_\Gamma^s$ (as a density-valued meromorphic function of $s$)
of $ \underline{\tilde u}^s_\Gamma$ onto
$Y_\mathcal{P}\setminus \mathcal{E}.$
\begin{thm} \label{thm:factorization} Let $\mathcal{N}$ be a $\mathcal{P}$-nested set in $\mathcal{C}_{div}(\Gamma)$ $(\mathcal{C}_{sing}(\Gamma)),$
and $\mathcal{B}=\{y^i_e: \, e\in E(t),\, i=0,\ldots,d-1\}$ an
adapted basis with marked elements $y^{i_A}_A,\,
A\in\mathcal{N}.$ Then, in the chart
$U^\mathcal{B}_\mathcal{N},$
\begin{equation}\label{eq:factorization}
\beta^\ast \underline{u}_\Gamma(\{y^i_e\}) = f_\Gamma(\{y^i_e\})\prod_{A\in\mathcal{N}} (y^{i_A}_A)^{n_A}
\end{equation}
where $f_\Gamma\in L^1_{loc}(U^\mathcal{B}_\mathcal{N})$
$(C^\infty(U^\mathcal{B}_\mathcal{N})),$ and $n_A\in-2\N.$ More precisely
\begin{equation}\label{eq:na}
n_{A_\gamma} = (2-d) | E(\gamma_s) |.
\end{equation}
In addition, $f_\Gamma$ is $C^\infty$ in the variables $y^{i_A}_A,$
$A\in\mathcal{N}.$
\end{thm}
Note: $\gamma_s$ is the subgraph defined in \dref{dfn:saturated}. Divergent subgraphs are saturated (\pref{pro:saturated}).
We write $f_\Gamma(\{y_e^i\})$ for $f_\Gamma(y_{e_1}^0,\ldots,y_{e_{|E(t)|}}^{d-1})$ etc.\\\\
\emph{Proof. } Recall from section
\ref{ss:propwonderful} that the map $\beta$ is given in the
chart $U^\mathcal{B}_\mathcal{N}$ by $\rho$ (see
(\ref{eq:blowupcoordinates})):
\begin{equation*}
\rho: \sum_{j=0}^{d-1} \sum_{e\in E(t)}y_e^j b_e^j \mapsto \sum_{j=0}^{d-1} \sum_{e\in E(t)} \prod_{y_e^j\preceq y_{e'}^k} y^k_{e'}b_e^j
\end{equation*}
where $\preceq$ is the partial order on the basis
$\mathcal{B}=\{y_e^j\}$ of $(M^{V_0})^\vee$ adapted to
$\mathcal{N}.$ Consequently, using (\ref{eq:beta}),
\begin{eqnarray} \label{eq:produos}
\beta^\ast \underline{u}_\Gamma(\{y_e^j\})&=&
u_0^{\otimes E(\Gamma)}i_\Gamma^{\oplus d}\rho(\{y_e^j\}) \nonumber\\
&= &\prod_{e\in E(\Gamma)} u_0 \left(\left\{ \Sigma_{e'\leadsto e}\Pi_{y_{e'}^j\preceq y_{e''}^k} y^k_{e''} \right\}_{j=0}^{d-1}\right).
\end{eqnarray}
By \pref{pro:blowupcoord} (iv), each $\xi_e^j=\sum_{e'\leadsto
e} x_{e'}^j=\Sigma_{e'\leadsto e}\Pi_{y_{e'}^j\preceq y_{e''}^k} y^k_{e''}$ is a product $x_A^{i_A} P_{\xi_e^j}(\{y_j^i\})$
where $A=p(\xi_e^j)\in\mathcal{N}$ (Special case: $x_A^{i_A}=1$ if $p(\xi_e^j)\not\in\mathcal{N}).$ As $u_0$ is homogeneous
(\ref{eq:homogenu0}), the factor $x_A^{i_A}= \prod_{A \subseteq
B\in\mathcal{N}}{y_B}^{i_B},$ can be pulled out, supplied with
an exponent $2-d.$ Since $x_A^{i_A}= \prod_{A \subseteq
B}{y_B}^{i_B},$ the factor $(y_{A_\gamma}^{i_{A_\gamma}})^{2-d}$ occurrs
once for each $e\in E(\Gamma)$ such that $A_e\subseteq
A_\gamma,$ in other words for each $e\le cc(\gamma).$ Hence
(\ref{eq:na}). We finally show that the remaining factor
\begin{equation}\label{eq:fgamma}
f_\Gamma(\{y_i^j\}) = \prod_{e\in E(\Gamma)} u_0 (\{P_{\xi_e^j}(\{y_i^k\})\}_{j=0}^{d-1})
\end{equation}
of $\beta^\ast \underline{u}_\Gamma$ satisfies $f_\Gamma\in
L^1_{loc}(U^\mathcal{B}_\mathcal{N})$ if the divergent
arrangement was resolved or $f_\Gamma\in
C^\infty(U^\mathcal{B}_\mathcal{N})$ if the singular
arrangement was resolved, respectively. The set
$U_\mathcal{N}^\mathcal{B}$ contains by definition (see section \ref{ss:propwonderful}) no point
with coordinates $y_i^j$ such that for any building block $B\in
\mathcal{P}$ all $P_x(\{y_i^j\})=0,$ $x\in B.$ In the case of
$\mathcal{C}_{sing}(\Gamma),$ all $A_e\in \mathcal{P},$ $(e\in
E(\Gamma)),$ since they are irreducible, see
\pref{pro:allbuildingsets}. On the other hand, $A_e$ is spanned
by the $\xi_e^j,$ $j=0,\ldots,d-1.$ Therefore for no $e\in E(\Gamma)$ all $d$ of the
$P_{\xi_e^j}$ $(j=0,\ldots,d-1)$ in (\ref{eq:fgamma}) vanish on
$U_\mathcal{N}^\mathcal{B}.$ Hence, using
(\ref{eq:singsuppu0}), $f_\Gamma\in
C^\infty(U_\mathcal{N}^\mathcal{B}).$ In the case of
$\mathcal{C}_{div}(\Gamma),$ let $\gamma$ be divergent. By
\pref{pro:irreddiv} we may assume without loss that $A_\gamma$
is irreducible. Therefore $A_\gamma\in \mathcal{P}$ as in the
first case. By the same argument as above, not all the
$P_{\xi_e^j}$ in the arguments of $\prod_{e\in E(\gamma)}u_0$
can vanish at the same time on $U_\mathcal{N}^\mathcal{B},$
whence this product is now locally integrable. In order to see
that $f_\Gamma$ is $C^\infty$ in the $y_A^{i_A},$ it suffices to
show that not all $d$ of the $P_{\xi_e^j}(\{y_i^k\})\rightarrow
0$ (for $j=0,\ldots,d-1)$ as the $y_A^{i_A}\rightarrow 0$ while
the other coordinates are fixed. From \pref{pro:blowupcoord}
(iv) we know that every $P_x$ is linear in the $y_A^{i_A},$ if
therefore all $P_{\xi_e^j}$ vanished at some $y_A^{i_A}=0$ they
would have $y_A^{i_A}$ as a common factor. This contradicts
\pref{pro:blowupcoord} as then $p(\xi_e)\subseteq A.$ \hfill $\Box$  \\\\
In the preceding theorem, $\underline{u}_\Gamma$ was pulled
back along $\beta$ as a distribution. The next corollary
clarifies the situation for the density $\beta^\ast \tilde
u_\Gamma = \beta^\ast (\underline{u}_\Gamma |dx|).$ We write
$|dy|$ for $|dy_{e_1}^0 \wedge \ldots\wedge dy_{e_k}^{d-1}|.$
\begin{cor}\label{cor:factorization} Under the assumptions of \tref{thm:factorization},
\begin{equation}\label{eq:factorization2}
\beta^\ast \tilde u_\Gamma(\{y^i_e\}) |dy|= f_\Gamma(\{y^i_e\})\prod_{A\in\mathcal{N}} |y^{i_A}_A|^{m_A} |dy|
\end{equation}
where
\begin{equation}\label{eq:ma}
m_{A_\gamma} = 2|E(\gamma_s)|-d\dim H_1(\gamma_s) -1 \ge -1.
\end{equation}
In the case of the divergent arrangement
$\mathcal{C}_{div}(\Gamma),$ all $m_{A_\gamma}=-1,$ and
moreover
\begin{equation}\label{eq:factorizations}
\beta^\ast \tilde u_\Gamma^s(\{y^i_e\}) |dy|= f_\Gamma^s(\{y^i_e\})\prod_{A\in\mathcal{N}} |y^{i_A}_A|^{-d_A s+d_A-1} |dy|
\end{equation}
where $d_A=\dim A.$
\end{cor}
We also write $d_\gamma=d_{A_\gamma}.$ \\\\
\emph{Proof.} Formally,
\begin{eqnarray*}
|dx|&=& |\bigwedge_{e\in E(t),j=0\ldots,d-1} dx_e^j|=|\bigwedge d\prod_{y_e^j\preceq y_{e'}^k}y_{e'}^k
|\\
&=&\prod_{A\in\mathcal{N}}|y_A^{i_A}|^{q_A} |\bigwedge dy^j_e|
\end{eqnarray*}
where the $q_A$ are determined as follows. Since the $x_e^j,$
$(j=0,\ldots,d-1)$ span $A_e,$ the factor
$y_{A_\gamma}^{i_{A_\gamma}}$ appears from all $dx_e^j$ such
that $e\le cc(\gamma),$ except one, namely
$dx_{A_\gamma}^{i_{A_\gamma}}$ itself which corresponds to the
marking. Since $t$ is an adapted spanning tree, the set $\{e\in
E(t): e\le cc(\gamma)\}$ defines a spanning forest of $\gamma,$
and one concludes using \pref{pro:spanningbasis} that
$q_{A_\gamma}=d_\gamma-1.$ Finally note that $\dim
H_1(\gamma_s)=|E(\gamma_s)|-d_{\gamma}/d$ and $\Gamma$ is at
most logarithmic. \hfill $\Box$
\subsection{Combinatorial description of the Laurent coefficients}\label{ss:poles}
Let $V=V(\Gamma),$ $E=E(\Gamma)$ and $p: V\rightarrow V'$ a map
of sets which is not injective. In the dual this defines a map
$p^\vee: \R^{V'}\rightarrow \R^{V}$ sending $\sum_{v\in V'}
\alpha_{v'}v'$ to $\sum_{v\in V} \alpha_{p(v)} v.$ Let
$E(\gamma)\subseteq E(\Gamma).$ Then the graph $\gamma_p$ with
vertex set $V(\gamma_p)=V'$ and set of edges
$E(\gamma_p)=E(\gamma)$ such that
$\delta_{\gamma_p}=\delta_\gamma\circ p^\vee:
\R^{V(\gamma_p)}\rightarrow \R^{E(\gamma_p)}$ (see
(\ref{eq:newexactseq}))
is called \emph{the graph $\gamma$ contracted along p}. \\\\
Note: The graph contracted along $p$ may have loops. It is not necessarily a subgraph of $\Gamma$ anymore. \\\\
We assume, as in (\ref{eq:config}), a distinguished vertex
$v_0\in V(\Gamma)$ such that $V_0=V(\Gamma)\setminus \{v_0\}.$ Let now
$t$ be a spanning tree of $\Gamma$ and $s\subseteq t$ a
subforest of $t.$ This defines a map $p_{t,s}:
V(\Gamma)\rightarrow V(\Gamma)$ as follows: Let $v\in
V(\Gamma)$ be given. Since $t$ is a spanning tree of $\Gamma,$
there is a unique path $t_v$ in $t$ from $v_0$ to $v.$ Let
$p_{t,s}(v)$ be the unique vertex which is connected to $v$ by
edges of $s$ only and is nearest to $v_0$ on the path $t_v.$
\begin{figure}
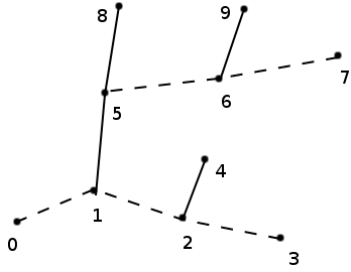

\begin{equation*} \pfade \end{equation*}
\caption{\label{fig:pfade} The edges of $s$ are broken lines, the edges of $t\setminus s$ full lines. $
p_{t,s}(\{v_0,v_1,v_2,v_3\})=v_0,$ $p_{t,s}(v_4)=v_4,$ $p_{t,s}(\{v_5,v_6,v_7\})=v_5,$ $p_{t,s}(v_8)=v_8,$
$p_{t,s}(v_9)=v_9.$
}
\end{figure}
See figure \ref{fig:pfade} for an example. This gives us a
graph $\Gamma_{p_{t,s}}.$ It is obvious from the construction that $t\setminus s$ is
a spanning forest of $\Gamma_{p_{t,s}}$ whereas all edges of $s$ are transformed into loops. \\\\
Let $\mathcal{N}=\{A_{\gamma_1},\ldots,A_{\gamma_k}\}$ be a
$\mathcal{P}$-nested set in $\mathcal{C}_{sing}(\Gamma)$ or
$\mathcal{C}_{div}(\Gamma).$ Let $t$ be an adapted spanning
tree. All $\gamma_i$ are assumed saturated. We define the graph
$\gamma_i//\mathcal{N}$ as follows. Let
$A_{\gamma_{j_1}},\ldots,A_{\gamma_{j_l}}$ be the maximal
elements $\subseteq A_{\gamma_i}.$ Let $s$ be the forest
defined by $E(s)=E(t)\cap (E(\gamma_{j_1})\cup\ldots\cup
E(\gamma_{j_l})).$ Then $\gamma_i//\mathcal{N}$ is the graph
with edges
$E(\gamma_i)\setminus \bigcup_{m=1}^l E(\gamma_{j_m})$ contracted along the map $p_{t,s}.$ \\\\
The graph $\gamma_i//\mathcal{N}$ obviously depends on $t,$
although only up to a permutation of the vertices, as is easily
verified.
\begin{lem} \label{lem:contrres} Under the assumptions above:
\begin{enumerate}
\item[(i)] The graph $\gamma_i//\mathcal{N}$ has no loops.
\item[(ii)] If $\gamma_i$ is connected, so is
    $\gamma_i//\mathcal{N}$ (wrt.~
    $V_{\operatorname{eff}}(\gamma_i//\mathcal{N})).$
\item[(iii)] In the case of the divergent collection
    $\mathcal{C}_{div}(\Gamma),$ let $\mathcal{N}$ be a
    maximal nested set. If $\gamma_i$ is connected,
    $\gamma_i//\mathcal{N}$ is at most logarithmic and
    primitive. Therefore $\operatorname{res}
    (\gamma_i//\mathcal{N})$ is defined (see
    (\ref{eq:defresidue})).
\item[(iv)] In this case $\operatorname{res}
    (\gamma_i//\mathcal{N})$ does not depend upon the
    choice of an adapted spanning tree $t.$
\end{enumerate}
\end{lem}
Note that for $\mathcal{P}=\mathcal{F}(\mathcal{C})$ every
$\gamma_i$ is connected (as it is irreducible).
For non-connected $\gamma_i,$ the statements hold for each component. \\\\
\emph{Proof.} (i) Suppose $e$ were a loop in
$\gamma_i//\mathcal{N}$ at the vertex $v.$ Since $\gamma_i$ has
no loops, $|p_{t,s}^{-1}(v)|> 1.$ However, $p_{t,s}$ moves only
the vertices adjacent to edges of $s.$ We conclude $e\in E(\gamma_{j_m})$
as the $\gamma_j$ are saturated, and have a contradiction. \\
(ii) By construction $p^\vee(\sum_{v'\in
V_{\operatorname{eff}}(\gamma_i//\NN)} v')= p^\vee(\sum_{v'\in
V(\gamma_i//\NN)}v')=\sum_{v\in V(\gamma_i)} v$ since the sum
is over \emph{all} vertices of
$V_{\operatorname{\operatorname{eff}}}(\gamma_i)$ (the vertices
not in $V_{\operatorname{eff}}$ map to $0$). On the other hand,
$p^\vee(x)$ of a sum $x=\sum_{v'\in U}v'$ where $U\subsetneq
V_{\operatorname{eff}}(\gamma_i//\NN),$ is not contained in
$\operatorname{span} \sum_{v\in V(\gamma_i)} v.$ Write
$\delta=\delta_{\gamma_i}$ and
$\delta_p=\delta_{(\gamma_i)_p}.$
\begin{equation*}
\please\begin{diagram}
0&\rTo &H^0(\gamma_i) & \rTo & \R^{V(\gamma_i)} & \rTo^{\delta} & \R^{E(\gamma_i)}\\
 &     &              &      & \uTo>{p^\vee}   &                 & \uTo             \\
0&\rTo & H^0(\gamma_i//\NN) & \rTo & \R^{V_{\operatorname{eff}}(\gamma_i//\NN)} & \rTo^{\delta_p} & \R^{E(\gamma_i)\setminus \cup_{m=1}^l E(\gamma_{j_m})}
\end{diagram}
\end{equation*}
 Note that $\delta_p$ as a map
into $\R^{E((\gamma_i)_p)}$ is the same as as a map into
$\R^{E(\gamma_i//\mathcal{N})}$ since the missing edges are all
loops. Consequently, if $x\in\ker\delta_p,$ then
$p^\vee(x)\in\ker\delta,$ by
definition of $(\gamma_i)_p.$ However, because $\gamma_i$ is connected,
$\ker\delta=\operatorname{span} \sum_{v\in V(\gamma_i)} v.$ Therefore
$\dim \ker\delta_p=1,$ if $\delta_p$ is restricted to $V_{\operatorname{eff}}(\gamma_i//\NN),$
and hence $\gamma_i//\NN$ connected. \\
(iii) By definition, a graph $\gamma$ on $V(\Gamma)$ is
divergent if and only if $\dim A_\gamma=(d-2)|E(\gamma)|.$ It
is convergent if $\dim A_\gamma>(d-2)|E(\gamma)|.$ We may
restrict ourselves to saturated subgraphs because the number of
edges increases the susceptibility to divergences, and every
divergent graph is saturated. Let $\gamma_p\subseteq
\gamma_i//\mathcal{N}$ be saturated as a subgraph of
$\gamma_i//\mathcal{N}.$ Therefore $E(\gamma_p)\subseteq
E(\gamma_i)\setminus  \bigcup_{m=1}^l E(\gamma_{j_m}).$ Let now
$\gamma_s$ be the saturated graph for $\gamma_p$ as a subgraph
of $\gamma_i.$ Since $p$ maps each component of $\gamma_{j_m}$
to a single vertex, $\gamma_i//\NN$ has $\sum_{m=1}^l \dim
A_{\gamma_{j_m}}$ components more than $\gamma_i.$ More
generally,
\begin{equation*}
\dim A_{\gamma_p} =
\dim A_{\gamma_s} - \dim A_{s\cap \gamma_s}.
\end{equation*}
On the other hand,
\begin{equation*}
|E(\gamma_p)| = |E(\gamma_s)|-|E((s\cap\gamma_s)_s)|.
\end{equation*}
Therefore $(d-2)|E(\gamma_p)|\le \dim A_{\gamma_p},$ and
equality only if $\gamma_s=\gamma_i$ (equivalently $\gamma_p=\gamma_i//\NN)$ by the maximality of $\mathcal{N}.$
It follows that $\gamma_i//\mathcal{N}$ is divergent, and proper subgraphs
$\gamma_p$ of $\gamma_i//\mathcal{N}$ are convergent,
divergent, worse than logarithmically divergent if and only if
they are as subgraphs of $\gamma_i;$ whence
$\gamma_i//\mathcal{N}$ is also at most logarithmic and primitive. \\
(iv) Let $t,t'$ be two choices of an adapted spanning tree. Then $t\setminus s$ and $t'\setminus s'$ are spanning trees of $\gamma_i//\NN,$ and by the argument in the proof of \tref{thm:primitive} (ii) $\operatorname{res}\gamma_i//\mathcal{N}$ is independent of the basis chosen. \hfill $\Box$\\\\
We will shortly use this lemma in connection with the following
theorem, which helps understand the geometry of the divisor
$\mathcal{E}$ in $Y_\mathcal{P}.$
\begin{thm} (see \cite[Theorem 3.2]{deCPro}) \label{thm:intersections}
Let $\beta: Y_\mathcal{P}\rightarrow V$ be a wonderful model.
\begin{enumerate}
\item[(i)] The divisor is
    $\mathcal{E}=\bigcup_{A\in\mathcal{P}} \mathcal{E}_A$
    with $\mathcal{E}_A$ smooth irreducible and
    $\beta(\mathcal{E}_A)=A^\perp.$
\item[(ii)] The components
    $\mathcal{E}_{A_1},\ldots,\mathcal{E}_{A_k}$ have
    nonempty intersection if and only if the set
    $\{A_1,\ldots,A_k\}$ is $\mathcal{P}$-nested. In this
    case the intersection is transversal.
\end{enumerate} \hfill $\Box$
\end{thm}
We also write $\mathcal{E}_{\gamma}$ for $\mathcal{E}_{A_\gamma}.$\\\\
We consider only the divergent case $\mathcal{C}_{div}(\Gamma)$
with arbitrary building set $\mathcal{P}$ and conclude for the
Laurent expansion at $s=1:$
\begin{thm}\label{thm:poles}
Let $\tilde w_\Gamma^s=\beta^\ast \tilde u_\Gamma^s$ as a
density.
\begin{enumerate}
\item[(i)] The density $\tilde w_\Gamma^s$ has a pole of
    order $N_{max}$ at $s=1,$ where $N_{max}$ is the
    cardinality of the largest nested set\footnote{We
    suspect, but this is not needed here, that in the
    divergent arrangement all maximal nested sets have
    (equal) cardinality $N_{max}.$}.
\item[(ii)] Let
\begin{equation}\label{eq:polecoeff}
\tilde w_\Gamma^s = \sum_{k=-N_{max}}^\infty \tilde a_{\Gamma,k}(s-1)^k.
\end{equation}
Then, for $k\le -1,$
\begin{equation*}
\operatorname{supp} \tilde a_{\Gamma,k} =\bigcup_{|\mathcal{N}|=-k}\bigcap_{A_\gamma\in\mathcal{N}}
\mathcal{E}_{\gamma},
\end{equation*}
which is a subset of codimension $-k$. The union is over
$\mathcal{P}$-nested sets $\mathcal{N}.$
\item[(iii)] Let
    $\mathcal{P}=\mathcal{F}(\mathcal{C}_{div}(\Gamma)).$ Recall that $\underline{1}$ denotes the constant function 1. Then
\begin{equation}\label{eq:st2}
\tilde a_{\Gamma,-N_{max}}[\underline{1}] = \sum_{|\mathcal{N}|=N_{max}}\prod_{A_\gamma\in \mathcal{N}} \operatorname{res} (\gamma//\mathcal{N}).
\end{equation}
where all $\gamma$ are assumed saturated.
\end{enumerate}
\end{thm}
Recall from \tref{thm:factorization} that $f_\Gamma$ is $C^\infty$
in the $y_A^{i_A}.$ Therefore the canonical regularization can
be used consistently (see (\ref{eq:mult})). The identity
(\ref{eq:st2}) is known as a consequence of the scattering
formula in \cite{CK3} in a momentum space context. More
general identities for the higher coefficients can be obtained but are not necessary for the purpose of this paper.\\\\
\emph{Proof.} (i) From (\ref{eq:factorizations}), $\tilde
w_\Gamma^s |dy| = f_\Gamma^s
\prod_{A\in\mathcal{N}}|y_A^{i_A}|^{(d_A-1)-d_As}|dy|$ in local
coordinates. By the results of section \ref{ss:regularization},
in particular (\ref{eq:xsa3}),
\begin{equation}\label{eq:localpoles}
\tilde w_\Gamma^s |dy|= f_\Gamma^s \prod_{A\in\mathcal{N}}\left(-\frac{2\delta_0(y_A^{i_A})}{d_A(s-1)}+|y^{i_A}_A|_{fin}^{(d_A-1)-d_A s}\right)|dy|,
\end{equation}
whence the first statement. \\
(ii) This follows from
(\ref{eq:localpoles}), using that $\mathcal{E}_{\gamma}$ is
locally given by
$y_{A_\gamma}^{i_{A_\gamma}}=0.$ \tref{thm:intersections} (ii) shows that the codimension is $k.$ \\
(iii) Throughout this proof we assume all $\gamma$ defining the
nested set are saturated. By \tref{thm:intersections} (ii), for
$|\mathcal{N}|=N_{max},$ the set $\cap_{\gamma\in \mathcal{N}}
\mathcal{E}_{\gamma}$ intersects no other
$\mathcal{E}_{\gamma'},$ $\gamma'\not\in\mathcal{N}.$ Using
(ii), $\tilde a_{\Gamma,-N_{max}}$ is in fact supported on a
disjoint union subsets of codimension $N_{max},$ and we may compute
$\tilde a_{\Gamma,-N_{max}}[\underline{1}]$ on each of them and
sum the results up. It suffices, therefore, to show
\begin{equation}\label{eq:st3}
(-2)^{N_{max}}\int f_\Gamma \prod_{A_{\gamma}\in\mathcal{N}}\delta_0(y_{A_\gamma}^{i_{A_\gamma}})/d_\gamma |dy| = \prod_{A_\gamma\in \mathcal{N}} \operatorname{res}(\gamma//\mathcal{N})\quad \mbox{ (in }U_\mathcal{N}^\mathcal{B})
\end{equation}
for all maximal nested sets $\mathcal{N}.$ Integration inside
one chart suffices since there is no other nested set
$\mathcal{N'}$ such that $j(U_\mathcal{N'})$ covers
$\cap_{A_\gamma\in\mathcal{N}}\mathcal{E}_\gamma$ and charts
from another choice of marked basis need not be considered, see
the argument in the proof of \tref{thm:primitive} (ii). Recall
(\ref{eq:beta}) on $M^{V_0}$ and (\ref{eq:produos})
\begin{equation*}
w_\Gamma(\{y_e^j\})=(\beta^\ast \underline{u}_\Gamma)(\{y_e^j\})=
\prod_{e\in E(\Gamma)} u_0 (\{ \sum_{e'\leadsto e}\prod_{y_{e'}^j\preceq y_{e''}^k} y^k_{e''}\}_{j=0}^{d-1}).
\end{equation*}
in $U_\mathcal{N}^\mathcal{B}.$ In order to study
$f_\Gamma|_{y_{A_\gamma}^{i_{A_\gamma}}=0}$ one observes that
all products $\prod_{y_{e'}^j\preceq y_{e''}^k} y^k_{e''}$
vanish at $y_{A_\gamma}^{i_{A_\gamma}}=0,$ once $e'\in
E(\gamma).$ If all $d$ components
$x^0_{e'},\ldots,x^{d-1}_{e'}$ of all $e'\leadsto e$ vanish at
the same time, this does not affect $f_\Gamma,$ as it is taken
care of by a power of $y_A^{i_A}$ pulled out of $u_\Gamma$ in
(\ref{eq:factorization}). Consequently, for a fixed $e\in
E(\Gamma),$
\begin{equation*}
u_0(\{\sum_{e':e'\leadsto e}\prod_{y_{e'}^j\preceq y_{e''}^k} y^k_{e''}\}^{d-1}_{j=0})
\prod_{A_\gamma\in\NN,e\in E(\gamma)} (y_{A_\gamma}^{i_{A_\gamma}})^{d-2}
\prod_{A_\gamma\in\NN} \delta_0(y_{A_\gamma}^{i_{A_\gamma}})
\end{equation*}
\begin{equation*}
= u_0(\{\sum_{\onatop{e': e'\leadsto e \operatorname{and}\,\forall A_\gamma\in\mathcal{N}}{e'\in E(\gamma)
\Rightarrow e\in E(\gamma)}} \prod_{y_{e'}^j\preceq y_{e''}^k} y^k_{e''}\}_{j=0}^{d-1})
\prod_{A_\gamma\in\NN, e\in E(\gamma)} (y_{A_\gamma}^{i_{A_\gamma}})^{d-2}.
\end{equation*}
On the other hand, consider the graph $\gamma//\NN$ where
$\gamma\in\NN.$ Write $p=p_{t_\gamma,s_\gamma}$ where
$E(t_\gamma)=E(t)\cap E(\gamma),$ $t$ is the chosen adapted
spanning tree for $\Gamma$ and $s_\gamma$ the subforest defined
by the maximal elements of the nested set contained in
$\gamma.$ Since $\gamma$ is connected, $t_\gamma$ is a spanning
tree of $\gamma.$ A vertex $v_{0,\gamma}\in
V_{\operatorname{eff}}(t_\gamma)$ is chosen. For each component
$c$ of $s_\gamma$ there is a unique element $v_c\in
V_{\operatorname{eff}}(c)$ which is nearest to $v_{0,\gamma}$
in $t_\gamma.$ By definition,
\begin{equation*}
p^\vee (v) = \left\{\begin{array}{ll}
\sum_{v'\in V_{\operatorname{eff}}(c)}v' & \mbox{ if } v=v_c,\\
0 & \mbox{ if } v\in V_{\operatorname{eff}}(s_\gamma)\setminus  \bigcup \{v_c\},\\
v & \mbox{ if } v\in V(\Gamma)\setminus  V_{\operatorname{eff}}(s_\gamma). \end{array}\right.
\end{equation*}
Let $x=\sum_{e\in E(t_\gamma)} x_e b_e$ with $b_e =
(-1)^{Q_e}\sum_{v\in V_1} v$ as in \pref{pro:dualadapted}. One
finds $p^\vee(b_e)=(-1)^{Q_e}\sum_{v\in V_{1}\setminus
V_{1}\cap V_{\operatorname{eff}}(c)}v$ where $c$ is the
component of $s_\gamma$ which contains $e,$ and $c=\emptyset$
if $e\in E(t_\gamma\setminus  s_\gamma).$ In particular
$p^\vee(b_e)=b_e$ if $e\in E(t_\gamma\setminus  s_\gamma).$
Consequently
\begin{eqnarray*}
i_{\gamma//\NN}(x)&=& \delta p^\vee (x)\\
&=& \sum_{e\in E(\gamma//\NN)} \sum_{e'\in E(t_\gamma)} (-1)^{Q_{e'}}x_{e'} \sum_{v\in V_1\setminus  V_1\cap V_{\operatorname{eff}}(c)} (v:e) e\\
&=& \sum_{e\in E(\gamma//\NN)} \sum_{\onatop{e'\leadsto e}{e'\in E(t_\gamma\setminus  s_\gamma)}} x_{e'}e
\end{eqnarray*}
where $t_\gamma\setminus  s_\gamma$ is a spanning tree for
$\gamma//\NN.$ Therefore
\begin{eqnarray*}
\tilde a_{\gamma//\NN,-1} &=& \prod_{e\in E(\gamma//\NN)}u_0(\{\sum_{\onatop{e'\leadsto e}{e'\in E(t_\gamma\setminus  s_\gamma)}}
\prod_{y_{e'}^j\preceq y_{e''}^k} y^k_{e''}\})\\
&&\times \prod_{\gamma \subseteq \gamma'\in\NN} (y_{A_{\gamma'}}^{i_{A_{\gamma'}}})^{(d-2)|E(\gamma//\NN)|}|dy|.
\end{eqnarray*}
In a final step, define for each $e\in E(\Gamma)$ the minimal
element $A_{\gamma_e}\in\mathcal{N}$ such that $e\in
E(\gamma_e).$ We have $E(\Gamma)= \bigsqcup_{A_\gamma\in\NN}
\{e\in E(\Gamma): \gamma_e=\gamma\}$
$=\bigsqcup_{A_\gamma\in\NN} E(\gamma//\mathcal{N})$ as is
shown by a simple induction. Similarly
$E(t)=\bigsqcup_{A_\gamma\in\NN}\{e\in E(t):\gamma_e=\gamma\}$
$=\bigsqcup_{A_\gamma\in\NN} E(t_\gamma)\setminus  E(s_\gamma)$
is a decomposition into spanning trees since $t$ is adapted.
Write $|dy|= | \bigwedge_{\onatop{e\in E(t)}{j=0,\ldots,d-1}}
dy_e^j|$ and $|d\hat y| = |\bigwedge_{\onatop{e\in
E(t),j=0,\ldots,d-1}{y_e^j\neq y_A^{i_A}}} dy_e^j|.$ Then, in
$U^\mathcal{B}_\mathcal{N},$
\begin{eqnarray}\label{eq:shrink}
\tilde a_{\Gamma,-N_{max}} &=& \tilde w_\Gamma(\{y_e^j\})\prod_{A_{\gamma}\in\mathcal{N}}|y_{A_\gamma}^{i_{A_\gamma}}|
\delta_0(y_{A_\gamma}^{i_{A_\gamma}})|dy|\nonumber\\
&=&\prod_{e\in E(\Gamma)}u_0(\{\sum_{\onatop{e': e'\leadsto e \operatorname{and}\,\forall A_\gamma\in\mathcal{N}}{e'\in E(\gamma)
\Rightarrow e\in E(\gamma)}} \prod_{y_{e'}^j\preceq y_{e''}^k} y^k_{e''}\})\prod_{\onatop{A_\gamma\in\NN}{e\in E(\gamma)}}
(y_{A_\gamma}^{i_{A_\gamma}})^{d-2}|d\hat y|\nonumber\\
&=&\prod_{A_\gamma\in\NN} (y_{A_{\gamma}}^{i_{A_{\gamma}}})^{(d-2)|E(\gamma)|}\prod_{\onatop{e\in E(\Gamma)}{\gamma_e=\gamma}} u_0(\{\sum_{\onatop{e'\leadsto e}
{\gamma_{e'}=\gamma_e}} \prod_{y_{e'}^j\preceq y_{e''}^k} y^k_{e''}\}_{j=0}^{d-1})|d\hat y|\\
&=& \bigotimes_{A_\gamma\in\NN} \tilde a_{\gamma//\NN,-1}\nonumber
\end{eqnarray}
Consequently (\ref{eq:shrink}) integrates to the product of
residues as claimed.\hfill $\Box$ \\\\
\tref{thm:intersections} and \tref{thm:poles} (ii) implicitly
describe a stratification of $Y_\mathcal{P}.$ In the next
section we will show that all the information relevant for
renormalization is encoded in the geometry of $Y_\mathcal{P}.$
\section{Renormalization on the wonderful model}\label{s:renormalization}
In this section we describe a map that transforms $\tilde
w_\Gamma^s=\beta^\ast \tilde u_\Gamma^s$ into a renormalized
distribution density $\tilde w_{\Gamma,R}^s,$ holomorphic at
$s=1,$ such that $\tilde u_{\Gamma,R}=\beta_\ast \tilde
w_{\Gamma,R}^s |_{s=1}$ is an extension of $u_{\Gamma}$ onto all of $M^{V_0}$ and
satisfies the following (equivalent) physical requirements:
\begin{enumerate}
\item[(i)] The terms subtracted from $u_\Gamma$ in order to
    get $u_{\Gamma,R}$ can be rewritten as counterterms in
    a renormalized local Lagrangian.
\item[(ii)] The $u_{\Gamma,R}$ satisfy the Epstein-Glaser
    recursion (renormalized equations of motion,
    Dyson-Schwinger equations).
\end{enumerate}
One might be tempted to simply define $u_{\Gamma,R}$ by
discarding the pole part in the Laurent expansion of
$u_{\Gamma,R}^s$ at $s=1.$ However, unless $\Gamma$ is
primitive, this would not provide an extension satisfying those
requirements, and the resulting ''counterterms'' would violate
the locality principle. See \cite[Section 5.2]{Collins} for a
simple
example in momentum space. In order to get an extension
using local counterterms, one has to take into account the geometry of $Y_\mathcal{P}.$\\\\
The equivalence between (i) and (ii) is adressed in the
original work of Epstein and Glaser \cite{EG}, see also
\cite{BS,PS,BF}. We circumvent a number of technical issues by
restricting ourselves to logarithmic divergences of massless
graphs on Euclidean space-time throughout the paper.
\subsection{Conditions for physical extensions}
In this section we suppose as given the unrenormalized
distributions
$\underline{u}_\Gamma\in\mathcal{D'}(M^{V_0}\setminus
M^{V_0}_{div}(\Gamma)),$ and examine what the physical
condition (ii)
implies for the renormalized distribution $\underline{u}_{\Gamma,R}\in\mathcal{D'}(M^{V_0})$ to be constructed.\\\\
Let $V=\{1,\ldots,n\}$ be the vertex set of all graphs under
consideration. The degree of a vertex is the
number of adjacent edges. In the previous sections, $\Gamma$ was always supposed to be connected. Here we need disconnected graphs and sums of graphs. Therefore all graphs are supposed to be subgraphs of the $N$-fold complete graph $K^N_n$ on $n$ vertices with $N$ edges between each pair of vertices. $N$ can always be chosen large enough as to accomodate any graph, in a finite collection of graphs $\Gamma$ on $V,$ as one of its subgraphs.\\\\
We write $l_V=(l_1,\ldots,l_n)$ for an $\N_0$- multiindex
satisfying $\sum l_i\in 2\N_0.$ Also $l_V-k_V =
(l_1-k_1,\ldots,l_n-k_n),$ ${l_V\choose k_V}={l_1\choose
k_1}\ldots {l_n\choose k_n}$ etc. Let $V=I\sqcup J.$ Let
$\operatorname{Bip}(k_I,k_J)$ be the set of $(I,J)$-bipartite
graphs on $V,$ where the degree of the vertex $i$ is given by
$k_i.$ Finally, let $(p_{I,J})_{\emptyset \subsetneq
I\subsetneq V}$ be a partition of unity subordinate to the open
cover $\bigcup_{\emptyset \subsetneq I\subsetneq V} C_I$ of
$M^{V_0}\setminus \{0\}$ with
\begin{equation*}
C_I = M^{V_0}\setminus M^{V_0}_{sing}(K_{I,V\setminus I})
\end{equation*}
where $K_{I,J}$ is the complete $(I,J)$-bipartite graph (i.~e.~
the graph with exactly one edge between each $i\in I$ and each
$j\in J$). The set $M^{V_0}_{sing}(K_{I,J})$ is therefore the
locus where at least one $x_i-x_j=0$ for $i\in I,$ $j\in J.$
 \\\\
The Epstein-Glaser recursion for vacuum expectation values of
time-ordered products (see \cite[Equation (31)]{BF}) is given,
in a euclidean version, by the equality
\begin{equation}\label{eq:EGrec}
t^{l_V}_V = \sum_{V=I\sqcup J}\Phi^\ast p_{I,J}\sum_{\genfrac{}{}{0pt}{}{k_V=0}{\sum_{i\in I}
l_i-k_i=\sum_{j\in J}l_j-v_j}}^{l_V} {l_V \choose k_V} t^{k_I}_I t^{k_J}_J \sum_{\Gamma \in \operatorname{Bip}(l_I-k_I,l_J-k_J)}u_\Gamma
\end{equation}
on $M^{V}\setminus \Delta=\Phi^{-1}(M^{V_0}\setminus  \{0\}).$
The distributions $t^{l_V}_V$ therein, vaccuum expectation
values of time-ordered Wick products, relate to the single
graph distributions $u_\Gamma$ and their renormalizations
$u_{\Gamma,R}$ as follows:
\begin{eqnarray}\label{eq:dfntN}
t_V^{l_V} &=& \sum_{\Gamma\in \operatorname{Gr}(l_V)} c_\Gamma u_\Gamma \quad \mbox{ on } \Phi^{-1}(M^{V_0}\setminus M^{V_0}_{sing}(K_n)) \nonumber\\
t_V^{l_V} &=& \sum_{\Gamma\in \operatorname{Gr}(l_V)} c_\Gamma u_{\Gamma,R} \quad \mbox{ on } M^{V}
\end{eqnarray}
$\operatorname{Gr}(l_V)$ is the set of all graphs $\Gamma$ with
given vertex set $V(\Gamma)$ such that the degree of the vertex
$i$ is $l_i.$ There are no external edges and no loops (edges
connecting to the same vertex at both ends). The combinatorial
constants $c_\Gamma=\frac{\prod_{i=1}^n l_i!}{\prod_{i<j}
l_{ij}!}$ where $l_{ij}$ is
 the number of edges between $i$ and $j,$ are not needed in the following.
 See \cite[Appendix B]{Keller} for the complete argument.
\begin{pro}\label{pro:EGrec}
On the level of single graphs, a sufficient condition for
equation (\ref{eq:EGrec}) to hold is, for any $\Gamma,$
\begin{equation}\label{eq:EGrec2}
u_{\Gamma,R}= u_{\gamma_1,R} \cdot u_{\gamma_2,R} \cdot u_{\Gamma\setminus  (\gamma_1\sqcup\gamma_2)} \mbox{ on }\Phi^{-1}(M^{V_0}\setminus  M^{V_0}_{sing}(\Gamma\setminus  (\gamma_1\sqcup\gamma_2)))
\end{equation}
whenever $\gamma_1,\gamma_2$ are connected saturated subgraphs
of $\Gamma,$ such that $V_{\operatorname{eff}}(\gamma_1)\cap V_
{\operatorname{eff}}(\gamma_2)=\emptyset.$
\end{pro}
Note that $u_{\gamma_1,R} \cdot u_{\gamma_2,R}$ is in fact a
tensor product since $cc(\gamma_1)\cap cc(\gamma_2)=0.$ The
locus where the remaining factor $u_{\Gamma\setminus
(\gamma_1\sqcup\gamma_2)}$ is not $C^\infty$ is excluded by
restriction to $M^{V_0}\setminus
M^{V_0}_{sing}(\Gamma\setminus  (\gamma_1\sqcup\gamma_2)).$
The product is therefore well-defined. Note also that
(\ref{eq:EGrec2}) trivially holds on $M^{V_0}\setminus
M^{V_0}_{div}(\Gamma)$ by the very definition (\ref{eq:FGD})
of $u_\Gamma.$ \pref{pro:EGrec} implies, in particular, that if
$\Gamma$ is
a disjoint union ($\Gamma=\gamma_1\sqcup\gamma_2$ and
$V_{\operatorname{eff}}(\gamma_1)\cap V_
{\operatorname{eff}}(\gamma_2)=\emptyset),$ then $u_{\Gamma,R}
= u_{\gamma_1,R}\otimes u_{\gamma_2,R}$ everywhere.
\\\\
The system of equations (\ref{eq:EGrec2}) is called the
Epstein-Glaser recursion for $u_{\Gamma,R}.$
Recursive equations of this kind are also referred to as renormalized Dyson-Schwinger equations (equations of motion) in a momentum space context \cite{Kreimer2,BK2}. \\\\
\emph{Proof of \pref{pro:EGrec}. } Let all $u_{\Gamma,R}$
satisfy the requirement of (\ref{eq:EGrec2}). We only need the
case where $\{I,J\}$ with $I=V_{\operatorname{eff}}(\gamma_1),$
$J=V_{\operatorname{eff}}(\gamma_2)$ is a partition, i.~e.~
$I\sqcup J=V.$ Since
$M^{V_0}_{sing}(\Gamma\setminus (\gamma_1\sqcup\gamma_2))\subseteq
M^{V_0}_{sing}(K_{I,J}),$ (\ref{eq:EGrec2}) is valid in
particular on $C_I\supseteq \operatorname{supp} p_{I,J}.$
Furthermore, since $\gamma_1$ and $\gamma_2$ are saturated,
$\Gamma\setminus (\gamma_1\sqcup\gamma_2)$ is $(I,J)$-bipartite.
Therefore, $t_V^{l_V}$ as in (\ref{eq:dfntN}) with
(\ref{eq:EGrec2}) inserted, provides one of the terms on the
right hand side of (\ref{eq:EGrec}). Conversely, every graph
$\Gamma$ with prescribed vertex degrees can be obtained by
chosing a partition $I\sqcup J=V,$ taking the saturated
subgraphs $\gamma_i$ for $I$ and $\gamma_j$ for $J,$
respectively, and supplying the missing edges from the
$(I,J)$-bipartite graph. \hfill $\Box$
\subsection{Renormalization prescriptions}\label{ss:prescriptions}
We consider the divergent arrangement
$\mathcal{C}=\mathcal{C}_{div}(\Gamma)$ only, with building set
$\mathcal{P}$ minimal or maximal, that is
$\mathcal{P}=\mathcal{F}(\mathcal{C})$ or $\mathcal{C}.$ Let
$\mathcal{N}$ be a nested set which, together with an adapted
spanning tree $t$ and a marking of the corresponding basis
$\mathcal{B},$ provides for a chart $U_\mathcal{N}^\mathcal{B}$
for $Y_\mathcal{P}.$ \\\\
By \tref{thm:poles} (ii) the subset of codimension 1 where
$\tilde w_\Gamma^s$ has only a simple pole at $s=1$ is covered
by those charts $U^\mathcal{B}_\mathcal{N}$ where
$\mathcal{N}=\{A_\gamma\}$ with $\gamma$ any divergent (and
irreducible if $\mathcal{P}=\mathcal{F}(\mathcal{C}))$ graph.
From (\ref{eq:localpoles}) one has
\begin{equation*}
\tilde w_\Gamma^s |dy|= f_\Gamma^s \left(-\frac{2\delta_0(y_{A_\gamma}^{i_{A_\gamma}})}{d_\gamma(s-1)}+|y^{i_{A_\gamma}}_{A_\gamma}|_{fin}^{(d_\gamma-1)-d_\gamma s}\right)|dy|
\end{equation*}
In these charts, one performs one of the following subtractions
in order to get a renormalized, i.~e.~ extended, distribution. In the first
case, only the pole is removed
\begin{equation}\label{eq:localms}
\tilde w_{\Gamma}^s|dy| \mapsto \tilde w_{\Gamma,R_0}^s |dy|= f_\Gamma^s |y^{i_{A_\gamma}}_{A_\gamma}|_{fin}^{d_\gamma s-(d_\gamma-1)}|dy|
\end{equation}
One might call this \emph{local minimal subtraction}. Other extensions differ from this one by a distribution
supported on $\mathcal{E}_\gamma.$ Here is an example of another renormalization prescription, producing a different extension:\\\\
For each $A_\gamma\in\NN$ let $A_{\gamma_1},\ldots,A_{\gamma_k}\in\NN$ be the maximal
elements contained in $A_\gamma$ (where all graphs are assumed saturated). Choose a
$\nu_{A_\gamma}\in C^\infty(Y_\mathcal{P})$ such that
$\nu_{A_\gamma}|_{y_{A_\gamma}^{i_{A_\gamma}}=0}=1$ and
$\nu_{A_\gamma}$ depends only on the coordinates $y_e^j,$ $e\in
E(t)\cap (E(\gamma)\setminus E(\cup_{j=1}^k\gamma_j))$ in $U_\mathcal{B}^\mathcal{N},$ and has compact support in the
associated linear coordinates $x_e^j,$ $e\in E(t)\cap (E(\gamma)\setminus E(\cup_{j=1}^k\gamma_j)).$ The
$\nu_{A_\gamma}$ are called \emph{renormalization conditions}. In practice, the $\nu_{A_\gamma}$ will be chosen as described at the end of section \ref{ss:residues}.\\\\
The second renormalization prescription is then
\begin{eqnarray}\label{eq:localmom}
\tilde w_{\Gamma}^s |dy| &\mapsto &\tilde w_{\Gamma,R_\nu}^s |dy| \nonumber \\
&&= \tilde w_\Gamma^s - |y^{i_{A_\gamma}}_{A_\gamma}|^{d_\gamma s-(d_\gamma-1)} [\nu_{A_\gamma}]_{p_{A_\gamma}} \delta_0(y^{i_{A_\gamma}}_{A_\gamma})
f_{\Gamma}^s|dy|,
\end{eqnarray}
which is called \emph{subtraction at fixed conditions}. The
notation $[\nu_A]_{p_A}$ means integration along the
fiber of the projection
\begin{equation*}
p_A: (y_{e_1}^{0},\ldots,y_{e_{|E(t)|}}^{d-1})\mapsto(y_{e_1}^{0},\ldots,\widehat{y_A^{i_A}},\ldots,y_{e_{|E(t)|}}^{d-1})
\end{equation*}
defined in (\ref{eq:partialint}). Both prescriptions provide us
local expressions holomorphic at $s=1$ in all charts
$U^\mathcal{B}_\mathcal{N}$ where $\mathcal{N}$ contains a single element. It remains to define them in the other charts.\\\\
In the charts $U_{\mathcal{N}}^\mathcal{B},$ for a general
nested set $\mathcal{N},$ where
\begin{equation*}
\tilde w_\Gamma^s |dy|= f_\Gamma^s\prod_{A\in\mathcal{N}} \frac{1}{|y^{i_A}_A|^{d_As-(d_A-1)}}|dy|
\end{equation*}
one applies the subtraction (\ref{eq:localms}) in every factor
(local minimal subtraction)
\begin{equation}\label{eq:genms}
\tilde w_{\Gamma,R_0}^s |dy|= f_\Gamma^s\prod_{A\in\mathcal{N}} |y^{i_A}_A|_{fin}^{(d_A-1)-d_As}|dy|.
\end{equation}
Similarly, by abuse of notation, in the same chart,
\begin{equation}\label{eq:genmom}
\tilde w_{\Gamma,R_\nu}^s |dy|= \tilde w_\Gamma^s \prod_{A\in\mathcal{N}} \left(1-\ldots[\nu_A]_{p_A} \delta_0(y_A^{i_A})\right)|dy|
\end{equation}
generalizing the subtraction at fixed conditions (\ref{eq:localmom}). A precise notation for (\ref{eq:genmom})
-- which disguises however the multiplicative nature of this operation -- is
\begin{eqnarray}\label{eq:genmomprec}
\tilde w_{\Gamma,R_\nu}^s |dy| &=& \sum_{\{A_1,\ldots,A_k\}\subseteq\mathcal{N}}(-1)^k \prod_{A\in\mathcal{N}} \frac{1}{|y^{i_A}_A|^{d_As-(d_A-1)}}
\left[\Pi_{j=1}^k \nu_{A_j}\right]_{p_{A_1,\ldots,A_k}}\nonumber\\
&&\times \prod_{j=1}^k \delta_0(y_{A_j}^{i_{A_j}})f_\Gamma^s|dy|
\end{eqnarray}
where $p_{A_1,\ldots,A_k}$ is the projection omitting the coordinates $y_{A_j}^{i_{A_j}}, $ $j=1,\ldots,k.$
\cref{cor:infraredsafe} shows that there are no infrared divergences when pushing forward along $\beta.$\\\\
Note that $\tilde w_{\Gamma,R_0}^s|_{s=1}|dy|$ defines a density on $Y_\mathcal{P},$ but
this is not true for general $s:$
\begin{pro}\label{pro:globallydefined} The local expressions $\tilde
w_{\Gamma,R_0}^s|_{s=1}|dy|$ given by (\ref{eq:genms}) define a density on
$Y_\mathcal{P}.$ The $\tilde w_{\Gamma,R_\nu}^s$ given by (\ref{eq:genmom},\ref{eq:genmomprec}) define
a density-valued function on $Y_\mathcal{P},$ holomorphic in a neighborhood of $s=1.$
\end{pro}
\emph{Proof.} Note that $\tilde w_{\Gamma}^s$ is by construction a density for all $s.$
Local minimal subtraction: The $|y_A^{i_A}|_{fin}^{-1}$ transform
like $|y_A^{i_A}|^{-1}$ under transition between charts. Subtraction at
fixed conditions: Each term in the sum (\ref{eq:genmomprec}) differs from
$\tilde w_{\Gamma}^s$ by a number of integrations in the $y_{A_j}^{i_{A_j}}$
and a product of delta distributions in the same $y_{A_j}^{i_{A_j}}.$ Under
transition between charts, the contribution to the Jacobian from the integrations
cancels the one from the delta distributions. It remains to show that $\tilde w_{\Gamma,R_\nu}^s$
has no pole at $s=1:$ Using that $\nu_A|_{y_A^{i_A}=0}=1,$ we have in local coordinates
\begin{eqnarray*}
\tilde w_{\Gamma,R_\nu}^s&=& \sum_{\{A_1,\ldots,A_k\}\subseteq \NN}(-1)^k \prod_{j=1}^k \left(\frac{-2\delta_0(y_{A_j}^{i_{A_j}})}
{d_{A_j}(s-1)}+|y_{A_j}^{i_{A_j}}|_{fin}^{d_{A_j}-1-d_{A_j}s}[\nu_{A_j}]_{p_{A_j}}\right.\\
&&\cdot\left.\delta_0(y_{A_j}^{i_{A_j}})\right)\prod_{A\in\NN\setminus\{A_1,\ldots,A_k\}}\left(
\frac{-2\delta_0(y_{A}^{i_{A}})}
{d_A(s-1)}+|y_{A}^{i_{A}}|_{fin}^{d_A-1-d_A s}\right)f_\Gamma^s.
\end{eqnarray*}
Combining this to a binomial power finishes the proof.
\hfill $\Box$
\begin{thm}\label{thm:fits} Let $\mathcal{P}=\mathcal{F}(\mathcal{C}_{div})$ for all graphs. Then both assignments
\begin{eqnarray*}
\Gamma & \mapsto &\tilde u_{\Gamma,R_0}=\beta_\ast \tilde w_{\Gamma,R_0}^s|_{s=1},\\
\Gamma& \mapsto &\tilde u_{\Gamma,R_\nu}=\beta_\ast \tilde w_{\Gamma,R_\nu}^s|_{s=1}
\end{eqnarray*}
(with consistent choice of the $\nu_A$) satisfy the locality condition (\ref{eq:EGrec2}) for graphs.
\end{thm}
The proof is based on the following lemmata. All building sets $\mathcal{P}$ are minimal. If
$A_\gamma\in\mathcal{P}$ then $\gamma$ is always supposed saturated.
\begin{lem}\label{lem:main1} Under the assumptions of \pref{pro:EGrec}, let $A_\gamma\in\mathcal{P}$
and $cc(\gamma)\not\le cc(\gamma_1\sqcup\gamma_2).$ Then
\begin{equation*}
\mathcal{E}_\gamma \subseteq \beta^{-1}(M^{V_0}_{sing}(\Gamma\setminus (\gamma_1\sqcup\gamma_2))).
\end{equation*}
\end{lem}
\emph{Proof.} If $cc(\gamma)\not\le cc(\gamma_1\sqcup\gamma_2),$
then $\gamma$ contains an edge $e\in
E(\Gamma\setminus (\gamma_1\sqcup\gamma_2)).$ Consequently
$A_\gamma^\perp = \bigcap_{e\in E(\gamma)} A_e^\perp \subseteq
\bigcup_{e\in E(\Gamma\setminus (\gamma_1\sqcup\gamma_2))}
A_e^\perp =
M^{V_0}_{sing}(\Gamma\setminus (\gamma_1\sqcup\gamma_2)).$ Since
$\beta^{-1}(A_\gamma^\perp) \supseteq \mathcal{E}_\gamma,$ the
result follows. \hfill $\Box$\\\\
Under the assumptions of \pref{pro:EGrec}, let
\begin{equation*}
\mathcal{G} = \{A_\gamma\in\mathcal{P}: cc(\gamma)\le cc(\gamma_1\sqcup\gamma_2)\}.
\end{equation*}
\begin{lem}\label{lem:main3} A subset $\mathcal{N}\subseteq\mathcal{G}$ is nested wrt.~ the minimal building set if and only
if $\mathcal{N}=\mathcal{N}_1\sqcup\mathcal{N}_2,$ where
$\mathcal{N}_i$ is a nested set wrt.~ the minimal building set
for the connected graph $\gamma_i$ with vertex set
$V_{\operatorname{eff}}(\gamma_i).$
\end{lem}
\emph{Proof.} Let
$\mathcal{P}(G)=\mathcal{F}(\mathcal{C}_{div}(G))$ for a graph
$G.$ First, since $V_{\operatorname{eff}}(\gamma_1)\cap
V_{\operatorname{eff}}(\gamma_2)=\emptyset,$ every connected
subgraph $\gamma$ of $\gamma_1\sqcup\gamma_2$ is either
contained in $\gamma_1$ or in $\gamma_2.$ Let now
$\mathcal{N}\subseteq\mathcal{G}$ be nested wrt.~
$\mathcal{P}(\Gamma).$ All irreducible graphs are connected. We
can therefore write
$\mathcal{N}=\mathcal{N}_1\sqcup\mathcal{N}_2$ where the
elements of $\mathcal{N}_i$ are contained in $\gamma_i.$ Since
$\gamma_i$ is saturated, a subgraph of $\gamma_i$ is
irreducible as a subgraph of $\gamma_i$ if and only if it is as
a subgraph of $\Gamma.$ Consequently the $\mathcal{N}_i$ are
$\mathcal{P}(\gamma_i)$-nested because
$\mathcal{P}(\gamma_i)\subseteq \mathcal{P}(\Gamma).$
Conversely, suppose $\mathcal{N}_1=\{A_{\gamma_i},i\in I\}$ and $\mathcal{N}_2=\{A_{\gamma_j},j\in J\}$ are
given. Let some $\gamma_{i_1},\ldots,\gamma_{i_l}\subseteq
\gamma_1$ and $\gamma_{j_1},\ldots,\gamma_{j_m}\subseteq
\gamma_2$ be pairwise noncomparable. Then the sum $\sum_{k=1}^l
A_{\gamma_{i_k}}+\sum_{n=1}^m A_{\gamma_{j_n}}$ is in fact a
decomposition into two terms and therefore not contained in
$\mathcal{P}(\Gamma),$ unless one of the two terms is zero. But
in this case, the other term is a nontrivial decomposition
itself, for it is not contained in $\mathcal{P}(\gamma_i).$
Therefore it is not contained in
$\mathcal{P}(\Gamma),$ and $\mathcal{N}_1\sqcup\mathcal{N}_2$ is nested wrt.~ $\mathcal{P}(\Gamma).$ \hfill $\Box$ \\\\
\emph{Proof of \tref{thm:fits}.} Let $\Gamma,\gamma_1,\gamma_2$
as in \pref{pro:EGrec}. Let $\phi\in\mathcal{D}(M^{V_0})$ such
that $\operatorname{supp} \phi\cap
M^{V_0}_{sing}(\Gamma\setminus (\gamma_1\sqcup\gamma_2))=\emptyset.$
In a first step, we study the compact set
$X=\operatorname{supp}\psi$ where $\psi= \beta^\ast \phi.$ By \lref{lem:main1},
$X$ does not intersect any $\mathcal{E}_\gamma$ where $\gamma\in \mathcal{P}\setminus\mathcal{G}$.
Therefore
\begin{equation*}
X\cap j^{\mathcal{B}}_{\mathcal{N}}(U_{\mathcal{N}}^\mathcal{B})\subseteq j^\mathcal{B}_{\mathcal{N}\cap\mathcal{G}}(U^\mathcal{B}_{\mathcal{N}\cap\mathcal{G}})
\end{equation*}
(where at the right hand side the marking of $\mathcal{B}$ is
restricted to $\mathcal{N}\cap \mathcal{G}).$ In order to test
(\ref{eq:EGrec2}), it suffices thus to consider the $U_\mathcal{N}^\mathcal{B}$ where $\mathcal{N}\in\mathcal{G}.$ Fix now such an $\mathcal{N}\in\mathcal{G}.$
In a second step, assume for simplicity that $V_{\operatorname{eff}}(\gamma_1)=\{1,\ldots,i\},$ $V_{\operatorname{eff}}(\gamma_2)=\{i+1,\ldots,i+j\}$ and write $V_r=n-(i+j)+1$ (By the remark in the proof of \pref{pro:EGrec} we really only need the case where $V_r=1$). Now
consider the map $\beta_{1,2}: Y_{\mathcal{P}(\gamma_1)}\times
Y_{\mathcal{P}(\gamma_2)}\times M^{V_r}\rightarrow M^{V_0}$ which is the
cartesian product of two wonderful models (with two minimal
building sets) for the graphs $\gamma_1$ and $\gamma_2,$ and
a factor corresponding to the remaining edges of an adapted spanning tree for $\Gamma$ where a spanning tree for $\gamma_1$ and $\gamma_2$ have been removed.
The map is the identity on this third factor.
If $U^{\mathcal{B}_i}_{\mathcal{N}_i}$ is a
chart for $Y_{\mathcal{P}(\gamma_i)},$ then
$U^{\mathcal{B}_1}_{\mathcal{N}_1}\times
U^{\mathcal{B}_2}_{\mathcal{N}_2}\times M^{V_r}$ is a chart for the product.
As the nested sets $\mathcal{N}_1$ and $\mathcal{N}_2$ and the
marking $\mathcal{B}_1$ and $\mathcal{B}_2$ of the basis vary,
one obtains an atlas for $Y_{\mathcal{P}(\gamma_1)}\times
Y_{\mathcal{P}(\gamma_2)}\times M^{V_r}.$ Similarly, let
$q^{\mathcal{B}_1,\mathcal{B}_2}_{\NN_1,\NN_2}=q^{\mathcal{B}_1}_{\mathcal{N}_1}\otimes
q^{\mathcal{B}_2}_{\mathcal{N}_2}\otimes id $ be a subordinate partition
of unity with compact support for the compact set
$X'=\operatorname{supp} \beta^{\ast}_{1,2} \phi$ in
$Y_{\mathcal{P}(\gamma_1)}\times Y_{\mathcal{P}(\gamma_2)}\times M^{V_r}.$ \\
In a third step, we use \lref{lem:main3} to identify
$\mathcal{P}(\Gamma)$-nested sets
$\mathcal{N}\subseteq\mathcal{G}$ with
$\mathcal{N}_1\sqcup\mathcal{N}_2,$ and to show that there is a
partition of unity $p_{\mathcal{N}}^{\mathcal{B}}$ for
$X\subset Y_\mathcal{P}$ subordinate to the atlas
$U^\mathcal{B}_\mathcal{N},$ which looks locally like
$q^{\mathcal{B}_1,\mathcal{B}_2}_{\mathcal{N}_1,\mathcal{N}_2}.$
Since
$U^\mathcal{B}_\mathcal{N}=U^{\mathcal{B}_1}_{\mathcal{N}_1}\times
U^{\mathcal{B}_2}_{\mathcal{N}_2}\times M^{V_r}\setminus \cup_{A\in\mathcal{P}\setminus \mathcal{G}}
Z_A,$ (see section \ref{ss:propwonderful}), with
$j^\mathcal{B}_\mathcal{N}=j^{\mathcal{B}_1}_{\mathcal{N}_1}\times
j^{\mathcal{B}_2}_{\mathcal{N}_2}\times id,$ the
$q^{\mathcal{B}_1,\mathcal{B}_2}_{\NN_1,\NN_2}$ provide indeed
such a partition of unity $p_{\mathcal{N}}^{\mathcal{B}}$ with compact support, because a small
enough neighborhood of $X$ does not intersect any $Z_A,$ $A\not\in \mathcal{G}.$  \\
Finally in a chart $U^\mathcal{B}_\mathcal{N},$ identified with
$U^{\mathcal{B}_1}_{\mathcal{N}_1}\times
U^{\mathcal{B}_2}_{\mathcal{N}_2}\times M^{V_r},$ by definition
(\ref{eq:genms},\ref{eq:genmom}), the renormalized
distributions satisfy
\begin{equation*}
\tilde w_{\Gamma,R}(y)|dy|= \tilde w_{\gamma_1,R} \tilde w_{\gamma_2,R} \tilde w_{\Gamma\setminus  (\gamma_1\sqcup\gamma_2)}(y)|dy|
\end{equation*}
where on the right hand side pullbacks along $\beta_{1,2}$ are understood. Let
$\psi_{1,2}=\beta^\ast_{1,2}\phi.$ Since also
$\beta=\beta_{1,2}$ in this chart, we have $\psi=\psi_{1,2}$ in
local coordinates. This finishes the proof.
\hfill $\Box$ \\\\
\emph{Remarks}. Local minimal subtraction is easily defined, but depends on the choice
of regularization in a crucial way. The subtraction at fixed conditions is independent of the
regularization and therefore the method of choice for the renormalization of amplitudes and
non-perturbative computations. \\
If one extends the requirement (\ref{eq:EGrec2}) to general decompositions
$A_\Gamma=A_{\gamma_1}\oplus A_{\gamma_2}$ into connected saturated subgraphs (the proof of
\tref{thm:fits} is easily adapted to this),
then it is obvious that the minimal model $(\mathcal{P}=\mathcal{F}(\mathcal{C}_{div}(\Gamma)))$
provides exactly the right framework for renormalization. On the other hand,
on the maximal model $(\mathcal{P}=\mathcal{C}_{div}(\Gamma)),$ for which \lref{lem:main3} usually fails
to hold, unnecessary
subtractions are required if there are disjoint or, more generally, reducible divergent subgraphs. Locality
must then be imposed by additional conditions.
It can be shown that local renormalization schemes such as local minimal subtraction
can also be applied on the maximal (and all intermediate) models, as will be reported elsewhere.
\subsection{Hopf algebras of Feynman graphs}\label{ss:Hopf}
In this section we relate our previous results to the Hopf
algebras introduced for renormalization by Connes and Kreimer
\cite{Kreimer,CK2}, and generalized in \cite{BlKr}.
This is not entirely straightforward, see also the remarks at
the end of this section. Isolating suitable polynomials in masses and space-time derivatives, position space Green functions can be chosen to have a perturbative expansion in terms of logarithmic divergent coefficients. Thus, in summary, as long as worse than logarithmic divergences are avoided, the Hopf algebras for
renormalization in momentum space \cite{BlKr} and position
space are the same.
\\\\
Only the divergent collection $\mathcal{C}_{div}(\Gamma)$ and
the minimal building set
$\mathcal{P}=\mathcal{F}(\mathcal{C}_{div}(\Gamma))$ is
considered at this stage, and
\emph{irreducible} and \emph{nested} refer to this setting.
\begin{dfn}\label{dfn:graphiso}
Two Feynman graphs $\Gamma_1,\Gamma_2$ are \emph{isomorphic} if
there is an isomorphism between their exact sequences
(\ref{eq:newexactseq}) for a suitable orientation of edges.
\end{dfn}
\begin{lem} \label{lem:prephopf}
Let $\gamma\subsetneq\Gamma$ be divergent graphs where $\Gamma$
is connected and at most logarithmic. Let $t$ be an adapted
spanning tree for the nested set
$\mathcal{N}=\{\Gamma,\gamma\}.$ Then the isomorphism class of
$\Gamma//\mathcal{N}$ is independent of $t$ and
$\Gamma//\mathcal{N}$ connected, divergent and at most
logarithmic.
\end{lem}
In this case we write $\Gamma//\gamma$ for the isomorphism class of $\Gamma//\mathcal{N}.$\\\\
\emph{Proof.} Follows from \lref{lem:contrres} (ii),(iii) and
the definition of the quotient graph using $p_{t,s}.$
 \hfill $\Box$ \\\\
Let $\mathcal{H}_{FG}$ be the polynomial algebra over $\mathbb{Q}$ generated by
the empty graph (which serves as unit) and isomorphism classes
of connected, at most logarithmic, divergent graphs. There is
no need to restrict to graphs of a specific interaction, but
this can obviously be done by introducing external (half-) edges and
fixing the degree of the vertices. All subgraphs are now
understood to have vertex set $V_{\operatorname{eff}}.$
Products of linear generators of $\mathcal{H}_{FG}$ are
identified with disjoint unions of graphs. One defines
\begin{equation}\label{eq:ourdelta}
\Delta(\Gamma)= \sum_{\gamma\subseteq \Gamma} \gamma\otimes \Gamma//\gamma
\end{equation}
where in the sum only divergent subgraphs $\gamma$ are
understood, including the empty graph. The quotient graph
$\Gamma//\gamma$ is well-defined and a generator of
$\mathcal{H}_{FG}$
by \lref{lem:prephopf}. One extends $\Delta$ as an algebra homomorphism onto all of $\mathcal{H}_{FG}.$ \\\\
By the analysis of \cite[Section 2.2]{BlKr}, the map
$\Delta:\mathcal{H}_{FG}\rightarrow \mathcal{H}_{FG}\otimes\mathcal{H}_{FG}$
is coassociative. Note that divergent and at most logarithmic
implies one-particle-irreducible (core) as in \cite{BlKr}:
\begin{dfn}\label{dfn:core}
A graph $\Gamma$ is called \emph{core (one-particle
irreducible)} if
 $\dim$ $H_1(\Gamma\setminus e)<\dim H_1(\Gamma)$ for any $e\in
E(\Gamma).$
\end{dfn}
\begin{pro}\label{pro:coreirred}
A divergent, at most logarithmic graph $\Gamma$ is core.
\end{pro}
\emph{Proof.} If $\dim H_1(\Gamma\setminus e)=\dim H_1(\Gamma)$ for some $e\in
E(\Gamma)$ then
$\Gamma\setminus e$ would be worse than logarithmically divergent. \hfill $\Box$ \\\\
One can divide $\mathcal{H}_{FG}$ by the ideal $\mathcal{I}$
generated by all polynomials $\gamma-\prod \gamma_j$ where
$A_\gamma=A_{\gamma_1}\oplus \ldots\oplus A_{\gamma_j}$ is an
irreducible decomposition, as in \cite[Equation (2.5)]{BlKr}.
Indeed, if $\gamma$ is connected and
$A_{\gamma}=A_{\gamma_1}\oplus A_{\gamma_2}$ a decomposition
then $\gamma$ is a join: $E(\gamma)=E(\gamma_1)\sqcup
E(\gamma_2)$ and $V_{\operatorname{eff}}(\gamma_1)\cap
V_{\operatorname{eff}}(\gamma_2)=\{v\}.$ We refer then to
\cite[Equation (2.5)]{BlKr} for the complete argument that
$\mathcal{I}$ is a coideal. The quotient Hopf algebra is
denoted
$\overline{\mathcal{H}}_{FG}=\mathcal{H}_{FG}/\mathcal{I},$ and
we will use only this Hopf algebra in the following. It corresponds to
the minimal building set. The
antipode is denoted $S$ and the convolution product of linear
endomorphisms $f\star g=m(f\otimes g)\Delta.$ Note that a
connected divergent graph $\Gamma$ is primitive in the sense of
\dref{dfn:primitive} if and only if
$\Delta(\Gamma)=\emptyset\otimes
\Gamma+\Gamma\otimes\emptyset.$
\begin{thm} \label{thm:antipode} If $\Gamma$ is irreducible,
\begin{eqnarray*}
S(\Gamma) &=& \sum_{A_\Gamma\in\mathcal{N}} (-1)^{|\mathcal{N}|} \prod_{A_\gamma\in\mathcal{N}} \gamma//\mathcal{N},\\
\end{eqnarray*}
where the sum is over nested sets $\mathcal{N}$ wrt.~
$\mathcal{F}(\mathcal{C}_{div}(\Gamma)).$
\end{thm}
\emph{Proof.} Since the antipode satisfies
$S(\emptyset)=\emptyset$ and
\begin{equation*}
S(\Gamma)=-\sum_{\gamma\subsetneq \Gamma} S(\gamma)\Gamma//\gamma,
\end{equation*}
for $\Gamma$ irreducible, $\gamma$ divergent, one has
$S(\Gamma)=-\Gamma$ if $\Gamma$ is primitive. Let now $\Gamma$
be general irreducible. The sum over nested sets $\mathcal{N}$
wrt.~ $\mathcal{F}(\mathcal{C}_{div}(\Gamma))$ containing
$A_\Gamma$ can be written as a sum over proper divergent
subgraphs $\gamma$ of $\Gamma$ and nested sets $\mathcal{N'}$
wrt.~ $\mathcal{F}(\mathcal{C}_{div}(\gamma))$ containing the
irreducible components of $A_\gamma$ such that
$\mathcal{N}=\mathcal{N}'\cup \{A_\Gamma\}.$ By
\lref{lem:prephopf}, $\Gamma//\gamma=\Gamma//\mathcal{N},$ and
the statement follows by
induction. \hfill $\Box$ \\\\
By \tref{thm:poles} (ii)-(iii), the antipode $S$ describes thus the stratification of
the divisor $\mathcal{E}$ of $Y_\mathcal{P}.$ A similar (but weighted) sum is given
by $S\star Y$ where $Y$ is the algebra homomorphism $Y:\mathcal{H}_{FG}\rightarrow \mathcal{H}_{FG},$
$Y(\Gamma)=\dim H_1(\Gamma) \Gamma,$ see for example \cite{CK3}. This provides the link between the
scattering formula of \cite{CK3} and \tref{thm:poles} (iii), and we refer to future work for the details. \\\\
In the case of dimensional regularization and minimal subtraction, one considers algebra homomorphisms
from $\mathcal{H}_{FG}$ into an algebra of Laurent series in the regulator, and a projector onto the finite
part of the series, in order to describe the renormalization process \cite{Kreimer,CK2,CK3}.
In our framework, the Hopf algebra is encoded in the geometry of the divisor.
The renormalization process is simply to approach the divisor and perform the simple subtraction along the
irreducible components, and to take the product of the subtracted factors where the components intersect.
Therefore the renormalization schemes studied here (\ref{eq:genms})-(\ref{eq:genmomprec}) can again be described
by the antipode twisted with a subtraction operator. The latter depends however on local information as opposed to global
minimal subtraction. A comprehensive discussion of the difference between local renormalization schemes as
described here and (global) minimal subtraction is reserved for future work.
\\\\
\emph{Remarks.} The role of the Connes-Kreimer Hopf algebras in
Epstein-Glaser renormalization was previously discussed in
\cite{GBL}, \cite{Pinter2} and \cite{BK}. The third paper,
which is about entire amplitudes and uses rooted trees,
relies on a quite symbolic notation which is now justified by the results of the previous sections.
A general flaw in the first paper \cite{GBL} is revealed in
the introduction of \cite{Pinter2}. On the other hand the
coproduct in the second paper \cite{Pinter2} does not seem to be
coassociative the way it is defined. As a counterexample
consider the cycle on four vertices plus two additional edges
between a pair of vertices. This can be repaired by
introducing irreducible, core or at most logarithmic and
saturated subgraphs as it is done here. See \cite[Section
2.2]{BlKr} for a general discussion for which classes
$\mathcal{P}$ of graphs the map
$\Delta(\Gamma)=\sum_{\onatop{\gamma\subseteq\Gamma}{\gamma\in\mathcal{P}}}\gamma\otimes
\Gamma//\gamma$ has a chance of being coassociative.
\subsection{Amplitudes, non-logarithmic divergences and regulators} \label{ss:quadratic}
In this section we briefly sketch how to extend our
previous results, which are so far confined to single graphs
with at most logarithmic divergences, to a more general class
of graphs. Indeed, if one considers amplitudes, or vacuum
expectation values of time-ordered products in the
Epstein-Glaser framework, one wants to regularize and
renormalize sums of Feynman distributions simultaneously, and
some of them will obviously have worse than
logarithmic singularities. \\\\
For an introductory discussion of non-logarithmic divergences
the reader is referred to \cite[Section 7.4]{BlKr},
\cite[Section 5]{Collins}. The general philosophy is to reduce
seemingly non-logarithmic (quadratic etc.) divergences to
logarithmic ones by isolating contributions to different terms
in the Lagrangian (such as wave function renormalization, mass
renormalization); and by projecting onto a subspace of
distribution-valued meromorphic functions where local terms
with infrared divergences are discarded. This shall only be
sketched at the example of the primitive graph
\begin{equation*}
\Gamma = \gammaeiner, \quad \underline{u}_\Gamma(x)|d^6x|=\frac{|d^6x|}{x^8}
\end{equation*}
in $d=6$ dimensions, which is quadratically divergent. By
(\ref{eq:xsa}), $\underline{u}^s_\Gamma$ has relevant
poles\footnote{Just as in dimensional regularization, the
(linear) divergence at $s=7/8$ is not detected
 by the regulator.} at $s=\frac{3}{4}$ and $s=1.$
Indeed, by (\ref{eq:xsa}),
\begin{equation}\label{eq:wgammaquadr}
\tilde w_\Gamma^s|dy| = \frac{f^s_\Gamma(y)|dy|}{|y^0|^{8s-5}}  = -\left(\frac{\delta_0(y^0)}{4s-3}+
\frac{\delta''_0(y^0)}{8(s-1)}-|y^0|^{5-8s}_{fin}\right)f^s_\Gamma(y)|dy|.
\end{equation}
Note that neither the residue at $s=\frac{3}{4}$ nor
$|y^0|^{5-8s}_{fin}f^s_\Gamma$ is globally defined as a
distribution density. One would like to work in a space of
distributions where $w_\Gamma$ is equivalent to a linear
combination of distribution densities with at most logarithmic
singularities, having only a pole at $s=1.$ If one disposes of
an infrared regulation such that the so-called adiabatic limit vanishes
\begin{equation}\label{eq:dimregtrick}
u^s_\Gamma[\underline{1}]=0
\end{equation}
one can subtract $u_\Gamma^s[\underline{1}]\delta_0$ from
(\ref{eq:wgammaquadr}) without changing it:
\begin{eqnarray*}
\tilde w_\Gamma^s|dy| &=& w_\Gamma^s- \delta_0(y^0)\int_{\mathcal{E}} \tilde w_\Gamma^s(z) |dz| \\
&=& -\left(\frac{\delta_0(y^0)}{4s-3}+
    \frac{\delta''_0(y^0)}{8(s-1)}-|y^0|^{5-8s}_{fin}\right)f^s_\Gamma(y)|dy|\\
&& -\delta_0(y^0)\left(-\frac{1}{4s-3}+\mbox{holomorphic terms}\right),
\end{eqnarray*}
which kills the pole at $s=\frac{3}{4}$ and leaves a linear
ultraviolet divergence. Using similar subtractions of zero the
linear divergence may then be reduced to logarithmic ones and
convergent terms, again at the expense of introducing infrared
divergent integrals which vanish however in a quotient space
where $u^s_\Gamma[\underline{1}]=0$ for all $\Gamma.$  We have
not worked out the general case, but dimensional regularization
suggests that it can be done consistently. Indeed, the idea
(\ref{eq:dimregtrick}) can be traced back to the ''identity''
\begin{equation}\label{eq:dimregmystery}
\int d^d k k^{2\alpha}= 0, \quad \alpha \mbox{ arbitrary}
\end{equation}
in momentum space dimensional regularization, see also
\cite[Sections 4.2, 4.3]{Collins}, \cite[Remark 7.6]{BlKr}.
Equation (\ref{eq:dimregmystery}) is a consequence of the fact
that dimensional regularization balances
ultraviolet and infrared divergences, using only one regulator $d.$ \\\\
A complete treatment of non-logarithmic singularities and
entire amplitudes is reserved for future work, as well as a
more general study of regularization methods, such as
dimensional regularization, in position space. Whereas
the analytic regularization used in this paper is based on raising the
propagator to a complex power, dimensional regularization
would replace $d$ by $d-2s,$ $s\in\C$ in (\ref{eq:dfnu0}). This can be seen
to lead to very similar expressions, simplifying the constants
in (\ref{eq:defresidue}), (\ref{eq:factorizations}) etc.
\section{Final remarks}
Pulling back the Feynman distribution onto a smooth model with normal
crossing divisor seems an obvious thing to do for an algebraic geometer.
Less obvious is maybe the question which kind of smooth model is useful
and how renormalization depends on the choice of a model.
Before addressing this question let us first point out what changes if
spherical instead of projective blowups are used (as in \cite{AS} and in the figures in section
\ref{s:models}) -- this choice is possible since we are only interested in blowing up a real
locus. In a spherical blowup of a point in some $\R^m,$ the exceptional locus is a codimension one
sphere instead of a codimension one projective space. In order to adjust to this different situation,
one simply introduces for example around equation (\ref{eq:simplechart}) twice the number of charts,
say $\rho_i^\pm: U_i^\pm\rightarrow\mathcal{M}\times S^{m-1},$ where now $U_i^\pm\subset U_i$ is the
half-space $y_i\ge 0$ resp.~ $y_i\le 0,$ and replaces $[z_1,\ldots,z_n]$ by $[z_1,\ldots,z_n]_+$ which means that
only an action of $\R_+$ is divided out. (In fact, the choice of sign made in (\ref{eq:simplechart}) is exactly the one obtained from identifying
two antipodal charts of the spherical blowup in the right way so as to have $\mathcal{E}$ oriented.)
This makes the spherical blowup $Y$ a manifold with boundary. \\\\
As is well-known, the spherical De Concini-Procesi models, and in particular the spherical Fulton-MacPherson
compactification \cite{AS}, are manifolds with corners since they are submanifolds of a product of manifolds
with boundary (compare (\ref{eq:closure})). Equations (\ref{eq:closure}),
(\ref{eq:blowupcoordinates}) etc. have to be modified accordingly. The corners are the expense to be paid in order
to get orientability, and one does not seem to gain or lose much by trading one for the other. For the simple kind
of propagator $u_0$ studied in this paper the analysis is more or less the same, taking into account that since
the sphere is the double cover of the projective space, the spherical residues come with a factor 2 compared to the
projective residues. \\\\
It is obvious that the Fulton-MacPherson compactification $M[n]$ (minimal De Concini-Procesi model for $M^{V_0}_{sing}(K_n)$)
is good for all Feynman distributions at the same time, and therefore for entire amplitudes, which are sums of
Feynman distributions. The combinatorics of the nested sets for $M[n]$ resemble the Hopf algebra of rooted trees \cite{Kreimer,Kreimer3}.
We chose to work with the graph-specific models because we wanted to make the connection to
the Hopf algebra of Feynman graphs and to Zimmermann's forest formula explicit.
One difference in renormalizing a Feynman graph $\Gamma$ on $M[n]$ and on the other hand on $Y_{\mathcal{F}(C_{div}(\Gamma))}$ is that
in the first case (\ref{eq:EGrec2}) really holds only for disjoint unions of subgraphs $\gamma_1,$ $\gamma_2$, whereas in the
second case an implicit renormalization condition ''(\ref{eq:EGrec2}) also for more general decompositions (joins) $V_{\operatorname{eff}}(\gamma_1)\cap V_{\operatorname{eff}}(\gamma_2)=\{v\}$'' is introduced. See also the corresponding remark (v) in \cite[1.3]{BlKr}. If one does not like this
condition, one can use instead a non-minimal, intermediate building set where certain reducibilities are allowed.\\\\
In the recent paper \cite{BlKr}, which studies the Schwinger parametric representation of Feynman integrals,
a toric compactification of the complement of certain coordinate linear spaces is used in order to understand the renormalized
Feynman distribution as a period of a limiting mixed Hodge structure.
We also mention \cite{Marcolli,BognerWeinzierl1} for recent related research in the parametric representation,
\cite{Hollands} with regard to the operator-product expansion, and \cite{Nikolov} for cohomological aspects.\\\\
Beyond the open problems already mentioned there arise three immediate questions. The first is to find the right
analytic framework in order to generalize our results to arbitrary propagators
on manifolds, with a more versatile notion of regularization than the ad-hoc analytic
regularization used here. The second question is how the motivic description of renormalization
in \cite{BlKr} is related to our approach. And finally it remains to carry out a general study of finite
renormalization and the renormalization group in the geometric context we have introduced.
\bibliographystyle{alpha}
\bibliography{lit}
\end{document}